\documentclass[prb,amsmath,amssymb,superscriptaddress,twocolumn]{revtex4}
\usepackage{graphicx}
\usepackage[usenames]{color}
\usepackage {amsmath}
\usepackage {amsfonts}
\usepackage {amssymb}
\usepackage {mathrsfs}
\usepackage {bm}
\usepackage {mathbbol}
\usepackage[usenames]{color}
\usepackage {array}

\newcommand{\br}{{\bf r}}

\newcommand{\bk}{{\bf k}}

\newcommand{\bt}{{\bf t}}

\newcommand{\bp}{{\bf p}}

\newcommand{\bA}{{\bf A}}

\newcommand{\bL}{{\bf L}}
\newcommand{\bM}{{\bf M}}

\newcommand{\bP}{{\bf P}}
\newcommand{\bR}{{\bf R}}
\newcommand{\bS}{{\bf S}}
\newcommand{\bX}{{\bf X}}

\DeclareMathAlphabet{\mathpzc}{OT1}{pzc}{m}{it}

\newcommand {\rmd}{{\rm d}}

\newcommand{\be}{\begin{eqnarray}}
\newcommand{\ee}{\end{eqnarray}}
\newcommand{\bw}{\begin{widetext}}
\newcommand{\ew}{\end{widetext}}

\newcommand{\diag}{{\rm diag}}

\newcommand{\bbone}{{\mathbb 1}}
\newcommand{\DDelta}{{\mathbb \Delta}}

\newcommand {\thalf}{{\tfrac 1 2}}
\newcommand {\btau}{{\boldsymbol \tau}}
\newcommand {\bdelta}{{\boldsymbol \delta}}

\begin{document}
\title{Space group symmetry, spin-orbit coupling and the low energy effective Hamiltonian for iron based superconductors}
\author{Vladimir Cvetkovic}
\author{Oskar Vafek}
\affiliation{National High Magnetic Field Laboratory and Department
of Physics,\\ Florida State University, Tallahasse, Florida 32306,
USA}

\date{\today}
\begin{abstract}
We construct the symmetry adapted low energy effective Hamiltonian for the electronic states in the vicinity of the Fermi level
in iron based superconductors. We use Luttinger's method of invariants, expanding about $\Gamma $ and $\bM$ points in the
Brillouin zone corresponding to two iron unit cell, and then matching the coefficients of the expansion to the 5- and 8-band models.
We then use the method of invariants to study the effects of the spin-density wave order parameters on the electronic spectrum, with and
without spin-orbit coupling included. Among the results of this analysis is the finding that
the nodal spin-density wave is unstable once spin-orbit coupling is included.
Similar analysis is performed for the $A_{1g}$ spin singlet superconducting state. Without spin-orbit coupling there is one
pairing invariant near the $\Gamma$ point, but two near the $\bM$ point. This leads to an isotropic spectral gap at the hole
Fermi surface near $\Gamma$, but anisotropic near $\bM$. The relative values of these three parameters determine whether
the superconducting state is $s_{++}$, $s_{+-}$, or nodal.
Inclusion of spin-orbit coupling leads to additional mixing of spin triplet pairing, with one additional pairing parameter near
$\Gamma$ and one near $\bM$. This leads to an anisotropic spectral gap near {\it both} hole and electron Fermi surfaces, the latter no longer cross, but rather split.
\end{abstract}

\maketitle

\section{Introduction}

The parent state of most iron based superconductors is an itinerant spin-density wave which, upon
doping \cite {KamiharaJACS2006, KamiharaJACS2008, ChenNature2008, SefatPRL2008, SefatPRB77_2008, SefatPRB78_2008, ChuPRB2009}
or pressure \cite{TakahashiJPhysScnJpn2008, HamlinJPhysCondMat2008}, gives rise to superconductivity.
Unlike their copper oxide counterparts, which are doped Mott insulators with an odd number of electrons per
unit cell, such iron based superconductors are compensated semi-metals with an even number of electrons
per unit cell \cite{CvetkovicTesanovicEPL2009, TesanovicPhysics2009}.
The correlation effects in the iron based superconductors appear to be more significant than in conventional metals,
but not as severe as in copper oxide superconductors \cite{QazilbashNatPhys2009}. Unlike in conventional superconductors,
the mechanism of superconductivity is believed to be different from the electron-phonon interaction driven pairing \cite{BoeriPRL2008, MazinPRL2008, SubediPRB2008, WangScience2011, BasovChubukovNatPhys2011}.

This paper is motivated by the need to develop a low energy effective theory which successfully describes the
electronic states in the vicinity of the Fermi level. Such low lying states which are responsible for a number of physical
characteristics of these materials, such as electrical and thermal conductivity, low temperature specific heat and
magnetic susceptibility, are most affected by spin or charge ordering, or by superconductivity.
Currently, the models which successfully capture these states are either based on the 5-band tight-binding approach
or by starting with it, diagonalizing, and then working in the resulting "band basis" as if it were a continuum model. The
former has the disadvantage of being impractical for studying the effects of an externally applied magnetic field, or, simply,
because it necessitates working with a large matrix. The latter also cannot be minimally coupled to the external magnetic field.
Moreover, it has an additional disadvantage, which stems from the $\bk$-space degeneracy of the hole bands at the
$\Gamma$-point and the electron bands at the $\bM$-point, and which results in $\bk$-space non-analyticity of the
single particle wave-functions. Such non-analyticity translates into non-local effective electron-electron interaction,
making the method impractical for studying interaction effects.

In contrast, the approach which we develop here allows minimal coupling to the external magnetic vector potential,
and maintains locality of the electron-electron interactions, provided they start out being of finite range in the lattice model. The
method is based on the theory of invariants used by Luttinger \cite{Luttinger1956} to study Si and Ge. The connection to the
methods used to study semiconductors \cite{BirPikus} is natural given that the parent state of iron-based superconductors is a multi-band
semi-metal. The interesting new aspect is the manifest presence of a broken symmetry, which can be readily included
within this approach. Furthermore, the space group of the iron-pnictogen or iron-chalcogen plane (we use ``iron plane'' from now on)
is non-symmorphic because it contains an $n$-glide plane: a mirror reflection about the iron plane followed by the translation
along the {\it half} of the unit cell diagonal. This has profound consequences on the nature of the irreducible representations
of the group of the wave-vectors at the Brillouin zone edges or corners, precisely where the electron pockets appear
\cite{LebeguePRB2007, MazinPRL2008, HaulePRL2008, GraserPRB2010,  YinNatMat2011,
DingEPL2008, TerashimaPNAS2009, LuShenPhysicaC2009, CarringtonPhysicaC2009, LiuNatComm2012}.
As we show below, the group of the wave-vector at $\bM$ has only two dimensional irreducible representations, which means that,
not including spin, at $\bM$ all Bloch states are doubly degenerate. This is unlike $\bk$-points inside the Brillouin zone, such as $\Gamma$-point
where the hole pockets appear, where the irreducible representations correspond to a known point group and therefore always
have one dimensional representations --- if such point group is non-abelian, then of course it also has higher dimensional,
irreducible representations, but it always has {\it some} one dimensional irreducible representations.
The states which cross the Fermi level near $\Gamma$ derive from the Fe $d_{xz}$ and $d_{yz}$ orbitals and transform under the $E_g$
representation of the group of the wave-vector which is isomorphic to ${\bf D}_{4h}$. This representation is two dimensional, guaranteeing
the degeneracy of these two states above any structural or magnetic transition.

The method used here allows us to analyze the effects of the atomic spin-orbit interaction, $\lambda \bL \cdot \bS$, which, in iron,
has been reported \cite{TiagoPRL2006} to be $\lambda \sim 80meV$. Such a value is larger than the temperature scale
associated with both the magnetic and superconducting ordering. We find that once the spin-orbit coupling is included,
one cannot avoid using the two iron unit cell: the two electron Fermi surfaces near the $\bM$-point no longer intersect, but
rather split due to avoided level crossing. Moreover, the degeneracy at the $\Gamma$-point is lifted through a spin-orbit term
which is analogous to the one written by Kane and Mele in graphene \cite{KaneMelePRL2005}. When the time reversal
symmetry and a center of inversion are present, as in the normal non-magnetic state, all bands are Kramers degenerate. At the
$\bM$-point, the states are four-fold degenerate, due to a single four dimension double-valued irreducible representation of the space
group \cite{BradleyCracknell}. In the immediate vicinity of such four-fold degeneracy, the bands disperse linearly in momentum, as
for a massless Dirac particle. For larger momentum deviation, the bands disperse upward.

We also analyze the electronic spectrum in the presence of the collinear spin density wave \cite{delaCruzNature2008,GoldmanPRB2008}. Such a
state breaks time reversal and inversion symmetry, but not their product. Therefore, the bands remain Kramers degenerate even
below the spin-density wave ordering transition temperature. For weak spin-density wave order, we find that the electron
and hole Fermi surfaces reconstruct, leaving behind several smaller pockets. In the absence of spin-orbit coupling, we find 6
nodal points below the Fermi level, similar to Y. Ran {\it et.al.} \cite{RanPRB2009}. However, once the spin-orbit interaction
is included, the degeneracy at the nodal points is lifted. This can be understood as a consequence of (a generalization of)
the Wigner-von Neuman argument for Kramers degenerate bands. As a result, we conclude that for a putative strong spin-density
wave, where Y.\ Ran {\it et.al.}\cite{RanPRB2009} predicted the gapless spectrum with nodal (Dirac) points, the spin-orbit
coupling results in a fully gapped spectrum.
In the case of co-planar spin-density wave with four-fold symmetry, recently reported in Ref.\ \onlinecite{AvciarXiv2013},
the center of inversion is lost and the Kramers degeneracy is lifted. The resulting spectrum therefore displays split Fermi surfaces.
Based on the symmetry invariants that can be constructed, one can readily see that a transition
from a normal non-magnetic state into a  collinear magnetic state necessarily induces orthorhombic distortion, but not
vice versa,  in agreement with, for example, Refs.\ \onlinecite{YildirimPRL2008, FangPRB2008, XuPRB2008, BarzykinGorkovPRB2009,
FernandesPRB2012}. A transition from a collinear spin-density wave into a coplanar spin-density wave with
4-fold symmetry is necessarily of first order, in agreement with Ref.\ \onlinecite{AvciarXiv2013}.
The absence of the Kramers degeneracy in the coplanar spin-density wave state  suggests interesting
repercussions on the microscopic coexistence of such a state with superconductivity, similar to Ref.\ \onlinecite{GorkovRashbaPRL2001}.

This method also allows us to analyze the symmetry of the superconducting pairing states and the Fermion bilinear terms they
induce in the Hamiltonian. In the absence of spin-orbit coupling, at the $\Gamma$ point there is a single, momentum independent,
spin singlet $A_{1g}$ invariant, whose strength we parameterize by $\Delta_\Gamma$, which apart from charge $U(1)$, does not
break any other symmetry of the crystal. This term leads to an isotropic spectral gap along the two (anisotropic) hole Fermi
surfaces. Similarly, at the $\bM$ point, there are two such invariants, whose strength we parameterize by $\Delta_{\bM 1}$
and $\Delta_{\bM 3}$. Depending on the relative ratio of $\Delta_{\bM 1}$ and $\Delta_{\bM 3}$ we find the spectrum to be
either gapped or nodal at the electron Fermi surfaces. Once spin-orbit coupling is included, there is an additional, momentum
independent, invariant at the $\Gamma$ point. We parameterize it by $\Delta_{\Gamma t}$. While being spin triplet, it nevertheless
respects all the symmetries respected by $\Delta_{\Gamma}$. The inclusion of this term makes the spectral gap on the hole
Fermi surfaces anisotropic. At the $\bM$-point, the spin-orbit coupling also permits one additional momentum
independent triplet invariant, which contributes to the anisotropy of the spectral gap along the two, now split, Fermi surfaces.

This paper is organized as follows: in the next subsection we summarize the key results and point the reader to the equations corresponding to them.
In Section \ref{SecP4nmm}, we present an extensive analysis of the irreducible representation of the space group of iron based superconductors,
provide the product tables of the irreducible representations, illustrate the construction of the symmetry adapted functions
at the $\bM$-point, and analyze the symmetry of the Bloch states in the vicinity of the Fermi level.
In Section III, we construct our low energy effective Hamiltonian in terms of the six component Fermi ``spinor'' and determine
the coefficients of the symmetry allowed terms to the 5-band tight-binding models of Ref.\ \onlinecite{KurokiPRL2008} and the 8-band
model of Ref.\ \onlinecite{CvetkovicTesanovicEPL2009}. The resulting Hamiltonian is very easy to handle and
effectively requires diagonalization of only a $2\times 2$ matrix. We also include the effects of spin-orbit
coupling and critically compare our approach to the two- and three-orbital models used in the literature.
In addition, we use our machinery to build the symmetry allowed four-Fermion contact interaction terms.
In Section \ref{SecSDW} we study the consequences of the spin-density wave ordering on the electronic
spectra, and similarly in Section V we study the superconductivity.
Section VI is devoted to discussion. The mathematical details are presented in the appendices.

\subsection {Summary of the key results}

The most important results of Section \ref{SecP4nmm} are the irreducible representation of the space group $P4/nmm$
at the $\Gamma$- and $\bM$-points. These are summarized in Tables \ref {tab:Gamma generators irreps} and \ref {tab:M generators irreps}.
The rest of the paper builds on these results.

Our key results for the low-energy effective theory are the definition of the low energy effective ``spinor'' $\psi$ in Eq. \eqref{eq:spinor}
whose components transform under the irreducible representations discussed in Section \ref{SecP4nmm}. Additionally,
Eqs.\ \eqref{eq:hMpm}-\eqref{eq:hGamma} (no spin-orbit), and
Eqs.\ \eqref {eq:hGammaso}-\eqref {eq:hMso} (with spin-orbit), correspond to the continuum description --- consistent with
the underlying crystalline symmetry --- of the electronic states which are most affected by either magnetic or superconducting ordering.
In the same section, the contact four-fermion interaction term is given in Eqs.\ \eqref {Hint0dir}-\eqref {HintM}.

Section \ref{SecSDW} presents the symmetry analysis of spin density wave order parameters together with
the electronic spectrum and the Fermi surfaces for the collinear spin-density wave, Fig.\ \ref{FigSDWdispersion},
and co-planar four-fold spin-density wave, Fig.\ \ref{FigC4dispersion}. In Subsection \ref{SubsecKramers} the conditions for
the Kramers degeneracy are analyzed. In Subsection \ref{SubsecWvN} we provide the generic, symmetry based, arguments
for the presence or absence of degeneracies in the spectrum using a generalization of Wigner-von Neumann analysis.

The main finding in section \ref {SecSuperconductivity} is that an $s$-wave superconducting state can be well described with only
three  pairing terms, Eqs.\ \eqref {HSCGamma} and \eqref {HSCM}. Although these terms
are $\bk$-independent, we find that while the pairing gap on the hole Fermi surfaces is isotropic, Eq.\ \eqref{EkGammaSCsinglet}, it is
$\bk$-dependent and anisotropic on the
electron Fermi surfaces with the possibility of gap nodes, Eq.\ \eqref {DeltaSCband1}. With the spin-orbit coupling, two additional, $\bk$-independent, spin triplet pairing
terms, Eqs.\ \eqref {HSCGammatriplet} and \eqref {HSCMtriplet}, are  allowed. These lead to gap anisotropy on the hole Fermi surfaces, as seen in
Eq.\ \eqref{DeltaGammaFS}. The  Eq.\ \eqref{DeltaMFSseries} gives the gap on electron Fermi surfaces in the presence of
spin-orbit coupling and the concomitant spin triplet admixture.

\section {Irreducible representation of the space group $P4/nmm$}
\label{SecP4nmm}

Our main analytical tool in this paper is the space group symmetry  together with its irreducible representations.
We use the method of C.\ Herring \cite {Herring1942} to construct the irreducible representations of
(non-symmorphic \cite {InuiTanabeOnodera}) $P4/nmm$ at the $\Gamma$- and $\bM$-points,
and provide the key steps in what follows. The results thus obtained  are first used to study the symmetry properties of the physical
states and operators in iron-pnictides; they are then used to construct the
low-energy effective model, the order parameters, and to study the physical consequences
of spin-orbit coupling on both the spin-density wave state and the superconducting state.

Throughout this paper, we use the Seitz notation for symmetry operations, $\{ g | \btau \}$, where $g$ is a point group operation
(rotation, reflection, etc.), which keeps the coordinate center invariant, followed by a translation
by vector $\btau$. The product rule is
\be
  \{ g_1 | \btau_1 \} \{ g_2 | \btau_2 \} = \{ g_1 g_2 | \btau_1 + g_1 \btau_2 \}, \label{Seitzproduct}
\ee
where $g_1 \btau_2$ is the result of the point group element $g_1$ acting on vector $\btau_2$. The group of all
{\em full unit cell} (or integer) translations,
\be
  {\mathcal T} = \Big \{ \left \{ e | m_1 m_2 m_3 \right \} \Big | m_1, m_2, m_3 \in {\mathbb Z} \Big \}, \label{Tgroup}
\ee
is an invariant subgroup of the space group, i.e., if we take any element $\{ g | \tau \}$ of $P4/nmm$, then for any
$m_i$'s, the product $\{ g | \btau \} \{ e | m_1 m_2 m_3 \} \{ g | \btau \}^{-1}$ is also an integer translation.
The three numbers $m_i$ represent a translation by $\bt  = \sum_{i=1}^3 m_i \hat a_i$.

$P4/nmm$ is a non-symmorphic group. Therefore, it is impossible to choose a coordinate center such that every
symmetry operation can be taken to be a point group operation followed by an integer translation, i.e., as
$\{ g | m_1m_2 m_3 \}$ with all $m_i$ integers. This is because of the presence of the $n$ glide plane
which involves a fractional (non-integer) translations by $\btau_0 = \left ( \thalf \thalf 0 \right )$, combined with the $ab$-plane mirror.

\begin{figure}[h]
\begin{center}
\includegraphics[width=0.48\textwidth]{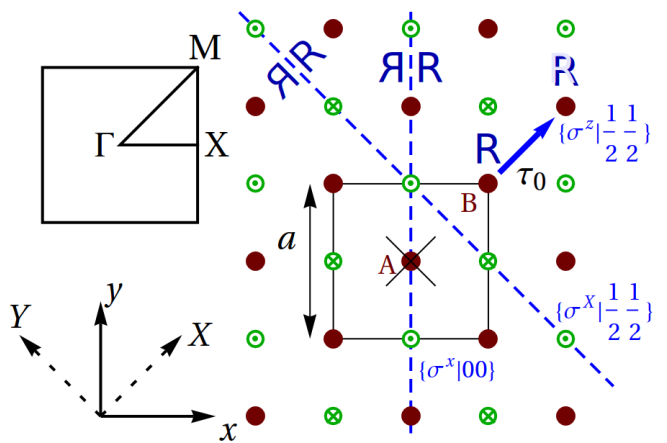}
\end{center}
  \caption{Bottom left: the coordinate systems used in this paper. Right: the lattice of iron based superconductors, represented here by a single layer.
    Iron atoms (dark red) are split into two sublattices, A and B. Pnictide atoms (green) sit at
    iron-plaquette centers; one sublattice of the pnictide atoms is puckered above, and the other sublattice
    is puckered below the layer. One unit cell, centered on the coordinate origin (crosshairs), is outlined by the black square.
    The three generators of $\bP_\Gamma$ are shown in blue. The vertical mirror $\{ \sigma^x | 00 \}$  passes
    through the coordinate origin. The vertical mirror $\sigma^X$ also passes through the coordinate origin, however,
    the symmetry of the lattice requires this mirror to be followed by a translation by $\btau_0$. The
    combined operation, $\{ \sigma^X | \thalf \thalf \}$, is equivalent to a mirror which passes through pnictide atoms. The third generator is an $n$-glide
    mirror, $\{ \sigma^z | \thalf \thalf \}$ and it has no fixed points. The action of each mirror is illustrated on an `R' symbol. Under the operation of
    any of the first two generators, a mirror image of `R' is created. The third generator, acting on the `R' above the plane,
    glides this symbol and puts it below the plane, which we represent by a hollow `R'. Top left: the first Brillouin zone for this lattice.}\label{FigUnitCell}
\end{figure}

There are 8 point group operations, $\{ g | 000 \}$, centered on an iron atom (see Fig.\ \ref{FigUnitCell}),
forming the group ${\bf D}_{2d}$. The group can be generated by two elements: $\{ \sigma^x | 000 \}$, a mirror reflection about
the  $yz$-plane, and $\{ S_4 | 000 \}$, a $90^\circ$ rotation about the $z$ axis, followed by the mirror reflection about the $xy$-plane.
In addition to the eight point group operations, we also have an inversion followed by the fractional translation, $\{ i | \thalf \thalf 0 \}$.
Together they can be used to generate all the elements of $P4/nmm$.

When we combine the 8 elements of ${\bf D}_{2d}$, with the  8 symmetry operations obtained by
multiplying $\{ i | \thalf \thalf \}$ with the elements of ${\bf D}_{2d}$, we find that this set is not closed
under multiplication, and therefore, as is, it cannot form a group. (From now on, to simplify the notation, we omit the translation along the c-axis since it is $0$ throughout the paper). However, if we think of the product as
being defined modulo integer translations, then these 16 operations form a group, which we denote by $\bP_\Gamma$.
More precisely \cite{InuiTanabeOnodera}, $\bP_\Gamma$ is the factor group $(P4/nmm) / {\mathcal T}$.

In order to find the irreducible representations of the space group, we first note that
the irreducible representations for the Abelian subgroup of integer translations, ${\mathcal T}$, are specified by wave-vector $\bk$, such that
$D_\bk \left ( \{ e | \bt \} \right ) = e^{i \bk \cdot \bt}.$
As usual, two irreducible representations, $D_{\bk_1}$ and $D_{\bk_2}$, are equivalent if vectors $\bk_1$ and $\bk_2$
differ by a reciprocal lattice vector, hence $\bk$'s can be taken only in the first Brillouin zone.

Since any element of $P4/nmm$ can be casted as a product of an integer translation and an element in $\bP_\Gamma$,
it is sufficient to determine how an object (e.g., a Bloch state) with momentum $\bk$ transforms under all
the elements of $\bP_\Gamma$. For each $\bk$, we first define the star of $\bk$ as the set of all distinct momenta $g \bk$, where
$\{ g | \btau \}$ are all the elements of $\bP_\Gamma$. The construction of irreducible representations of the space group is
different when $\bk$ lies inside or on the border of the Brillouin zone.

When $\bk$ is inside the Brillouin zone, the little
co-group of wave-vector $\bk$, which we denote a  $\bP_\bk$,  is defined  as the subgroup of $\bP_\Gamma$ which keeps
vector $\bk$ invariant up to a reciprocal lattice vector. The irreducible representations of the space group are then
labeled by wave-vector $\bk$ and another label for the irreducible representation of $\bP_\bk$. Any element
of the space group which is a product of an integer translation and an element of $\bP_\bk$ is represented as
\be
  D_{\bk, i} \left ( \{ g | \btau = \bt + \btau' \} \right ) &=& e^{i \bk \cdot \bt} D_{\bk, i} \left ( \{ g | \btau' \} \right ) \nonumber \\
    &=& e^{i \bk \cdot \btau} D_i  \left ( \{ g | \btau' \}  \right ), \label{spacegrouphomoinside}
\ee
where $D_i$ is a representation of $\bP_\bk$, and $\bt$ is an integer translation vector chosen such that
$\{ g | \btau' \}$ is an element of $\bP_\bk$. Therefore, in order to determine symmetry properties of a Bloch
state with momentum $\bk$, it is sufficient to determine how it transforms under the members of the
little co-group $\bP_\bk$, i.e., which irreducible representation of $\bP_\bk$ it belongs to. Notice that
the symmetry operations which do not keep $\bk$ invariant change the momentum of a Bloch
state from $\bk$ to some $\bk'$ which is in the star of $\bk$. Hence an irreducible representation
at $\bk$ also determines the symmetry properties for states with $\bk'$ in the star of $\bk$, and
once we find all the irreducible representations at some $\bk$, we have automatically found the
irreducible representations at all other wave-vectors that belong to the star of $\bk$. (This is also
true when $\bk$ is at the border of the Brillouin zone.)

If the momentum $\bk$ labeling an irreducible representation of the space group lies at the border
of the Brillouin zone, and the space group is symmorphic, then the construction of the irreducible
representations follows the same steps as when $\bk$ is inside the Brillouin zone.

When the momentum $\bk$ is at the border of the Brillouin zone, and the space group is non-symmorphic,
the irreducible representations of the space group at such a wave-vector cannot be constructed as
described previously due to the presence of fractional translations. To demonstrate this, consider
two symmetry operations, $\{ g_1 | \btau_1 \}$ and $\{ g_2 | \btau_2 \}$ and an irreducible representation
$D_{\bk, i}$. The representation of the product of the two symmetry operations, according to Eq.\ \eqref {spacegrouphomoinside}, is
\begin {align}
  D_{\bk, i}& \left ( \{ g_1 | \btau_1 \} \{ g_2 | \btau_2 \} \right ) = D_{\bk, i} \left ( \{ g_1 g_2 | \btau_1 + g_1 \btau_2 \} \right ) \nonumber \\
  &= e^{i \bk \cdot (\btau_1 + g_1 \btau_2 )} D_i \left ( \{ g_1 g_2 | \btau_1 + g_1 \btau_2 \} \right ). \label{Dg1g2}
\end {align}
On the other hand, if we take the representations for each symmetry operation and multiply them, we get
\begin {align}
  D_{\bk, i} & \left ( \{ g_1 | \btau_1 \} \right ) D_{\bk, i} \left ( \{ g_2 | \btau_2 \} \right ) \nonumber \\
  &= e^{i \bk \cdot \btau_1} D_{i} \left ( \{ g_1 | \btau_1 \} \right ) e^{i \bk \cdot \btau_2} D_{i} \left ( \{ g_2 | \btau_2 \} \right ) \nonumber \\
  &= e^{i \bk \cdot ( \btau_1 + \btau_2 )} D_i \left ( \{ g_1 g_2 | \btau_1 + g_1 \btau_2 \} \right ). \label{Dg1Dg2}
\end {align}
These two expressions, Eqs.\ \eqref {Dg1g2} and \eqref {Dg1Dg2}, must be equal, which is true if
\be
  e^{i \bk \cdot (\btau_1 + g_1 \btau_2 )} = e^{i \bk \cdot ( \btau_1 + \btau_2 )} \Leftrightarrow e^{i \btau_2 \cdot (\bk - (g_1)^{-1} \bk )} = 1, \label{expsymmorphic}
\ee
for any $g_1$ and $\btau_2$. This is the case when the space group is symmorphic ($\tau_2 = 0$), or $\bk$ lies inside the Brillouin zone
(then $(g_1)^{-1} \bk = \bk$). However, when the space group is non-symmorphic, {\em and} the wave-vector $\bk$ sits on
the Brillouin zone border, then $(g_1)^{-1} \bk$ and $\bk$ can differ by a reciprocal lattice vector, in which case Eq.\ \eqref {expsymmorphic}
is violated for some $\btau_2$'s.

The irreducible representations of $P4/nmm$ for $\bk$'s on the Brillouin zone boundary are
therefore constructed differently \cite {Herring1942}. We discuss the irreducible representations of the space group at wave-vector
$\bM$ in Subsection \ref {SubsecPM}, while the full construction of these is delegated to Appendix \ref {AppPM}. The irreducible
representations at the other $\bk$-points sitting the Brillouin zone edge are enumerated in Subsection \ref {SubsecIRlow}.

\subsection{Group $\bP_\Gamma$ and its irreducible representations}

The  16 elements of the group $\bP_\Gamma$ have been introduced earlier in this section. To repeat,
8 elements are given by $\{ g | 00 \}$ where $g \in {\bf D}_{2d}$, the other 8 elements are obtained from these
by $\{ i | \thalf \thalf \} \{ g | 00 \} = \{ i g | \thalf \thalf \}$. These 16 elements form a closed group $\bP_\Gamma$, where the Seitz multiplication
rule, Eq.\ \eqref {Seitzproduct}, is defined modulo an integer translation.

The group $\bP_\Gamma$ is isomorphic to ${\bf D}_{4h}$. The mapping from $\bP_\Gamma$ onto ${\bf D}_{4h}$ is performed
by `stripping-off' the translation part from the symmetry operation, $\{ g | \btau \} \to g$. Under the inverse mapping,
an element $g \in {\bf D}_{4h}$ is mapped onto $\{ g | 00 \}$ if $g \in {\bf D}_{2d}$, and onto $\{ g | \thalf \thalf \}$ otherwise.
We use this isomorphism when providing the irreducible representation tables, which are widely available \cite{Tinkham}, in this subsection.

\begin{table}[h]
\begin{center}
  \begin{tabular}{| c || c | c | c | c |}
    \hline
    $\bP_\Gamma$ & $\{\sigma^X| \thalf \thalf \}$ &
      $\{\sigma^z| \thalf \thalf \}$ & $\{\sigma_x| 00 \}$ & f({\bk}) \\
    \hline \hline
    $A_{1g/u}$ & $\pm1$ & $\pm1$ & $\pm1$ & $g: \bk^2$  \\
    $A_{2g/u}$ & $\mp1$ & $\pm1$ & $\mp1$ & $-$  \\
    $B_{1g/u}$ & $\mp1$ & $\pm1$ & $\pm1$ & $g: k_x^2 - k_y^2$  \\
    $B_{2g/u}$ & $\pm1$ & $\pm1$ & $\mp1$ & $g: 2 k_x k_y$  \\
    $E_{g/u}$ & $\left[\begin{array}{cc} \pm1 & 0 \\ 0 & \mp1 \end{array}\right]$ &
    $\left[\begin{array}{cc} \mp1 & 0 \\ 0 & \mp1\end{array}\right]$ & $\left[\begin{array}{cc} 0 & \mp1 \\ \mp 1 & 0\end{array}\right]$ &
      ${ \begin {array}{c} u: (\pm k_x  + k_y ), \\ (\pm k_x^3 + k_y^3), \\ k_x k_y (k_x \pm k_y) \end {array} }$  \\
     \hline
   \end{tabular}
 \end{center}
\caption{The complete list of the irreducible representations at the $\Gamma$-point for the 3 point group generators.  The additional column
  on the right shows the symmetry properties of  $\bk$-polynomials at $k_z=0$, up to the third order in $\bk$. Please note that the upper and the lower components of the axial vector representation $E_g$ transform as $Yz$ and $-Xz$, respectively.}
  \label{tab:Gamma generators irreps}
\end{table}

By the virtue of the isomorphism, we know that group $\bP_\Gamma$ has three generators which, for future convenience, can be chosen to be
$\{\sigma_X|\thalf \thalf \}$, $\{\sigma_z| \thalf \thalf \}$, and $\{\sigma_x|00\}$.  The irreducible representations for the
three generators are given in Table \ref{tab:Gamma generators irreps}. The irreducible representations for any other element
of $\bP_\Gamma$ can be obtained from these by multiplication. We use the names for the irreducible representations
according to the isomorphism to ${\bf D}_{4h}$. The last column on the right in
Table \ref {tab:Gamma generators irreps} contains the classifications of the $\bk$-polynomials.
Finally, Table \ref{TabD4hProduct} shows the multiplication identities for the irreducible representations of $\bP_\Gamma$.

\begin {table}
\begin {tabular}{|  >{$} c<{$} || >{$} c<{$} | >{$} c<{$} | >{$} c<{$} | >{$} c<{$} | >{$} c<{$} |}
  \hline
  ~ & A_{1b} & A_{2b} & B_{1b} & B_{2b} & E_b \\
  \hline
  \hline
  A_{1a} & A_{1c} & A_{2c} & B_{1c} & B_{2c} & E_c \\
  A_{2a} & A_{2c} & A_{1c} & B_{2c} & B_{1c} & E_c \\
  B_{1a} & B_{1c} & B_{2c} & A_{1c} & A_{2c} & E_c \\
  B_{2a} & B_{2c} & B_{1c} & A_{2c} & A_{1c} & E_c \\
  E_a & E_c & E_c & E_c & E_c & A_{1c} \oplus A_{2c} \oplus B_{1c} \oplus B_{2c} \\
  \hline
\end {tabular}
\caption {The product table for the irreducible representations of  $\bP_\Gamma \cong {\bf D}_{4h}$.
    The parity of the product is even, $c=g$, if the parities of the two multiplying irreducible representations
    are the same, $a=b$; if the parities are opposite, $a\neq b$, the product has odd parity, $c=u$.}
\label{TabD4hProduct}
\end {table}

\subsection{Group  $\bP_\bM$ and its irreducible representations}
\label{SubsecPM}

The corner of the Brillouin zone, $\bM$, has a particular importance in iron based superconductors; the
electron Fermi surfaces are centered around this point, as found theoretically \cite {LebeguePRB2007, MazinPRL2008, HaulePRL2008, GraserPRB2010,  YinNatMat2011}
and in experiments \cite {DingEPL2008, TerashimaPNAS2009, LuShenPhysicaC2009, CarringtonPhysicaC2009, LiuNatComm2012}.
Following Eq.\ \eqref {expsymmorphic},  we have demonstrated that the irreducible representations at the $\bk$-points
at the Brillouin zone boundary, such as $\bM$, are not given by the little co-group $\bP_\bk$, a subgroup of $\bP_\Gamma$.
Instead, there is a separate procedure, due to C.\ Herring \cite {Herring1942}; we outline the procedure here
and give the final result: the list of the irreducible representations of $P4/nmm$ at the wave-vector $\bM$.
The technical parts of the construction are shown in Appendix \ref {AppPM}.

The basic idea of the approach is to notice that the naively constructed little co-group at $\bM$, which has
16 elements of $\bP_\Gamma$ is not closed under the multiplication as demonstrated in Eq.\ \eqref {expsymmorphic}.
Therefore, the group $\bP_\bM$ must contain additional elements. In this case, we construct group $\bP_\bM$ as a
quotient group of $P4/nmm$ with the invariant subgroup of even translations
\be
  {\mathcal T}_\bM = \Big \{ \{ e | \bt \} \Big | \exp \left ( i \bM \cdot \bt \right ) = 1 \Big \}. \label{TbM}
\ee
Group $\bP_\bM$ defined in this manner is closed under the Seitz product Eq.\ \eqref {Seitzproduct} modulo even
translations. It contains 16 elements originally found in $\bP_\Gamma$, and additional 16 elements obtained by
a multiplication with an odd translation, $\{ e | 10 \} \{ g | \btau \} = \{ g | 10 + \btau \}$.

Any element of $P4/nmm$ can be written as a product of an integer translation and an element of $\bP_\bM$. Since
all even translations are represented by the unity per definition Eq.\ \eqref {TbM}, it is sufficient to find the irreducible
representations of $\bP_\bM$ in order to determine the representation for any element of $P4/nmm$. There is,
however, one subtlety: the odd translation, $\{ e | 10 \}$, which is among the elements of $\bP_\bM$, must be
represented by
\be
  D_{\bM, i} ( \{ e | 10 \} ) = e^{i a \bM \cdot \hat {\bf x}} \bbone = - \bbone. \label{DMe10}
\ee
Therefore, only the irreducible representations of $\bP_\bM$ for which Eq.\ \eqref{DMe10} is satisfied are physical at the
$\bM$-point. All other irreducible representations are unphysical at $\bM$ and should be disregarded.

\begin{table}[h]
\begin{center}
  \begin{tabular}{| c || c | c | c | >{$}c<{$} |}
    \hline
    $\bP_\bM$ & $\{\sigma_X|\frac{1}{2}\frac{1}{2}\}$ & $\{\sigma_z|\frac{1}{2}\frac{1}{2}\}$ & $\{\sigma_x| 00 \}$ & T^{a}/T^b \\
    \hline\hline
    $E_{\bM 1}$ & $\left[\begin{array}{cc} -1 & 0 \\ 0 & -1 \end{array}\right]$ &
      $\left[\begin{array}{cc} -1 & 0 \\ 0 & 1\end{array}\right]$ & $\left[\begin{array}{cc} 0 & 1 \\ 1 & 0\end{array}\right]$ & \cos / \sin \\
    $E_{\bM 2}$ & $\left[\begin{array}{cc} 1 & 0 \\ 0 & 1 \end{array}\right]$ &
      $\left[\begin{array}{cc} -1 & 0 \\ 0 & 1\end{array}\right]$ & $\left[\begin{array}{cc} 0 & 1 \\ 1 & 0\end{array}\right]$ & \sin / \cos \\
    $E_{\bM 3}$ & $\left[\begin{array}{cc} 1 & 0 \\ 0 & -1 \end{array}\right]$ &
      $\left[\begin{array}{cc} -1 & 0 \\ 0 & 1\end{array}\right]$ & $\left[\begin{array}{cc} 0 & 1 \\ 1 & 0\end{array}\right]$ & \sin / \sin \\
    $E_{\bM 4}$ & $\left[\begin{array}{cc} -1 & 0 \\ 0 & 1 \end{array}\right]$ &
      $\left[\begin{array}{cc} -1 & 0 \\ 0 & 1\end{array}\right]$ & $\left[\begin{array}{cc} 0 & 1 \\ 1 & 0\end{array}\right]$ & \cos / \cos \\
    \hline
  \end{tabular}
\end{center}
\caption{The list of physical irreducible representations at the $\bM$-point for three group generators. Since $P4/nmm$ is non-symmorphic,
  all the physical irreducible representations
  are two dimensional. This implies the double degeneracy of any states at the $\bM$-point which are classified
  according to these irreducible representations. The last column in the table should be used in Eq.\ \eqref {symmetryadaptedbasis}
  in order to construct the complete symmetry adapted basis of $\br$-functions which transform according to one
  of the $E_{\bM}$ representations. The illustrative examples of such functions are given in Figs.\ \ref{FigLowestHarmonics} and \ref{FigNextHarmonics}.}
\label{tab:M generators irreps}
\end{table}

The group $\bP_\bM$ constructed here is not isomorphic to any three-dimensional point groups, all of which
are listed in Ref.\ \onlinecite{Tinkham}. We are therefore left with the task of deriving the irreducible representations of $\bP_\bM$.
The full procedure, based on the method of induced representations \cite{InuiTanabeOnodera}, is spelled out in Appendix \ref {AppPM}.
Here we list only the physical irreducible representations of $\bP_\bM$ in Table \ref {tab:M generators irreps}. The complete list of the
irreducible representations is found in Appendix \ref {AppPM}.

All the physical irreducible representations at the $\bM$-point are two dimensional; the two components
of each doublet are related to each other by the mirror reflection $\{ \sigma^x | 00 \}$. One immediate consequence is that all the states with momentum
$\bM$ must be doubly degenerate; other consequences of this double degeneracy are discussed in the remainder of this subsection.

\begin {table}[h]
\begin{center}
  \begin {tabular}{| >{$}c<{$} || >{$}c<{$} | >{$}c<{$} | >{$}c<{$} | >{$}c<{$} |}
    \hline
    ~ & E_{\bM 1} & E_{\bM 2} & E_{\bM 3} & E_{\bM 4} \\
    \hline \hline
    E_{\bM 1} & \begin {array}{c} A_{1g} \! \oplus \! B_{2g}\\ \oplus  A_{2u} \! \oplus \! B_{1u}\end {array} &
      \begin{array}{c} A_{2g} \! \oplus \! B_{1g} \\ \oplus  A_{1u} \! \oplus \! B_{2u} \end{array} & E_g \oplus E_u & E_g \oplus E_u \\
    E_{\bM 2} & \begin{array}{c} A_{2g} \! \oplus \! B_{1g} \\ \oplus  A_{1u} \! \oplus  \! B_{2u} \end {array} &
      \begin{array}{c} A_{1g} \! \oplus \! B_{2g} \\ \oplus A_{2u} \! \oplus \! B_{1u} \end {array} & E_g \oplus E_u & E_g \oplus E_u \\
    E_{\bM 3} & E_g \oplus E_u & E_g \oplus E_u & \begin{array}{c} A_{1g} \! \oplus \! B_{2g} \\ \oplus A_{1u} \! \oplus \! B_{2u} \end {array} &
      \begin{array}{c} A_{2g} \! \oplus \! B_{1g} \\ \oplus A_{2u}  \! \oplus  \! B_{1u} \end {array} \\
    E_{\bM 4} & E_g \oplus E_u & E_g \oplus E_u & \begin{array}{c} A_{2g} \! \oplus \! B_{1g} \\ \oplus A_{2u}\!  \oplus \! B_{1u} \end {array} &
      \begin{array}{c} A_{1g} \! \oplus \! B_{2g} \\ \oplus A_{1u} \! \oplus \! B_{2u} \end {array} \\
    \hline
  \end {tabular}
\end{center}
\caption{Product table for the physical irreducible representations of $\bP_\bM$. Since two vectors $\bM$ add to zero in the reciprocal space,
  the product has no momentum, hence it decomposes onto irreducible representations of $\bP_\Gamma \cong {\bf D}_{4h}$. } \label{TableExE}
\end {table}

\begin {table}[h]
\begin{center}
  \begin {tabular}{| >{$}c<{$} || >{$}c<{$} | >{$}c<{$} | >{$}c<{$} | >{$}c<{$} | >{$}c<{$} |
    >{$}c<{$} | >{$}c<{$} | >{$}c<{$} | >{$}c<{$} | >{$}c<{$} | >{$}c<{$} |}
    \hline
    ~ & E_{\bM 1} & E_{\bM 2} & E_{\bM 3} & E_{\bM 4} \\
    \hline \hline
    A_{1g} & E_{\bM 1} & E_{\bM 2} & E_{\bM 3} & E_{\bM 4} \\
    A_{2g} & E_{\bM 2} & E_{\bM 1} & E_{\bM 4} & E_{\bM 3} \\
    B_{1g} & E_{\bM 2} & E_{\bM 1} & E_{\bM 4} & E_{\bM 3} \\
    B_{2g} & E_{\bM 1} & E_{\bM 2} & E_{\bM 3} & E_{\bM 4} \\
    E_g &  E_{\bM 3} \oplus E_{\bM 4} &  E_{\bM 3} \oplus E_{\bM 4} & E_{\bM 1} \oplus E_{\bM 2} & E_{\bM 1} \oplus E_{\bM 2} \\
    A_{1u} & E_{\bM 2} & E_{\bM 1} & E_{\bM 4} & E_{\bM 3} \\
    A_{2u} & E_{\bM 1} & E_{\bM 2} & E_{\bM 3} & E_{\bM 4} \\
    B_{1u} & E_{\bM 1} & E_{\bM 2} & E_{\bM 3} & E_{\bM 4} \\
    B_{2u} & E_{\bM 2} & E_{\bM 1} & E_{\bM 4} & E_{\bM 3} \\
    E_u & E_{\bM 3} \oplus E_{\bM 4} &  E_{\bM 3} \oplus E_{\bM 4} & E_{\bM 1} \oplus E_{\bM 2} & E_{\bM 1} \oplus E_{\bM 2} \\
    \hline
  \end {tabular}
\end{center}
\caption{Product table between the irreducible representations of ${\bf D}_{4h}$ and the physical irreducible representations of $\bP_\bM$.} \label{TableExD4h}
\end {table}

The product table for the physical irreducible representations of $\bP_\bM$ is given in Table \ref {TableExE}. A product of two Bloch states,
each with momentum $\bM$, has no total lattice momentum, hence the products of two $E_{\bM}$ irreducible representations are
decomposed into the irreducible representations of $\bP_\Gamma$. The mixed product table, where
one irreducible representation is of $\bP_\Gamma$, while the other is a physical irreducible
representation of $\bP_\bM$, has the total momentum $\bM$. Therefore, such a product decomposes into $E_{\bM}$ irreducible representations,
as shown in Table \ref {TableExD4h}.

We use the physical irreducible representations of $\bP_\bM$ for the symmetry classification of two-dimensional Bloch functions with momentum $\bM$,
i.e., all the functions $f ( \br )$, where $\br = (x, y, 0)$, such that
\be
  f (\br) &=& e^{i \bM \cdot \br} u (\br) \nonumber \\
  &=&  e^{i \bM \cdot \br} \sum_{m_1, m_2 \in {\mathbb Z}} u_{m_1, m_2} e^{i \frac {2 \pi}a (m_1 x + m_2 y) } \nonumber \\
  &=& \sum_{n_1, n_2 \in {\mathbb Z} + 1/2} v_{n_1, n_2} e^{i \frac {2 \pi}a (n_1 x + n_2 y) }. \label{MFourier}
\ee
Here, $u (\br)$ is a lattice periodic function, with Fourier harmonics coefficients $u_{m1, m2}$. The function is alternatively decomposed
into the half-integer Fourier harmonics in the last line of Eq.\ \eqref {MFourier}. The half-integer Fourier harmonics form a complete
basis for the two-dimensional $\bM$, however, they are not adapted to the lattice symmetries. To remedy this,
we use the projector method on the half-integer Fourier harmonics,  and find a new basis such that each basis function
transforms as a particular component of one of the $E_{\bM}$'s. All the functions in this symmetry adapted basis come in
doublets, as they should, due to the two-dimensionality of each $E_{\bM}$. For any physical irreducible representation, the two doublet states
can be written as
\begin {align}
  f^{E_{\bM i}^X}
    =& T_i^a \left \lbrack \frac {(2 m_1 + 1) \pi} a (x+y) \right \rbrack
    T_i^b \left \lbrack \frac {2 m_2 \pi}a (-x+y) \right \rbrack, \nonumber \\
  f^{E_{\bM i}^Y}
    =& T_i^a \left \lbrack \frac {(2 m_1 + 1) \pi}a (-x+y) \right \rbrack
    T_i^b \left \lbrack \frac {2 m_2 \pi}a (x+y) \right \rbrack,  \label{symmetryadaptedbasis}
\end {align}
where integers $m_{1,2} \ge 0$, and $T_i^{a/b}$ are two trigonometric functions, listed in
Table \ref {tab:M generators irreps} for each physical irreducible representation.
The basis defined by Eq.\ \eqref {symmetryadaptedbasis} is complete.

\begin{figure}[h]
\begin{center}
\includegraphics[width=0.22\textwidth]{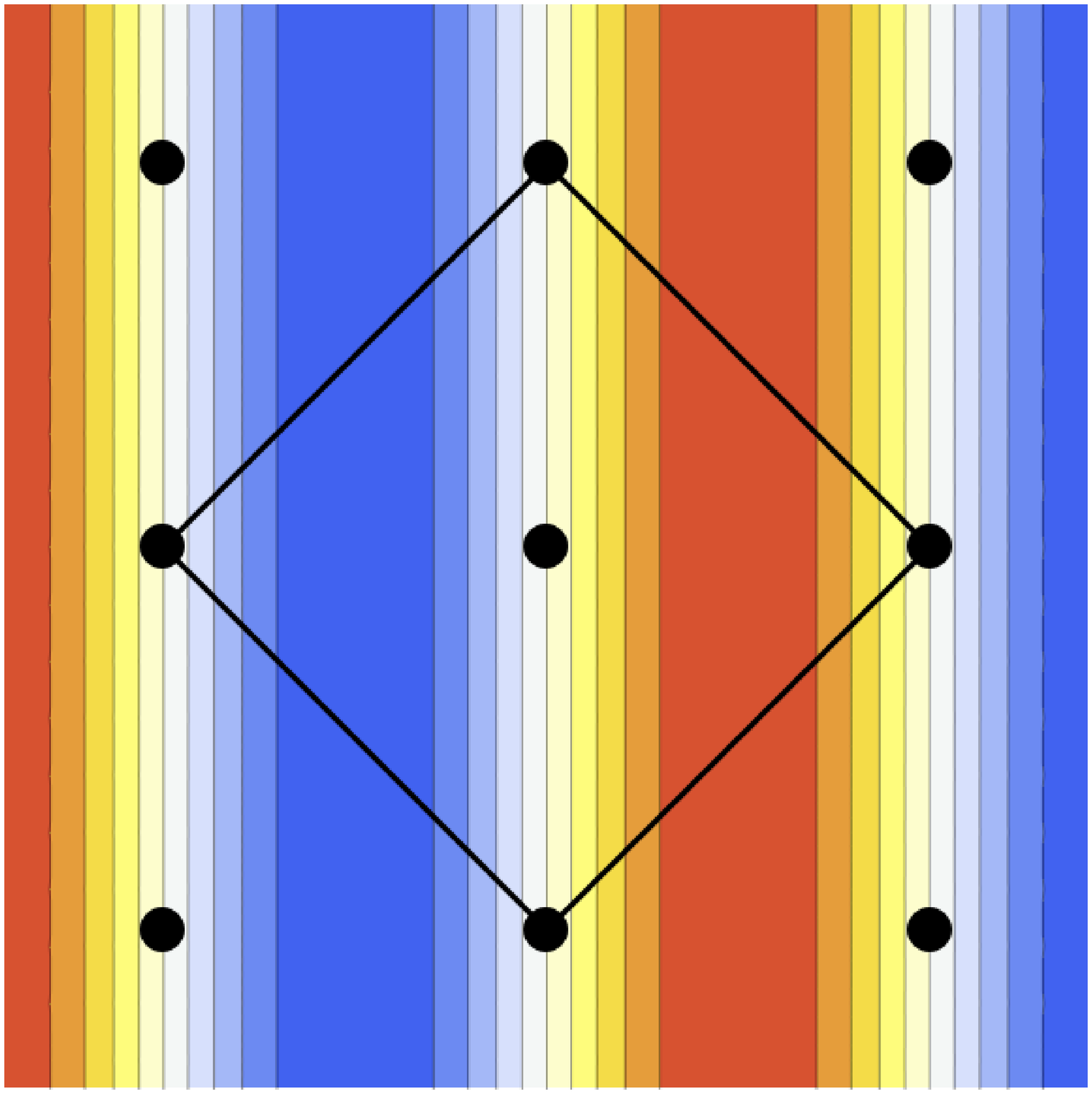} \includegraphics[width=0.22\textwidth]{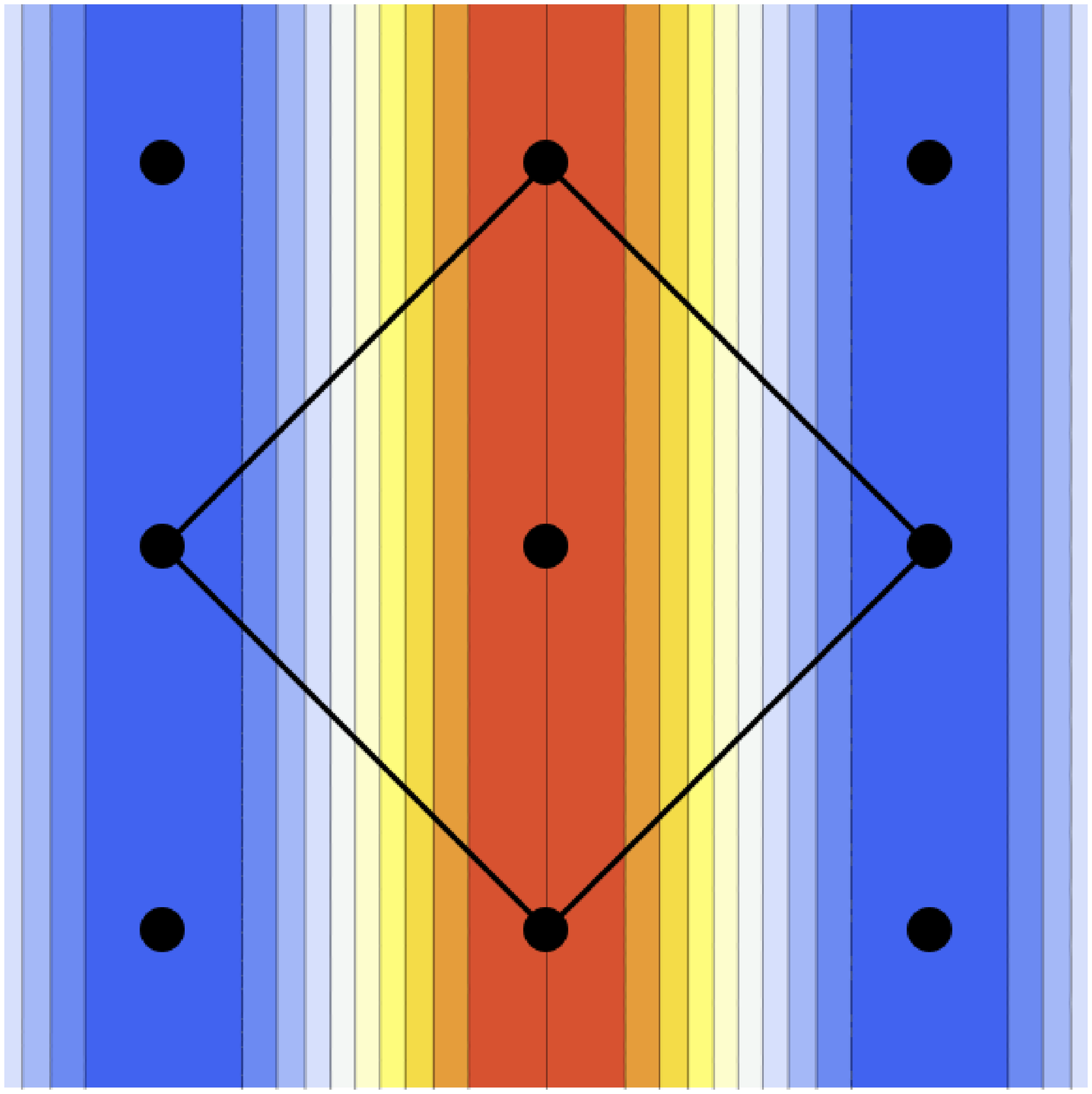}\\
{$E_{\bM 2}^X$ \hspace {0.18\textwidth} $E_{\bM 4}^X$} \\
\includegraphics[width=0.22\textwidth]{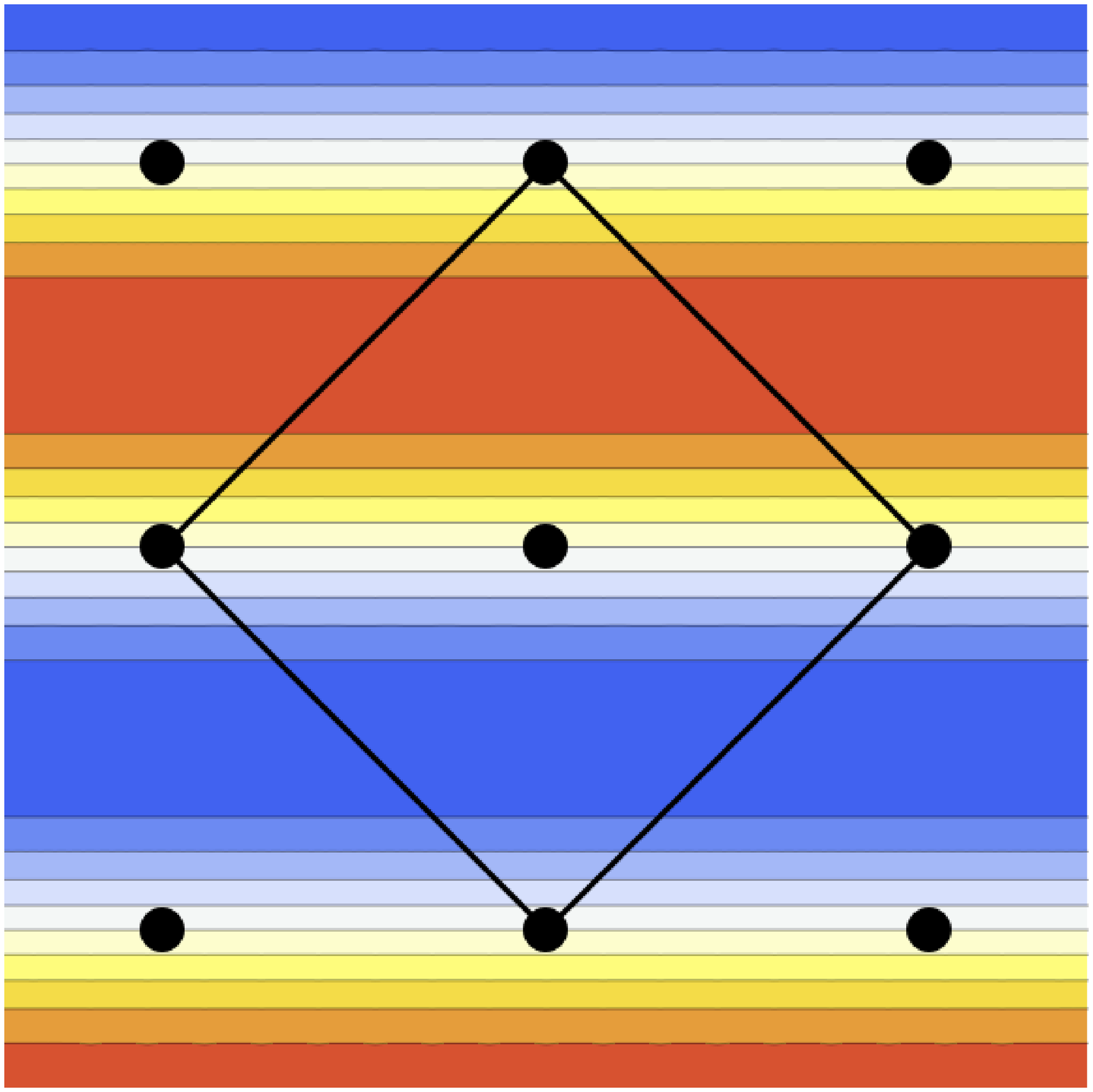} \includegraphics[width=0.22\textwidth]{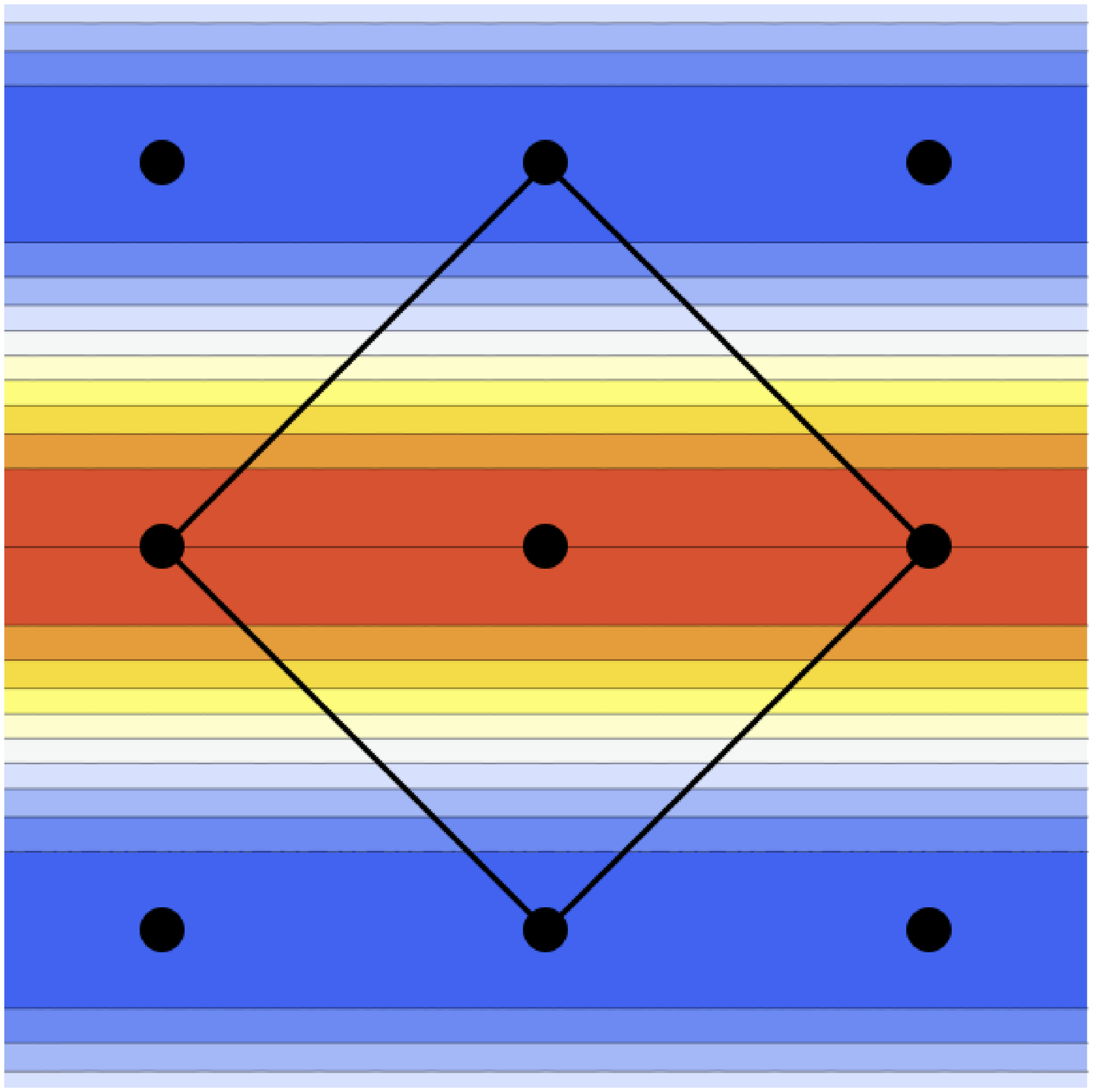}\\
{$E_{\bM 2}^Y$ \hspace {0.18\textwidth} $E_{\bM 4}^Y$}
\end{center}
  \caption{The lowest order harmonics ($m_1 = m_2 = 0$) in the symmetry adapted basis of functions given by
    Eq.\ \eqref {symmetryadaptedbasis}. Notice how the $E_{\bM 4}$ functions have a finite value at the
    positions of iron atoms and vanish on pnictide sites. Conversely, the $E_{\bM 2}$ functions are finite on
    pnictide atoms and vanish on iron atoms.}\label{FigLowestHarmonics}
\end{figure}

\begin{figure}[h]
\begin{center}
\includegraphics[width=0.115\textwidth]{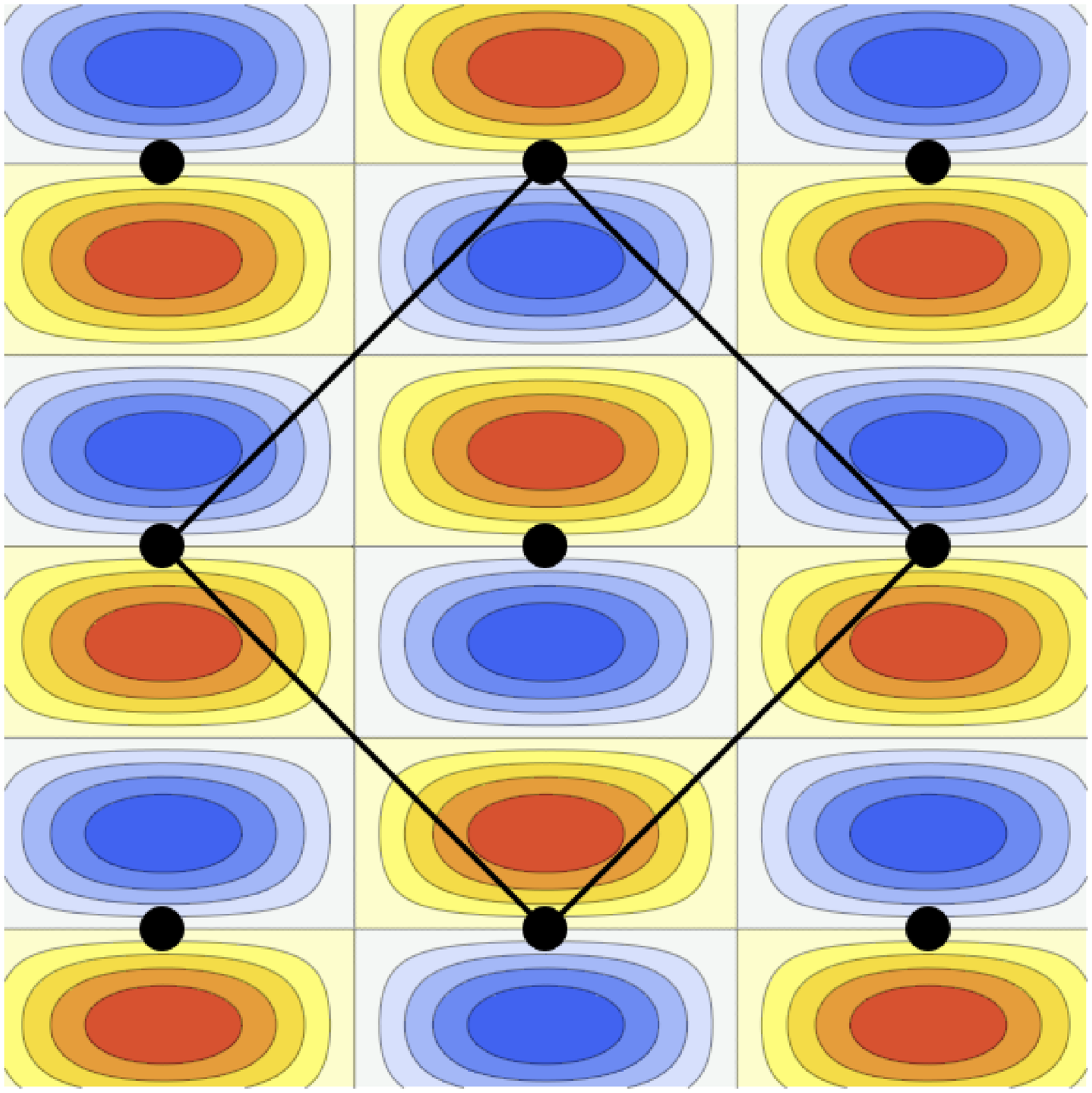}\includegraphics[width=0.115\textwidth]{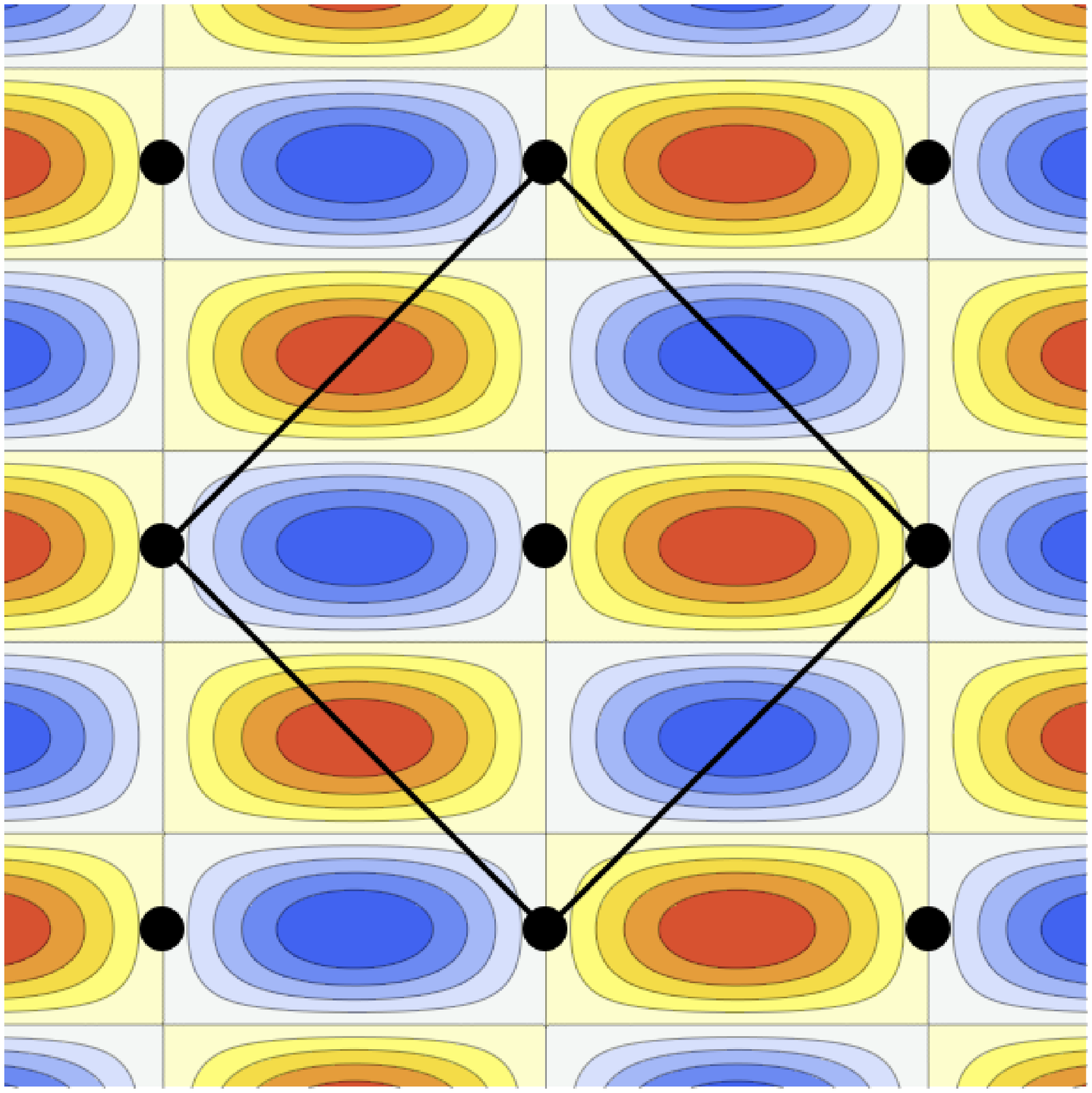}
\includegraphics[width=0.115\textwidth]{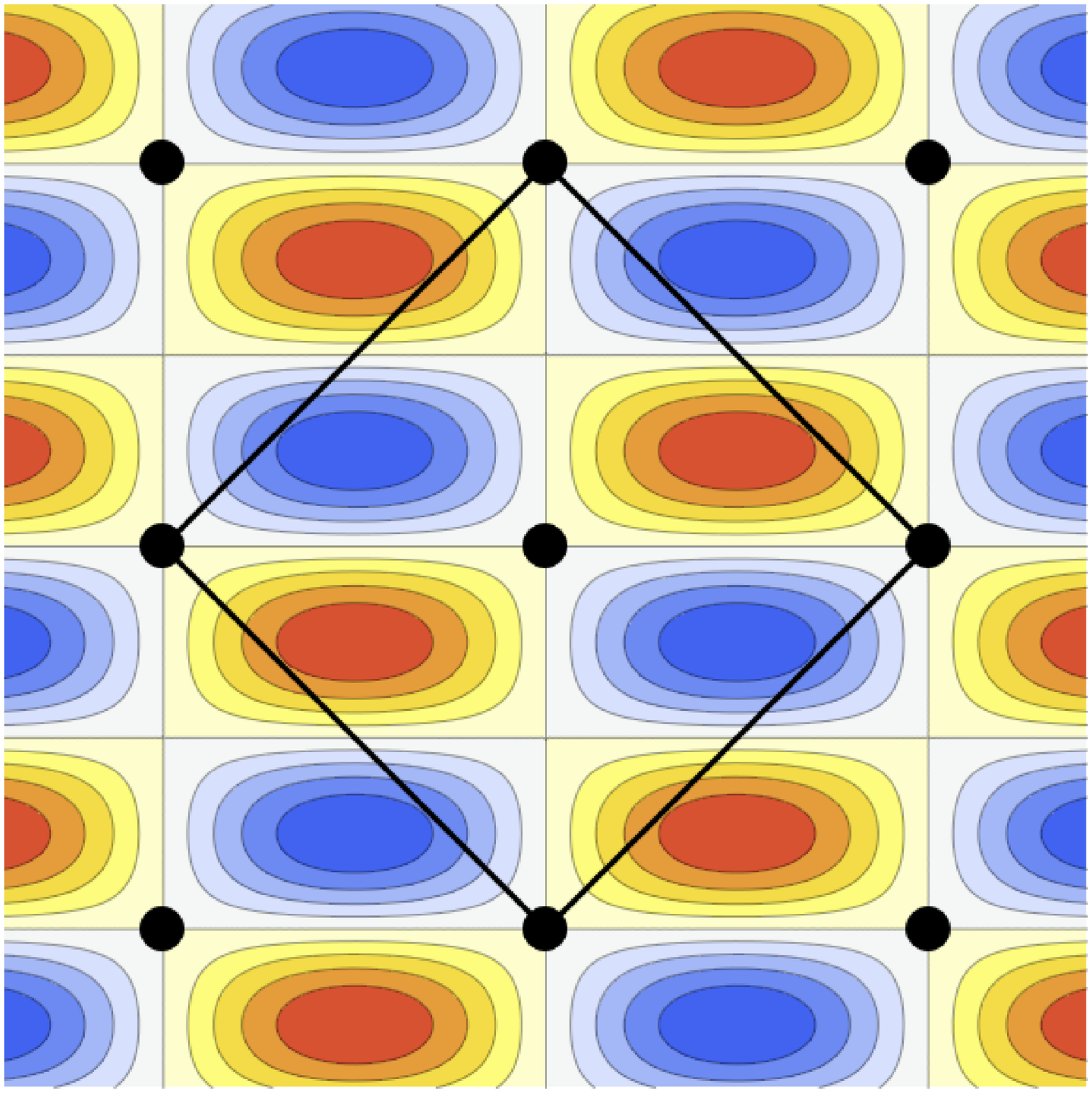}\includegraphics[width=0.115\textwidth]{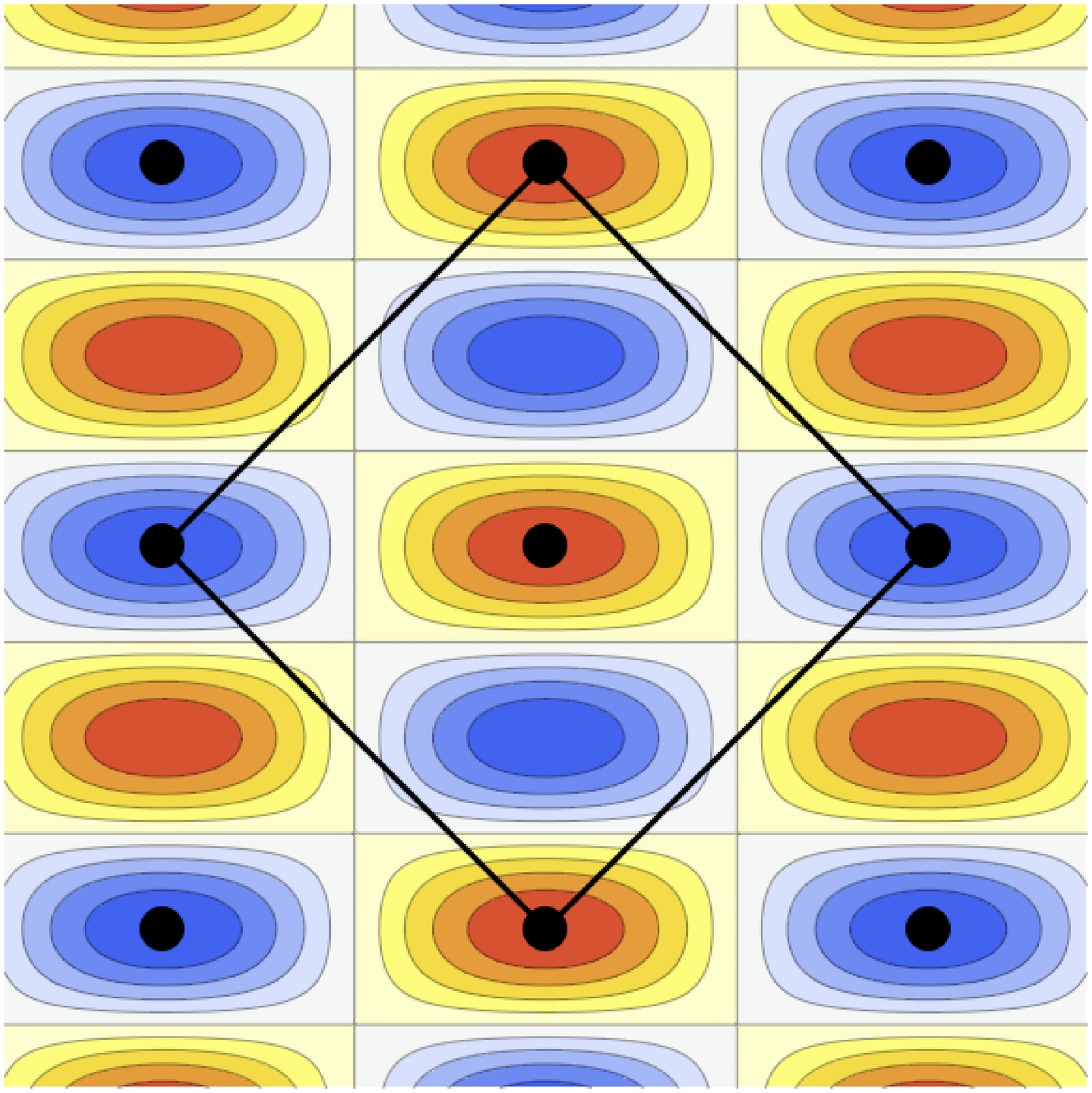} \\
{$E_{\bM 1}^X$ \hspace {0.07\textwidth} $E_{\bM 2}^X$ \hspace {0.07\textwidth} $E_{\bM 3}^X$ \hspace {0.07\textwidth} $E_{\bM 4}^X$} \\
\includegraphics[width=0.115\textwidth]{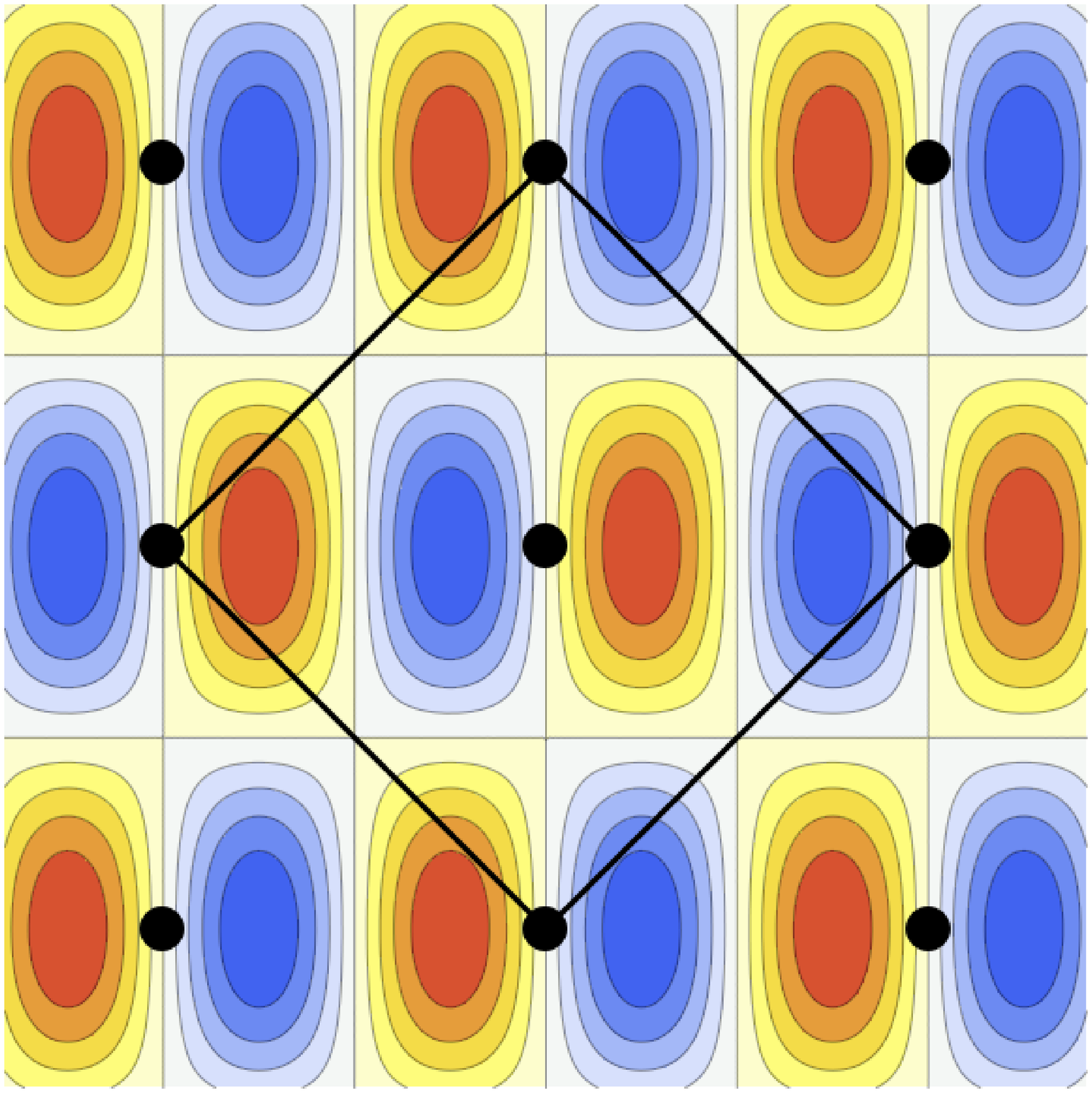}\includegraphics[width=0.115\textwidth]{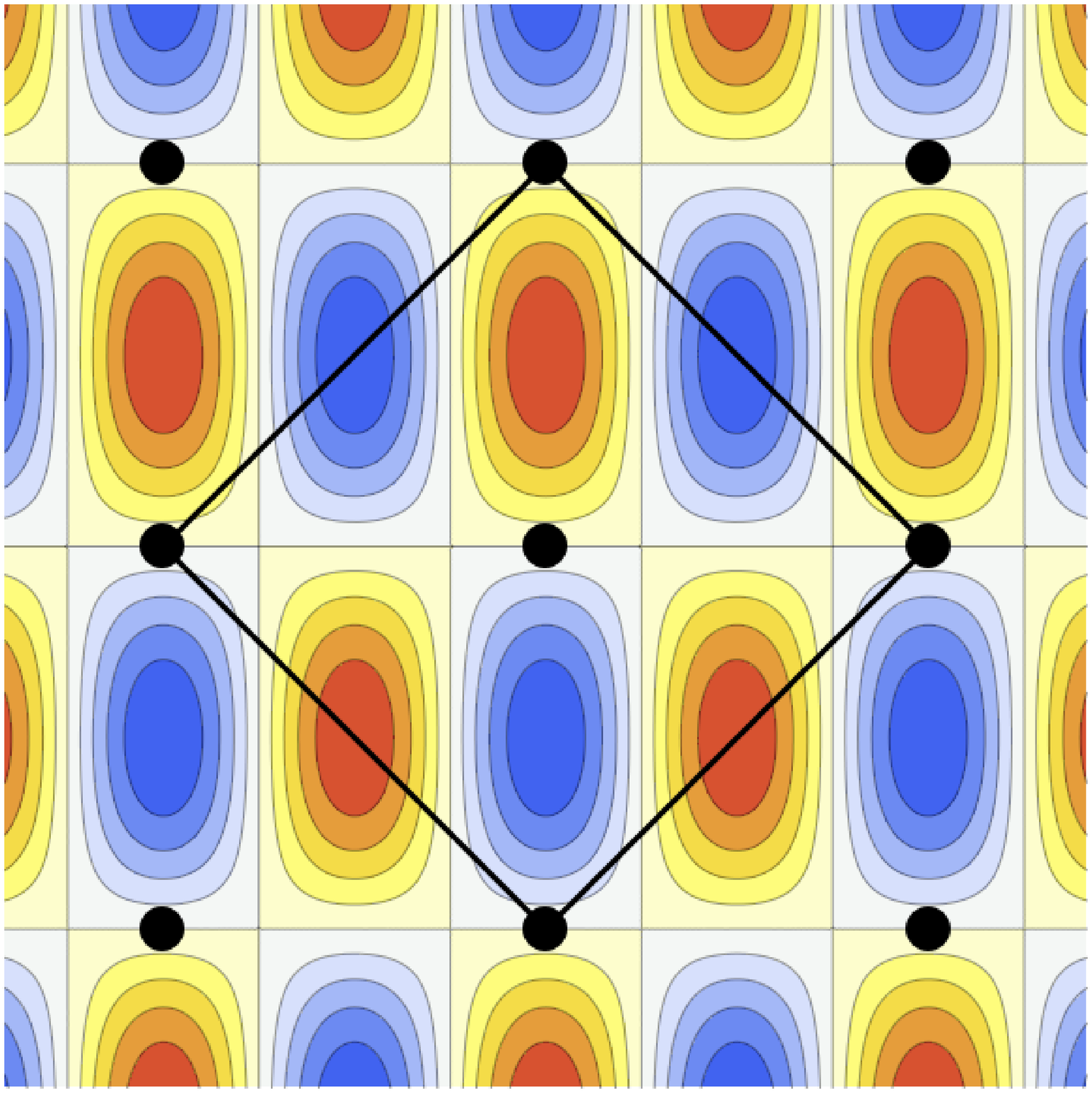}
\includegraphics[width=0.115\textwidth]{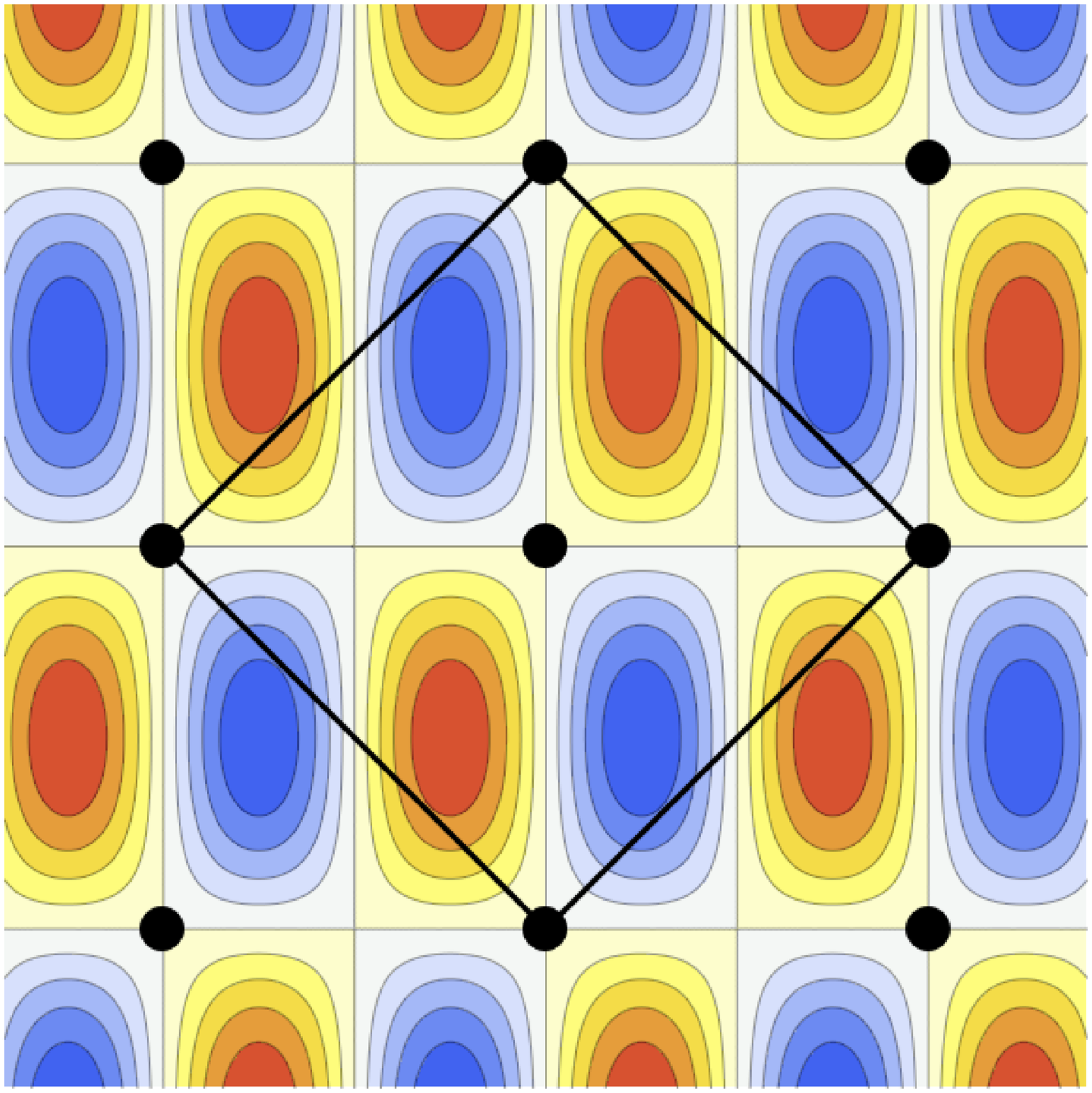}\includegraphics[width=0.115\textwidth]{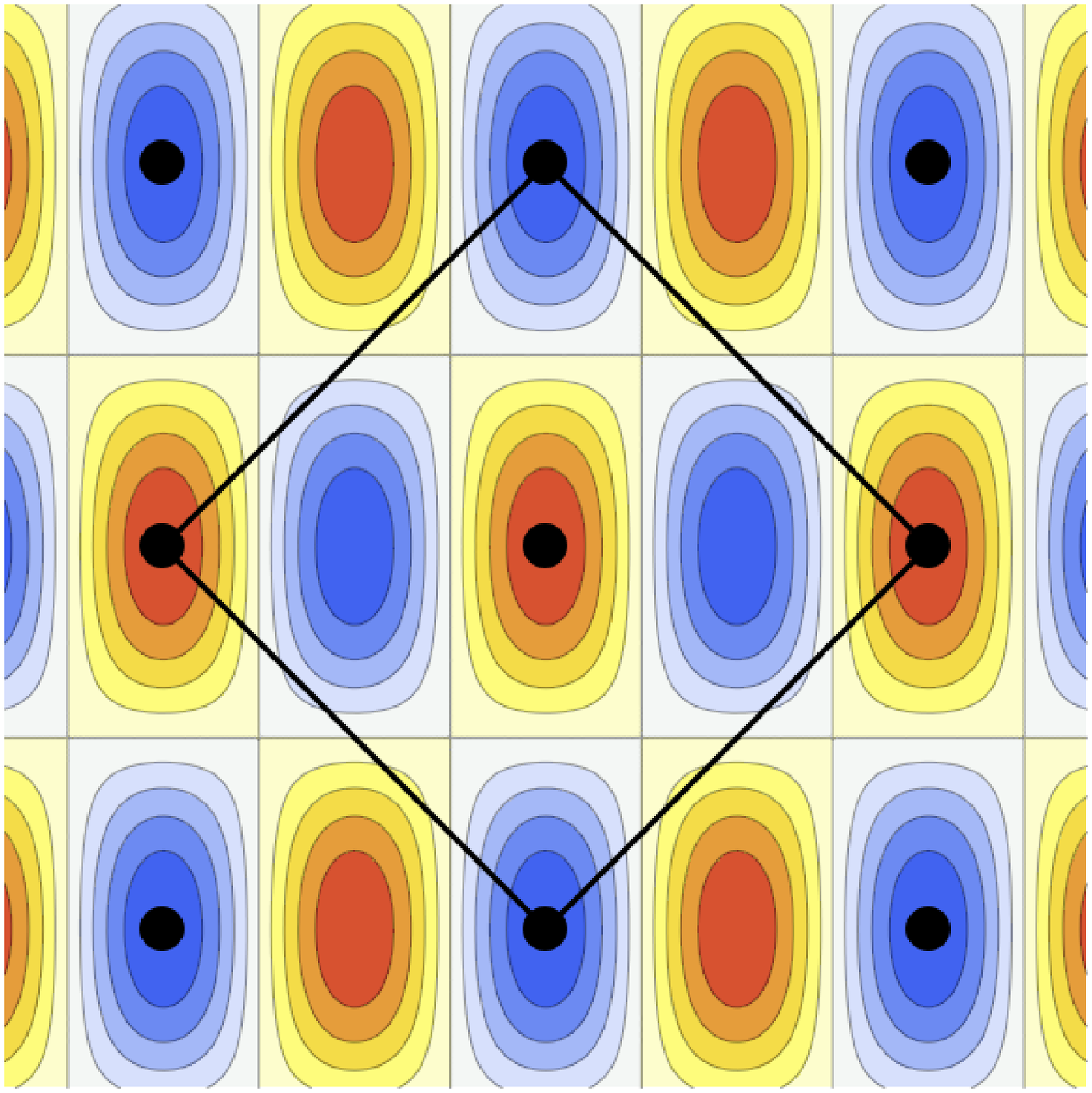}\\
{$E_{\bM 1}^Y$ \hspace {0.07\textwidth} $E_{\bM 2}^Y$ \hspace {0.07\textwidth} $E_{\bM 3}^Y$ \hspace {0.07\textwidth} $E_{\bM 4}^Y$} \\
\end{center}
  \caption{For each physical irreducible representation $E_{\bM i}$, we plot the symmetry adapted functions doublet. In this figure
    we show the next lowest order harmonics ($m_1=0$, $m_2 = 1$) given in Eq.\ \eqref {symmetryadaptedbasis}.
    Similarly to Fig.\ \ref {FigLowestHarmonics}, the only doublet with non-vanishing values on the iron sites is $E_{\bM 4}$; the only
    doublet with non-vanishing values on the pnictide sites is $E_{\bM 2}$.}\label{FigNextHarmonics}
\end{figure}

The lowest harmonics, $m_1 = m_2 = 0$, are non-zero only for $E_{\bM 2}$ an $E_{\bM 4}$. These four
functions are plotted in Fig.\ \ref {FigLowestHarmonics}. In the same plots, iron atoms are marked by solid
dots, and one unit cell is outlined as a guide to the eye. For any higher harmonics (i.e., $m_1 + m_2 > 0$), and for each $E_{\bM}$, there is
precisely one doublet of functions. For illustration, we plot the second lowest harmonics ($m_1 = 0$, $m_2=1$) in  Fig.\ \ref {FigNextHarmonics}.

In both Figs.\ \ref{FigLowestHarmonics} and \ref{FigNextHarmonics}, the $E_{\bM 4}$ harmonics
have minima and maxima on iron atoms and nodes on (the a-b plane projection of) the pnictide atoms. All other harmonics vanish
precisely on the iron atom sites. Similarly, the $E_{\bM 2}$ harmonics have the largest magnitude
on pnictide atom sites, while all the other harmonics vanish there. This is true for
any higher harmonics. It can be demonstrated from Eq.\ \eqref {symmetryadaptedbasis} that
only $E_{\bM 4}$ harmonics are nodeless on iron-sites; on pnictide sites only $E_{\bM 2}$
harmonics have no nodes.  Experimentally, the spin-wave density order is found to have
magnetic moments located on iron atoms \cite{delaCruzNature2008,GoldmanPRB2008}.
We conclude therefore that the orbital part of any spin-wave density wave order in iron based superconductors
must have the symmetry properties of $E_{\bM 4}$.

The last statement comes with one caveat. In the present exposition, we constrain ourselves to the analysis of a single layer. The experiments
also show that the spin-density wave has the opposite sign in two neighboring layers \cite{delaCruzNature2008,GoldmanPRB2008}. Therefore, we should consider the symmetry
adapted functions at ${\bf A} = \pi (1/a, 1/a, 1/c)$ instead when discussing the spin-density wave order found in the experiments.
Since ${\bf A}$ is at the edge of the Brillouin zone, $\bP_{\bf A}$ must be constructed according to the same procedure as $\bP_\bM$.
It turns out that these two groups are isomorphic, $\bP_{\bf A} \cong \bP_\bM$, and, as a consequence, their physical
irreducible representations must be equivalent. Therefore, the symmetry properties of spin-density waves in iron based superconductors
are, strictly speaking, governed by $E_{{\bf A} 4}$, which is given by the same matrices as $E_{\bM 4}$ in Table \ref {tab:M generators irreps}.

\subsection {Irreducible representations of $P4/nmm$ at lower-symmetry points in the Brillouin zone}
\label{SubsecIRlow}

In  this subsection we focus on the remaining symmetry indistinguishable $\bk$ points in the (two-dimensional) Brillouin zone.
These are $\bX = (\pi / a, 0, 0)$, lines $\Sigma = \Gamma-\bM$, $\Delta = \Gamma-\bX$, and $Y = \bX-\bM$, and the set of all
other $\bk$-points not sitting on any of the high-symmetry points or lines.  The expansions around the $\Gamma$-
and $\bM$-points, discussed in the previous two subsections, are, in principle, independent of the classification provided in this section.
Nevertheless, the results presented here
will be used in comparing our results to two- and three-orbital tight-binding models, as well as, in understanding the
appearance of nodal points in the symmetric, and in the collinear spin-density wave state.

The high symmetry point $\bX = (\pi / a, 0, 0)$ lies at the edge of the Brillouin zone.
Since $P4/nmm$ is a non-symmorphic group, the point group at $\bX$ is constructed in a similar fashion as that at $\bM$.
We do not give the details of the construction in the paper; instead, we only present the final result: group $\bP_\bX$ is
the subgroup of $\bP_\bM$ whose elements keep $\bX$ invariant. This group has 16 elements and is isomorphic to ${\bf D}_{4h}$.
An isomorphism between the two groups is defined through the generators,
\be
  \{ \sigma^x | 00 \} &\longrightarrow& \sigma^x, \nonumber \\
  \{ C_2^y | \thalf \thalf \} &\longrightarrow& \sigma^X, \nonumber \\
  \{ \sigma^y | 10 \} &\longrightarrow& \sigma^z. \label{PXisomorphism}
\ee
All the other elements of $\bP_\bX$ can be generated from these; as a consequence of this isomorphism
$\{ e | 10 \} \to C_2^z$. Just as it was the case with the $\bP_\bM$ group, the only physical irreducible representations of $\bP_\bX$
are those for which  $D \left ( \{ e | 10 \} \right ) = - \bbone$. Using Table \ref {tab:Gamma generators irreps},
we find that there are only two such representations, $E_{\bX g}$ and $E_{\bX u}$. Since both physical irreducible representations of $\bP_\bX$
are two-dimensional, one consequence is that all the bands at this point must be doubly degenerate. To distinguish between the
two irreducible representations, one easy method is to observe the parity of the Bloch states at $\bX$ under the mirror reflection $\{ \sigma^y | 00 \}$. The Bloch
states transforming according to $E_{\bX g}$ are even, those transforming according to $E_{\bX u}$ are odd under this operation.

The group $\bP_Y$ is the subgroup of $\bP_\bX$ which keeps a wave-vector $\bk \in Y$ invariant. This group is
isomorphic to ${\bf C}_{4v}$, with the isomorphism defined by the first two lines in Eq.\ \eqref {PXisomorphism}.
The mirror reflection $\{ \sigma^y | 00 \}$ is absent in $\bP_Y$. The $Y$-line constitutes the Brillouin zone boundary,
and there is only one physical irreducible representation, $E$, specified in Table \ref {TableC4v}.
The symmetry classification at the $Y$-line is therefore almost trivial since
everything transforms according to the same irreducible representations. However, one should notice that, since $E$
is two-dimensional, each doublet of Bloch functions at $\bk \in Y$ must be chosen in such a way that the two doublet components
transform into each other (up to a phase factor) under the $\{ \sigma^z | \thalf \thalf \}$ mirror reflection. One consequence of
this double degeneracy is that all the Bloch states at any $\bk \in Y$ are doubly degenerate.

\begin {table}[h]
\begin{center}
  \begin {tabular}{| >{$}c<{$} || >{$}c<{$} | >{$}c<{$}  |}
    \hline
    {\bf C}_{4v} & \sigma^x & \sigma^X \\
    \bP_Y & \{ \sigma^x | 00 \} & \{ C_2^y | \thalf \thalf \} \\
    \hline \hline
    E & \left[\begin{array}{cc} -1 & 0 \\ 0 & 1 \end{array}\right] & \left[\begin{array}{cc} 0 & -1 \\ -1 & 0 \end{array}\right] \\
    \hline
  \end {tabular}
\end {center}
\caption{The unique physical irreducible representation of the generators of point group $\bP_Y \cong {\bf C}_{4v}$.} \label{TableC4v}
\end {table}

The remaining $\bk$-points are located inside the Brillouin zone. For any of these, group $\bP_\bk$ is the little co-group of
$\bP_\Gamma$. The representations of an element of $P4/nmm$ are given by Eq.\ \eqref {spacegrouphomoinside}.

For $\bk \in \Sigma$, the little co-group $\bP_\Sigma$ is generated by two mirror reflections, $\{ \sigma^Y | \thalf \thalf \}$ and
$\{ \sigma^z | \thalf \thalf \}$. This group is isomorphic to ${\bf C}_{2v}$, with the isomorphism
\be
  \{ \sigma^Y | \thalf \thalf \} \to \sigma^x, \qquad \{ \sigma^z | \thalf \thalf \} \to \sigma^y.
\ee
The table of the irreducible representations and their product table are given in Table \ref {TableC2v}.

\begin {table}[h]
\begin{center}
  \begin {tabular}{| >{$}c<{$} || >{$}c<{$} | >{$}c<{$} |}
    \hline
    {\bf C}_{2v} & \sigma^x & \sigma^y \\
    \bP_\Sigma & \{ \sigma^Y | \thalf \thalf \} & \{ \sigma^z | \thalf \thalf \} \\
    \bP_\Delta & \{ \sigma^y | 00 \} & \{ \sigma^z | \thalf \thalf \} \\
    \hline \hline
    A_1 & 1 & 1 \\
    A_2 & -1 & -1 \\
    B_1 & -1 & 1 \\
    B_2 & 1 & -1 \\
    \hline
  \end {tabular}
\hspace {0.5cm}
  \begin {tabular}{| >{$}c<{$} || >{$}c<{$} | >{$}c<{$} | >{$}c<{$} | >{$}c<{$} |}
    \hline
    ~  & A_1 & A_2 & B_1 & B_2 \\
    \hline \hline
    A_1 & A_1 & A_2 & B_1 & B_2 \\
    A_2 & A_2 & A_1 & B_2 & B_1 \\
    B_1 & B_1 & B_2 & A_1 & A_2 \\
    B_2 & B_2 & B_1 & A_2 & A_1 \\
    \hline
  \end {tabular}
\end{center}
\caption{The irreducible representations for the generators of the isomorphic point groups ${\bf C}_{2v} \cong \bP_\Sigma \cong \bP_\Delta$.
  These are used in the classification of Bloch states with $\bk$ at $\Sigma = \Gamma - \bM$ or $\Delta = \Gamma- \bX$ lines in the Brillouin zone.
  The product table is given on the right. } \label{TableC2v}
\end {table}

For $\bk \in \Delta$, the little co-group $\bP_\Delta$ is generated by two mirror reflections, $\{ \sigma^y | 00 \}$ and
$\{ \sigma^z | \thalf \thalf \}$. This group is isomorphic to ${\bf C}_{2v}$, with the isomorphism
\be
  \{ \sigma^y | 00 \} \to \sigma^x, \qquad \{ \sigma^z | \thalf \thalf \} \to \sigma^y.
\ee
The table of the irreducible representations and their product table are given in Table \ref {TableC2v}.

At last, for a state with a momentum $\bk = (k_x, k_y, 0)$, lying on none of these high symmetry points or lines, there is still a single mirror reflection
$\{ \sigma^z | \thalf \thalf \}$ which keeps $\bk$ invariant. This is due to the fact that we contained ourselves to
the two-dimensional Brillouin zone and $k_z = 0$. This mirror defines group $\bP_\bk \cong {\bf C}_s$ for a
general $\bk$-point. Group ${\bf C}_s$ has two irreducible representations $A'$ and $A''$ corresponding to the states that are respectively even and odd
under the action of $\{ \sigma^z | \thalf \thalf \}$ mirror reflection, in sense of Eq.\ \eqref {spacegrouphomoinside}.

This list exhausts all the possible $\bk$-points in the two-dimensional Brillouin zone for iron-pnictides where $k_z=0$. It appears that we
omitted the irreducible representations of the $P4/nmm$ with wave-vector $\bk = ( k_x, k_y, \pi /c)$. These irreducible
representations are relevant for the Bloch states or order parameters which change the sign in alternating layers, as
mentioned in Subsection \ref {SubsecPM}. It turns out that, for the $P4/nmm$ group, the irreducible representations at
any $\bk = (k_x, k_y, \pi / c)$ are equivalent to those at $\bk = (k_x, k_y, 0)$. Therefore, the irreducible representations
at the $\Gamma$-point are also governing the symmetry properties at ${\bf Z} = (0, 0, \pi /c)$, the physical irreducible representations
at ${\bf R} = (\pi / a , 0, \pi / c)$ are the the same as those at the $\bX$-point, and most importantly, the physical irreducible representations
at ${\bf A} = (\pi / a, \pi / a, \pi / c)$ are equivalent to the four $E_{\bM}$'s. Similarly, the three high-symmetry lines $\Sigma$, $\Delta$, and $Y$, share
the symmetry classification with their $k_z= \pi /c$ counterparts, $U$-, $S$-, and $T$-line, respectively. For a general $\bk = (k_x, k_y, \pi /c)$
momentum, $\bP_\bk \cong {\bf C}_s$. Although $k_z = \pi / c$ implies that $\bk$ lies on the Brillouin zone border, there is no issue
with the non-symmorphicity of $P4/nmm$ here since  the fractional translation vector $\btau_0$ has no $z$-component.

\subsection {The classification of the Bloch states near Fermi level}

The first step in constructing an effective theory is recognizing which states, in the full model, are the most physically relevant and
which states bring only quantitative corrections and may be integrated out. Generally, states with the energy at, and in the vicinity of, the
Fermi level are the ones to be kept in the effective model. The  dispersion of these states should match the dispersion of the
states of the full model as closely as possible, but without overburdening the effective model with unnecessary details.
Perhaps even more importantly, the states of the effective model must have exactly the same symmetry properties as the states in the full model they originate from.
Otherwise, the model may not be able to capture the correct symmetries of physical effects intended to be described by the effective model.
For example, the expectation value of a bilinear  may not be associated with an order parameter of correct symmetry,
transitions violating selection rules may occur, etc.
If we are going to construct an effective theory for iron based superconductors, we ought to identify the symmetries of every state that will make a  part
of the theory.  That is the main result of this subsection.

The symmetries of states can, in principle, be determined from a `first principle' band structure calculation.
A direct output of such a calculation is the composition of eigenstates and from there, the
symmetry nature of each eigenstate may be derived.
An alternative, and the one that we use here, is the band structure obtained from a tight-binding model. We use two independently derived
tight-binding models. The details of the first model are given in Ref.\ \onlinecite {CvetkovicTesanovicEPL2009}. That tight-binding model uses 16 states per unit cell, five $d$ states per each iron atom and three $p$ states per each pnictide atom, and
is fitted to the band structure of undoped LaOFeP obtained in an ab initio calculation \cite {LebeguePRB2007}. The other
tight-binding model we use is presented in Ref.\ \onlinecite {KurokiPRL2008}.
That tight-binding model is fitted to a first principle band structure calculation for $x=0.1$ doped LaOFeAs. The
band structure calculation is presented in the same paper. This tight-binding model uses five maximally localized
Wannier states per iron atom, each of these states having orbital symmetries of
one of the iron $3d$ orbitals. Between the two tight-binding models, we find no difference in the symmetries of
the bands which cross the Fermi level.

We use the irreducible representation of $P4/nmm$, presented in this section, to classify the iron states at any $\bk$.
In a tight-binding model, an iron state at momentum $\bk$ is given by
\be
  \psi_\bk (\br) = \sum_{i} \sum_{\bdelta} e^{i \bk \cdot (\bR_i + \bdelta )} \sum_{\mu} \phi_\mu \left ( \br - \bR_i - \bdelta \right ) d_{\mu}^{\bdelta} (\bk). \label{kstate}
\ee
Here, $\bR_i$ are the positions of iron atoms in the A sublattice; $\bdelta$ takes two values, $0$ or $\btau_0$  corresponding to
the A or B lattice, respectively. $\phi_{\mu, \delta} (\br)$ is the orbital part of the wave function for $\mu$, an iron 3d orbital centered at $\br = 0$.
There are ten iron states for each $\bk$, two per iron atom in a unit cell, and their symmetry properties  are presented in Table \ref {TableLowering}.

These tables contains additional information related to the symmetry lowering in the vicinity of
high symmetry point or lines. Namely, if a band is classified according to a certain irreducible representation
at a high symmetry point, then its symmetry properties in the vicinity of that point can be deduced from there.
This is a consequence of the fact that the transformation properties of that band under the remaining symmetry
operations, away from the high symmetry point, are inherited from the properties at the high symmetry point.
Conversely, the symmetry properties at the high symmetry point may be seen as the same properties from its
vicinity augmented by additional symmetries at the high symmetry point. The same is true when the symmetry is lowered from
a high symmetry line to its vicinity. One dimensional irreducible representations at a high symmetry point
uniquely determine the symmetry properties in its vicinity. The two dimensional irreducible representations
usually, but not always, split into two one dimensional irreducible representations once the symmetry is
lowered.

\begin {table}[h]
\begin{center}
  \begin {tabular}{| >{$}c<{$} || >{$}c<{$} | >{$}c<{$} | >{$}c<{$} | >{$}c<{$} |}
    \hline
    \bP_\Gamma & \mbox {Fe states} & \bP_\Sigma & \bP_\Delta & \bP_\bk \\
    \hline \hline
    A_{1g} & d_{3z^2-R^2}^A + d_{3z^2-R^2}^B & A_1 & A_1 & A' \\
    A_{2g} & - & B_1 & B_1 & A' \\
    B_{1g} & d_{XY}^A + d_{XY}^B & B_1 & A_1 & A' \\
    B_{2g} & d_{X^2-Y^2}^A + d_{X^2-Y^2}^B & A_1 & B_1 & A' \\
    E_{g} & ( d_{Yz}^A + d_{Yz}^B, -d_{Xz}^A - d_{Xz}^B, ) & A_2 \oplus B_2 & A_2 \oplus B_2 & A'' \\
    A_{1u} & d_{X^2-Y^2}^A - d_{X^2-Y^2}^B & A_2 & A_2 & A'' \\
    A_{2u} & d_{XY}^A - d_{XY}^B & B_2 & B_2 & A'' \\
    B_{1u} & - & B_2 & A_2 & A'' \\
    B_{2u} & d_{3z^2-R^2}^A - d_{3z^2-R^2}^B  & A_2 & B_2 & A'' \\
    E_{u} & ( d_{Yz}^A - d_{Yz}^B, d_{Xz}^A - d_{Xz}^B, ) & A_1 \oplus B_1 & A_1 \oplus B_1 & A' \\
    \hline
  \end {tabular}\\
\vspace {0.2cm}
  \begin {tabular}{| >{$}c<{$} || >{$}c<{$} | >{$}c<{$} | >{$}c<{$} | >{$}c<{$} |}
    \hline
    {\bP}_{\bM} & \mbox {Fe states} &  \bP_\Sigma & \bP_Y & \bP_\bk \\
    \hline \hline
    E_{\bM 1} & ( d_{Xz}^A - d_{Xz}^B, d_{Yz}^A + d_{Yz}^B ) & A_2 \oplus B_1 & E & A' \oplus A'' \\
    E_{\bM 2} & ( d_{Yz}^A - d_{Yz}^B, d_{Xz}^A + d_{Xz}^B ) & A_1 \oplus B_2 & E & A' \oplus A'' \\
    E_{\bM 3} & ( d_{XY}^A + d_{XY}^B, d_{XY}^A - d_{XY}^B ) & B_1 \oplus B_2 & E & A' \oplus A'' \\
    E_{\bM 4} & \begin {array}{c} ( d_{X^2-Y^2}^A + d_{X^2-Y^2}^B, \\ - d_{X^2-Y^2}^A + d_{X^2-Y^2}^B ) \\
      ( d_{3z^2-R^2}^A + d_{3z^2-R^2}^B,\\ d_{3z^2-R^2}^A - d_{3z^2-R^2}^B ) \end {array} & A_1 \oplus A_2 & E & A' \oplus A'' \\
    \hline
  \end {tabular}\\
\vspace {0.2cm}
  \begin {tabular}{| >{$}c<{$} || >{$}c<{$} | >{$}c<{$} | >{$}c<{$} | >{$}c<{$} |}
    \hline
    \bP_\bX & \mbox {Fe states} & \bP_\Delta & \bP_Y & \bP_\bk \\
    \hline \hline
    E_{\bX g} & \begin {array}{c} d_{3z^2-R^2}^A \pm d_{3z^2-R^2}^B \\ d_{x^2-y^2}^A \pm d_{x^2-y^2}^B \\
      d_{xz}^A \mp d_{xz}^B \end {array}  & A_1 \oplus B_2 & E & A' \oplus A'' \\
    \hline
    E_{\bX u} & \begin {array}{c} d_{xy}^A \pm d_{xy}^B \\ d_{yz}^A \mp d_{yz}^B \end {array}  & A_2 \oplus B_1 & E & A' \oplus A'' \\
    \hline
  \end {tabular}
\end{center}
\caption{The symmetry classification of the tight-binding states, Eq.\ \eqref {kstate} at the three high symmetry points, $\Gamma$, $\bM$, and $\bX$,
  is shown in the second column of each table. In the vicinity of each high symmetry point, the symmetry is lower than it is at the
  high symmetry point. The remaining columns show the symmetry properties of the same states inherited from the symmetry
  properties (i.e., the irreducible representation) at the high symmetry point. Two-dimensional irreducible representations
  usually get split into two one-dimensional ones. Physically, that corresponds to an opening of a gap due to a reduced
  symmetry.} \label{TableLowering}
\end {table}

\begin{figure}[h]
\begin{center}
\includegraphics[width=0.48\textwidth]{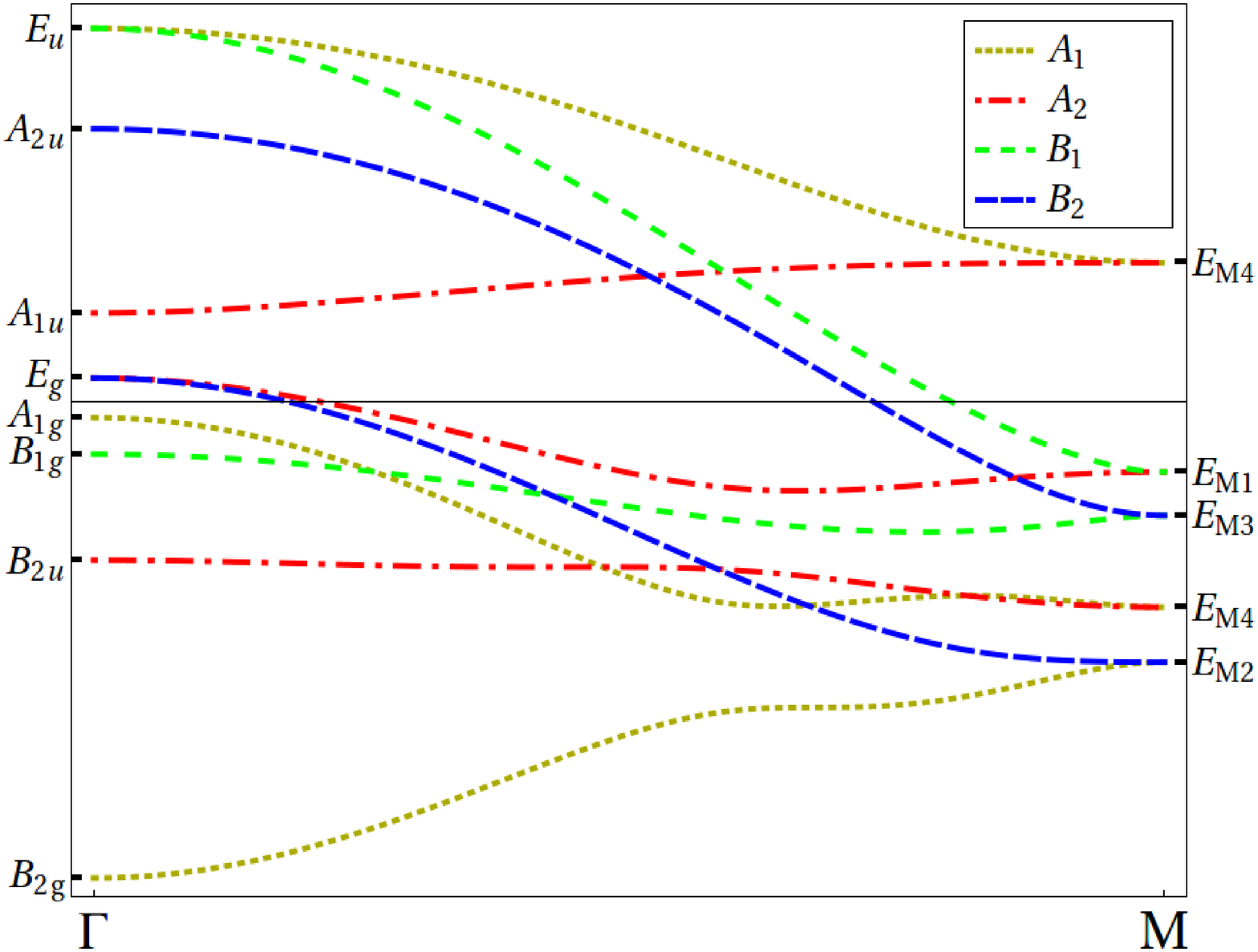} \\
a)
\includegraphics[width=0.48\textwidth]{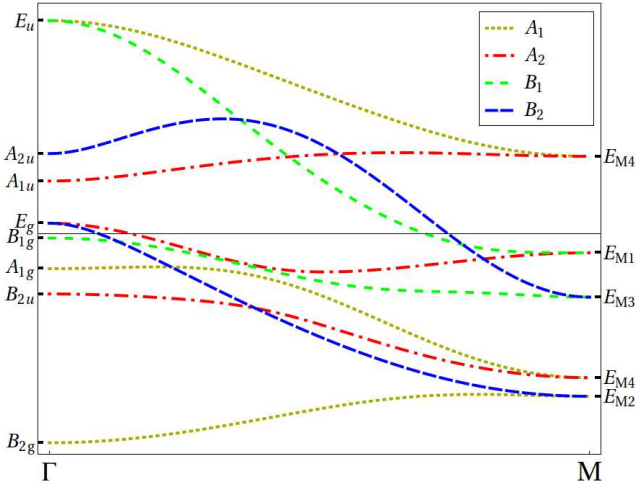} \\
b)
\end{center}
  \caption{The band structure along the $\Sigma = \Gamma - \bM$ line in the Brillouin zone (see Fig.\ \ref {FigUnitCell})
    reconstructed from the tight-binding model in Refs.\ a) \onlinecite {CvetkovicTesanovicEPL2009}, b) \onlinecite {KurokiPRL2008}.
    On the left-hand side of the plot, the symmetry properties of the bands at
    the $\Gamma$-point are shown; analogously, on the right-hand side, the symmetry properties for each band at the $\bM$-point are shown. In between,
    the color/dashing corresponds to the symmetry classification of the bands according to $\bP_\Sigma \cong {\bf C}_{2v}$ group.}\label{FigHamBands}
\end{figure}

\begin{figure}[h]
\begin{center}
\includegraphics[width=0.44\textwidth]{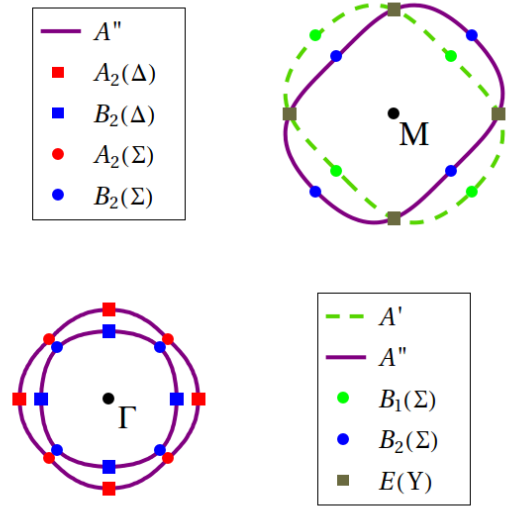}
\end{center}
  \caption{Fermi surfaces, reconstructed from the tight-binding model in Ref.\ \onlinecite {CvetkovicTesanovicEPL2009}.
    The color of each Fermi surface at an arbitrary momentum $\bk$ corresponds to the parity of states at the surface under the $\{ \sigma^z | \thalf \thalf \}$ mirror.
    Along the high symmetry directions the symmetry is augmented, and the states on the Fermi surfaces are classified accordingly.}\label{FigFermiSurfaces}
\end{figure}

Having recognized the symmetry properties of each tight-binding iron state, we diagonalize the Hamiltonians in Refs.\ \onlinecite {CvetkovicTesanovicEPL2009}
and \onlinecite {KurokiPRL2008} and establish the symmetry properties of each eigenstate based on its mixture of iron orbitals. We have
performed this classification for any $\bk$ in the Brillouin zone. For clarity, in Fig.\ \ref {FigHamBands} we present the band structure along
the $\Gamma-\bM$ (the $\Sigma$-line). As shown in the legend of the plot, the color/dashing coding for each band corresponds to the irreducible representation
of $\bP_\Sigma$ which gives the symmetry properties of that band. On the left- and right-hand side of the plot, the irreducible representations
for the bands at $\Gamma$- and $\bM$-points, respectively, are shown. Notice that the irreducible representations for each band at
the high symmetry points and along the $\Sigma$-line are matching in accordance with Table \ref {TableLowering}. The Fermi surfaces
for these two tight-binding models and the symmetry properties of the Fermi surface states are shown in Fig.\ \ref {FigFermiSurfaces}.
The bands and their symmetry properties shown in Figs.\ \ref{FigHamBands}a and  \ref{FigHamBands}b have the same outline. The major
difference is the order of bands at the $\Gamma$-point; however, it does not concern the bands which cross the Fermi level. These bands
have the same properties not only along the $\Sigma$-line, but at any momentum $\bk$, as seen from Fig.\ \ref {FigFermiSurfaces}.

The symmetry properties of each band crossing the Fermi level in iron-pnictides are now well understood. The two
hole bands originate at the $\Gamma$-point where they form a degenerate $E_g$-doublet. Along the
high symmetry lines, $\Sigma$ and $\Delta$, the doublet splits into one $A_2$ and one $B_2$ state. These two states are
not degenerate unless the Hamiltonian is fine-tuned. The upper band, responsible for the outer Fermi surface, has the symmetry properties governed
by $A_2$ irreducible representation. It is odd under $\{ \sigma^z | \thalf \thalf \}$ mirror, and
also odd under $\{ \sigma^x | 00 \}$ or $\{ \sigma^X | \thalf \thalf \}$ mirror. The other band, defining the inner hole Fermi surface, has
symmetry properties given by $B_2$ irreducible representation, hence it is odd under
the $\{ \sigma^z | \thalf \thalf \}$ mirror, and even under the other mirror of  $\bP_\Sigma$ or $\bP_\Delta$ group.
Away from the high symmetry lines, the states on the two hole bands are classified as $A''$, i.e.,
odd under $\{ \sigma^z | \thalf \thalf \}$ mirror, neglecting the phase due to the translation part, Eq.\ \eqref {spacegrouphomoinside}.

The symmetry properties of the electron bands in the vicinity of the $\bM$-point are quite different.
Let us first notice that the two electron bands in the vicinity of the $\bM$-point originate from the $E_{\bM 1}$ and $E_{\bM 3}$ doublet.
Along the $\Sigma$-line in the Brillouin zone the doublet with the higher energy at $\bM$, $E_{\bM 1}$, splits into one $A_2$ and
one $B_2$ state. The $A_2$ state is weakly dispersive and,  although it connects to the $E_g$ doublet at the $\Gamma$-point where it
crosses the Fermi level, we can consider that this state does not cross the Fermi level in the vicinity of the $\bM$-point. The other state,
$B_1$, has a steep dispersion such that it crosses the Fermi level close to the $\bM$-point. This is the state which corresponds to the
point on the minor axis of one of the electron Fermi surfaces. Observing the other doublet, $E_{\bM 3}$, we find that along the
$\Sigma$-line it splits into two states, $B_1$ and $B_2$. The $B_1$ state originating in the $E_{\bM 3}$ doublet has a nearly flat dispersion
and it does not cross the Fermi level. The other state, $B_2$, does cross the Fermi level, and this state corresponds to the point
on the major axis on the second elliptical Fermi surface near the $\bM$-point. As one moves away from the high symmetry line $\Sigma$,
the symmetry is reduced to ${\bf C}_s$. The two $B_1$ states are both even under $\sigma^z$ mirror, i.e., belong to the $A'$ irreducible representation.
The two other states, $A_2$ and $B_2$, are odd, i.e., transform according to $A_2$ irreducible representation. Notice that
these two states therefore mix at any $\bk$ which does not lie on the $\Sigma$-line. However, precisely at the $\Sigma$-line, these
two states split into two different irreducible representation and their respective bands must cross by the symmetry
constraints. Hence, these two bands form a Dirac point on the $\Sigma$-line.

By extending the analysis of the states to the edge of the Brillouin zone, the $Y$-line, we notice that the two upper bands
form one $E$-doublet that originates from the $E_{\bM 1}$-doublet at the $\bM$-point. It is the states of this doublet that
form two degenerate points where the Fermi surfaces intersect. One state is even and the other odd under the $\sigma^z$-mirror.
Similarly, the two lower bands, the ones that originate from the $E_{\bM 3}$-doublet and do not cross the Fermi level,
form an $E$-doublet along the $Y$-line. Crossing into the second Brillouin zone, we wish to connect the degenerate states
from the $Y$-line to the states on $\bM-\Gamma'$-line, where $\Gamma'=(2 \pi/a,0)$. It is important to notice that the splitting
of the $E_{\bM}$-doublets changes along this line as compared to the $\Sigma$-line. The reason is that along the $\bM-\Gamma'$-line,
the two symmetry mirrors are $\{ \sigma^X | \thalf \thalf \}$ and $\{ \sigma^z | \thalf \thalf \}$. Compare that to the $\Sigma$-line
where the first mirror was $\{ \sigma^Y | \thalf \thalf \}$. Consequently, the $E_{\bM 1}$-doublet splits into one $A_1$ and one $B_2$
state. The $B_2$ state is strongly dispersive and it crosses the Fermi level close to the $\bM$-point. The $A_1$ state is weakly
dispersive and it does not cross the Fermi level in the vicinity of the $\bM$-point. Even though the symmetry generating mirror
is different along the $\bM-\Gamma'$-line, the $E_{\bM }$ doublet is still split into one $B_1$ and one $B_2$ state. The difference,
with respect to the $\Sigma$-line, is that the $B_1$ state emanating from the $E_{\bM 3}$ crosses the Fermi level while the
$B_2$ state is weakly dispersive and it does not cross the Fermi level.

The bands with states that belong to the $A_1$ or $B_1$ irreducible representations  at the high symmetry line $\bM-\Gamma'$,
have all their states even under the $\{ \sigma^z | \thalf \thalf \}$ mirror, hence away from this line the states transform according to the $A'$
irreducible representation of ${\bf C}_s$ group. Conversely, the bands whose states at the $\bM-\Gamma'$-line transform
according to the $B_2$ irreducible representation are odd under the $\sigma^z$ mirror, and the states on these bands
transform as $A''$ at an arbitrary momentum $\bk$ away from the high symmetry lines.

\section{Low energy effective theory and spin-orbit coupling}

In the previous section we have identified the symmetry (i.e., the irreducible representation) of the states
within $\sim2eV$ of the Fermi level. This information, and the analogous classification of $\bk$-polynomials,
allows us to form all invariants that can appear in an effective Hamiltonian. We now construct such a low energy
effective Hamiltonian accurately describing the states which cross the Fermi level.

The non-interacting part of the Hamiltonian in the normal state
is
\be
\mathcal{H}_0=H_0+H_{so}.
\ee
The second term accounts for the spin-orbit coupling, and we discuss it in the next subsection. The first term is spin $SU(2)$ symmetric, and reads
\begin {align}
  H_0 = \sum_{\bk,\sigma=\uparrow,\downarrow} \psi_{\sigma}^\dagger(\bk) \left ( \begin {array}{c c c} h_\bM^+ (\bk) & 0 & 0 \\
    0 & h_\bM^- (\bk) & 0 \\ 0 & 0 & h_\Gamma (\bk) \end {array} \right ) \psi_{\sigma}(\bk), \label{eq:H0block}
\end{align}
where the six component (pseudo) spinor is
\be
\label{eq:spinor}
\psi_{\sigma}(\bk)=\left(\begin{array}{c}
\psi_{X,\sigma}(\bk) \\
\psi_{Y,\sigma}(\bk) \\
\psi_{\Gamma,\sigma}(\bk)
\end{array}\right).
\ee
For each spin projection, the upper component of $\psi_{X}$ transforms as $E^X_{\bM_1}$ and
the lower as $E^X_{\bM_3}$; the upper component of $\psi_{Y}$ transforms as $E^Y_{\bM_1}$ and the
lower as $E^Y_{\bM_3}$. Similarly, for each spin projection, the upper and the lower components of the spinor
$\psi_{\Gamma}(\bk)$ transform under the (axial vector) $E_g$ representation at the $\Gamma$ point as $Yz$ and $-Xz$, respectively.
The $2\times2$ blocks in Eq.\ (\ref{eq:H0block}) are
\begin{align}
  h_\bM^\pm ( \bk ) =&  \left ( \begin {array}{c c}
    \epsilon_1 + \frac {\bk^2}{2 m_1} \pm a_1k_x k_y
      & -i v_\pm ( \bk) \\
    i v_\pm ( \bk ) & \epsilon_3 + \frac {\bk^2}{2 m_3} \pm
      a_3 k_x k_y \end {array} \right ), \label{eq:hMpm} \\
  h_\Gamma ( \bk ) =&  \left ( \begin {array}{c c}
    \epsilon_{\Gamma} + \frac {\bk^2}{2 m_\Gamma} + b k_x k_y
    & c \left ( k_x^2 - k_y^2 \right ) \\
    c \left ( k_x^2 - k_y^2 \right ) & \epsilon_{\Gamma} + \frac {\bk^2}{2 m_\Gamma}
    - b k_x k_y \end {array} \right ), \label{eq:hGamma}
\end{align}
with
\be
  v_\pm (\bk) &=& v \left ( \pm k_x + k_y \right ) + p_1 \left ( \pm k_x^3 + k_y^3 \right ) \nonumber \\
  && ~ + p_2 k_x k_y \left ( k_x \pm k_y \right ). \label{vpm}
\ee
In the above, $\bk$ is in-plane and measured in units of the inverse lattice spacing.
The form of this Hamiltonian follows from the symmetry properties of $\psi$ and the product tables \ref {TabD4hProduct} and \ref {TableExE}
for the irreducible representations at $\Gamma$ and $\bM$.
Finding the spectrum of $H_0$ is now reduced to solving a simple $2\times2$ eigenproblem.

Note that we use the exact Bloch eigenstates at only two $\bk$-points, $\Gamma$ and $\bM$. By the
very nature of this $\bk\cdot\bp$ construction, we always deal with an {\it analytic} expansion of the
effective Hamiltonian in powers of the components of $\bk$. This is not the case within other approaches
to this problem \cite {ChubukovPRB2008, VorontzovPRB2009, CvetkovicTesanovicPRB2009, KorshunovPRL2009,
FernandesPRB2010, FernandesPRL2011, FernandesPRB2012, MaitiPRB2011, MaitiPRB2012,
KhodasPRB2012, LevchenkoarXiv2012, FernandesPRL2013, AvciarXiv2013}.
There one first constructs the exact Bloch eigenstates at {\it each} $\bk$, and then
uses such states as the basis for the expansion of the electron creation and annihilation operators. Among the
many advantages of our approach is its suitability in a study of the effects of a quantizing magnetic field,
since in Eqs.\ (\ref{eq:hMpm}) and (\ref{eq:hGamma}) the electro-magnetic vector potential $\bA$ can now be
minimally coupled \cite {Luttinger1956}. While
straightforward in principle, exploration of such effects, however, is beyond the scope of this paper.

To find the undetermined coefficients, we first diagonalize the tight-binding Hamiltonian at $\Gamma$ and
then use the resulting eigenvectors to construct the matrix elements of the Hamiltonian at an arbitrary $\bk$
away from $\Gamma$. The off-diagonal matrix elements vanish as $\bk$ vanishes. Therefore, we can integrate
out all the states except for the $E_g$ doublet (see Fig.\ \ref {FigHamBands}) and expand \cite{note:blgHeff} the resulting $2\times2$
effective Hamiltonian to second order in $\bk$. An analogous procedure was carried out at $\bM$, except we kept the
$E_{\bM 1}$ doublet and the $E_{\bM 3}$ doublet, making the effective Hamiltonian at the $\bM$-point $4\times4$.
Still, our choice of basis in Eq.\ (\ref{eq:spinor}), and our ignoring of any out-of-plane $k_z$-momentum dispersion,  allows
us to write such $4\times4$ Hamiltonian in a block diagonal form. Depending on the tight-binding Hamiltonian we use,
Ref.\ \onlinecite{CvetkovicTesanovicEPL2009} or Ref.\ \onlinecite {KurokiPRL2008}, such method  results in the values tabulated in
Table \ref {TableLoEff} and stated in ${\rm meV}$'s.

\begin {table}[h]
\begin {center}
  \begin {tabular}{| c || >{$}c<{$} | >{$}c<{$} | >{$}c<{$} | >{$}c<{$} | >{$}c<{$} | >{$}c<{$} |}
    \hline
    ~ & \epsilon_\Gamma & \epsilon_1 & \epsilon_3 & 1/ \left (2 m_\Gamma \right ) & 1 / \left ( 2 m_1 \right ) & 1 / \left ( 2 m_3 \right ) \\
    \hline \hline
    Ref.\ \onlinecite {CvetkovicTesanovicEPL2009} & 132 & -400 & -647 & -184 & 149 & 317 \\
    Ref.\ \onlinecite {KurokiPRL2008} & 100 & -180 & -600 & -462 & -65.9 & 322 \\
    \hline
  \end {tabular}

\vspace* {0.2cm}
  \begin {tabular}{| c || >{$}c<{$} | >{$}c<{$} | >{$}c<{$} | >{$}c<{$} | >{$}c<{$} | >{$}c<{$} | >{$}c<{$} |}
    \hline
    ~ & a_1 & a_3 & b & c & v & p_1 & p_2 \\
    \hline \hline
    Ref.\ \onlinecite {CvetkovicTesanovicEPL2009} & 419 & -533 & 56.5 & -62.3 & -243 & -40 & 10   \\
    Ref.\ \onlinecite {KurokiPRL2008} & 41.8 & -384 & 438 & 244 & 99.0 & 39.1 & 0.99   \\
    \hline
  \end {tabular}
\end {center}
\caption {The parameters of the low energy effective model, Eq.\ (\ref {eq:hMpm}-\ref {eq:hGamma}), derived from the
  two tight-binding models referenced in the text.} \label{TableLoEff}
\end {table}

\begin{figure}[h]
\begin{center}
\includegraphics[width=0.48\textwidth]{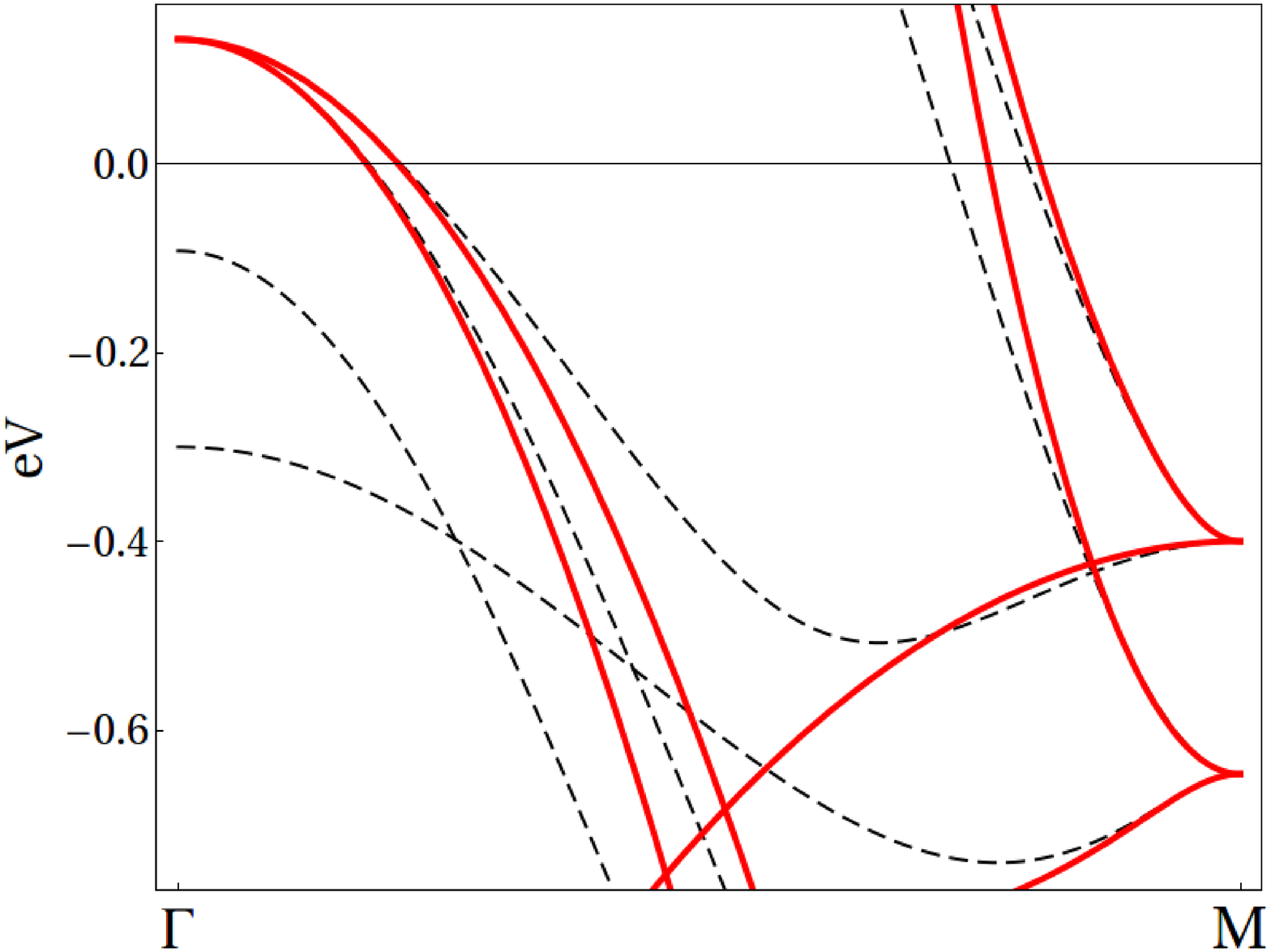} \\
a)
\includegraphics[width=0.48\textwidth]{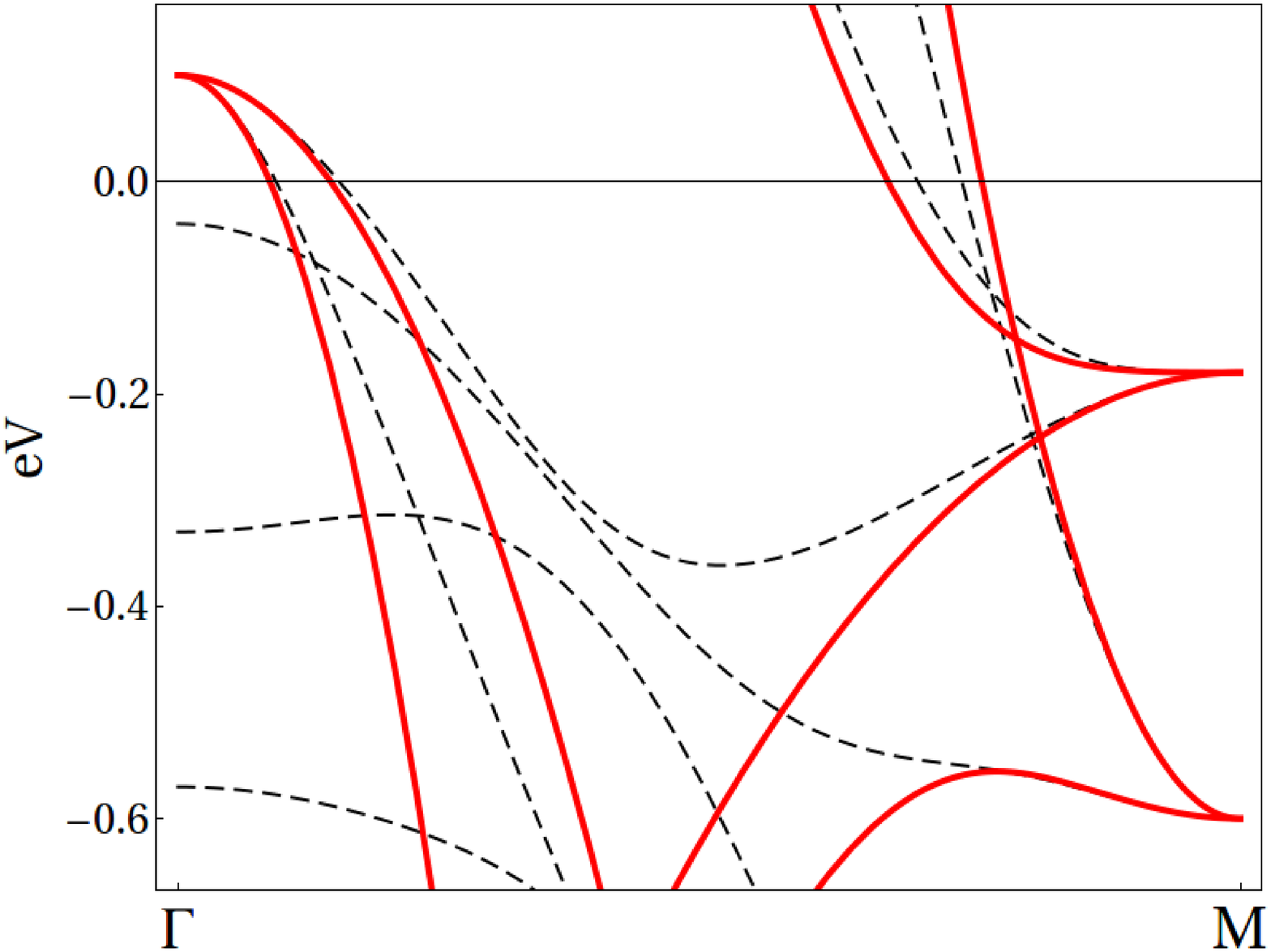} \\
b)
\end{center}
  \caption{The comparison between the band structures for the tight-binding model (black, dashed)
    and the low energy effective theory (red, solid) along the $\Sigma$-line in the Brillouin zone.
    The two panes correspond to a) Ref.\ \onlinecite {CvetkovicTesanovicEPL2009} and b) Ref.\ \onlinecite {KurokiPRL2008}.
    Comparing the band structures in the vicinity of the Fermi level, we find that the low energy effective theory
    matches the tight-binding band structure almost perfectly near the $\Gamma$-point and the two are close in
    the vicinity of the $\bM$-point.}\label{FigCompareTB2LoEff}
\end{figure}

\begin{figure}[h]
\begin{center}
\includegraphics[width=0.234\textwidth]{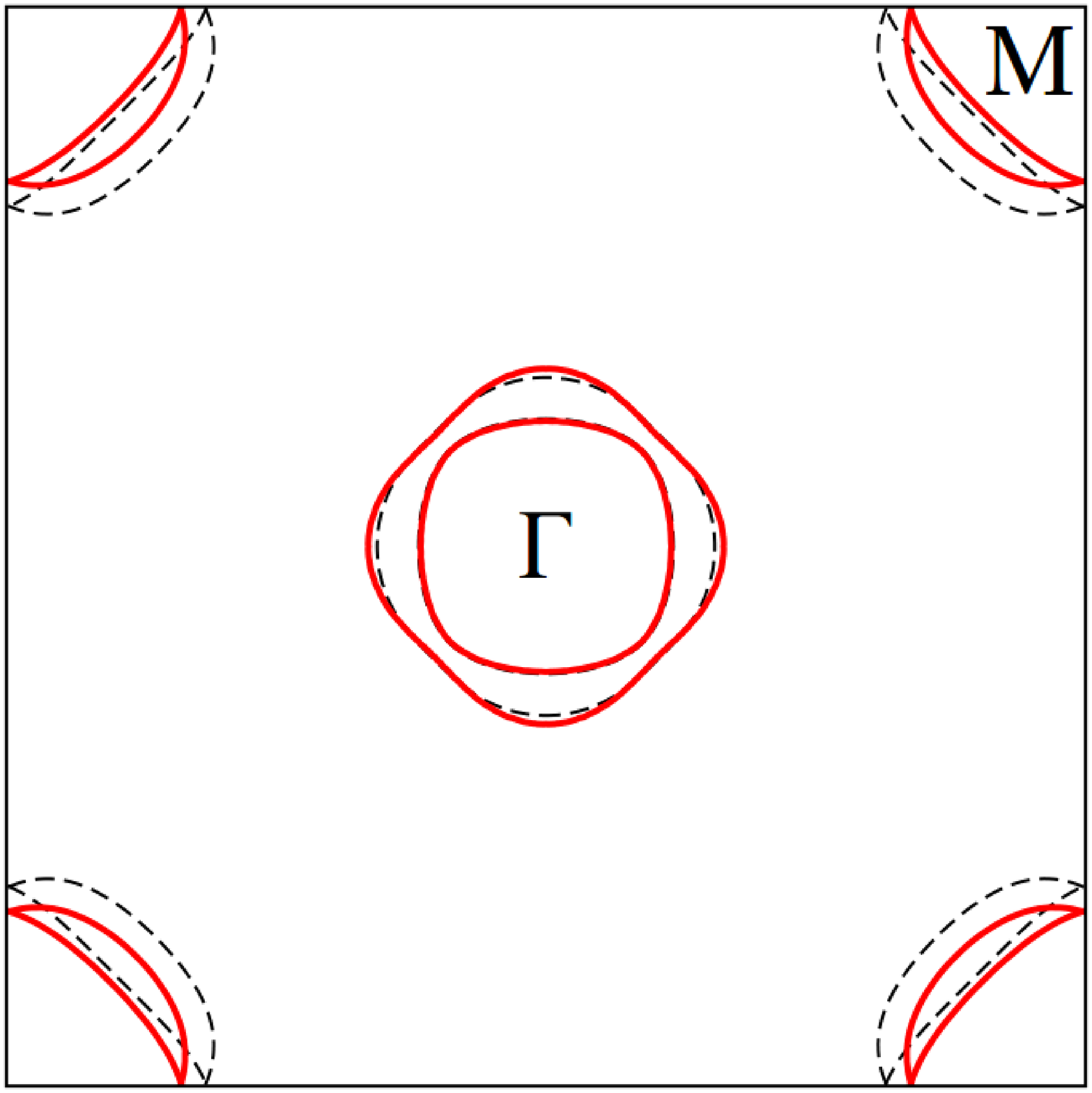}
\includegraphics[width=0.234\textwidth]{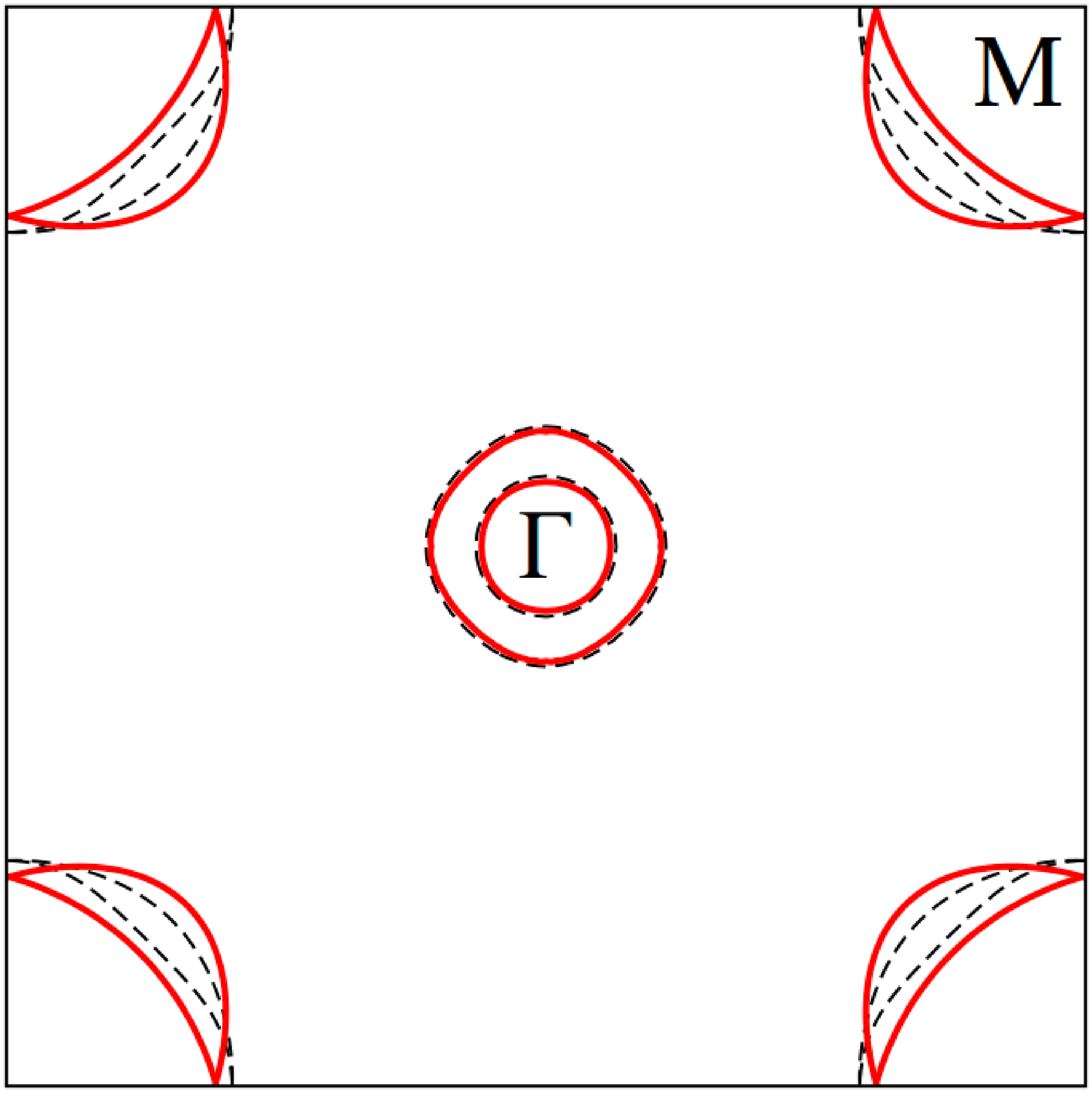} \\
a) \hspace {0.22\textwidth} b)
\end{center}
  \caption{The comparison between the Fermi surfaces in the tight-binding model (black, dashed)
    and the low energy effective theory (red, solid).
    The two panes correspond to a) Ref.\ \onlinecite {CvetkovicTesanovicEPL2009} and b) Ref.\ \onlinecite {KurokiPRL2008}.}\label{FigCompareFS}
\end{figure}

The spectrum obtained from Eq.\ (\ref{eq:H0block}) (red, solid) is compared to the full tight-binding
dispersion (black, dashed) along the $\Sigma$-line in Fig.\ \ref {FigCompareTB2LoEff}.
The Fermi surfaces are compared in Fig.\ \ref {FigCompareFS}.
For the hole bands, the Fermi surfaces are captured with a marked precision; the hole bands dispersion of the low-energy
effective model does not deviate significantly from the dispersions of the tight-binding models for the energies of order of few
hundred ${\rm meV}$'s from the Fermi level. For the electron bands, the shape of their Fermi surfaces is well
captured by the low-energy effective model, although there are small quantitative deviations.  The dispersions too have small deviations from the tight-binding bands in the vicinity of the Fermi level. The fit may be improved by
including higher order terms (quartic) in $\bk$ in Eqs.\ \eqref {eq:hMpm} and \eqref {eq:hGamma}.

\subsection{Spin-orbit coupling}

In order to include the effects of spin-orbit coupling, we begin by analyzing the transformations of charge
neutral Fermion bilinear operators under the symmetry operations $R = \{ g | \btau \}$. We have
\begin {align}
  &\sum_{c,\nu,d}D_{ac,\mu\nu,bd} (R) \left \lbrack \psi_\alpha^{c*} \sigma^\nu_{\alpha \beta}  \psi_\beta^d \right \rbrack = \nonumber\\
  & \ \left ( \sum_{c} D_{ac} (R)^* \psi_\alpha^{c*} \right)\!\!\!
   \left (\sum_{\nu=0}^3 D^{AV}_{\mu \nu} (R) \sigma^\nu_{\alpha \beta} \right)\!\!\! \left(\sum_{d} D_{bd} (R) \psi_\beta^d \right ).  \label{eq:Dspinbilinear}
\end {align}
In the above, the sum over $\alpha=\uparrow,\downarrow$ and $\beta=\uparrow,\downarrow$ should be understood.
Here, $\sigma^\mu$ is either a Pauli spin matrix $\mu = j = x, y, z$ or a unity matrix, $\sigma^0 = \bbone_2$. In
Eq.(\ref{eq:Dspinbilinear}) $a$, $b$, $c$ and $d$ may be any one of the six components the spinor. For each $R$, the non-zero elements
of matrix $D_{ab}(R)$ are $\left(\begin{array}{cc}D_{11}(R) & D_{13}(R)\\ D_{31}(R) & D_{33}(R)\end{array}\right)=E_{\bM_1}(R)$,
$\left(\begin{array}{cc}D_{22}(R) & D_{24}(R)\\ D_{42}(R) & D_{44}(R)\end{array}\right)=E_{\bM_3}(R)$,
and $\left(\begin{array}{cc}D_{55}(R) & D_{56}(R)\\ D_{65}(R) & D_{66}(R)\end{array}\right)=E_{g}(R)$. The transformation
properties of the Pauli spin matrices $\sigma^{\mu}$ follow from $\sigma^0\equiv \bbone$ corresponding to spin singlet,
and $\sigma^{1}$, $\sigma^{2}$ and $\sigma^{3}$ corresponding to spin triplet. Naturally, a spin singlet is
left invariant under all space group transformations, hence for each $R$ we have
$D_{00}^{AV} (R) = 1$ and $D_{0j}^{AV} (R) = D_{j0}^{AV} (R) = 0$. On the other hand, the spin triplet operators
transform according to the axial vector representation of the point group component $R$ in $R= \{ g | \btau \}$, and are left
unaffected by the translation by $\btau$. Therefore, $D_{ij}^{AV} (R) = g_{ij}^{AV}$.
The action of various symmetry operations in $P4/nmm$ on the spin operators $\sigma^j$ can now be readily obtained
from the action of the group generators. For $\mu=i$ and $\nu=j$,
\be
  D^{AV} \left ( \{ \sigma^X | \thalf \thalf \} \right ) &=& \diag (1,-1,-1), \label{DAVsigmaX} \\
  D^{AV} \left ( \{ \sigma^Z | \thalf \thalf \} \right ) &=& \diag (-1,-1,1), \label{DAVsigmaZ} \\
  D^{AV} \left ( \{ \sigma^x |  00 \} \right ) &=& \left ( \begin {array}{c c c} 0 & -1 & 0 \\ -1 & 0 & 0 \\ 0 & 0 & -1 \end {array}  \right ). \label{DAVsigmax}
\ee

In order to find the terms in $H_{so}$, i.e., the invariants of the form
$M^{\mu}_{a,b}\psi_{\alpha}^{a*}\sigma_{\alpha\beta}^{\mu}\psi_{_{\beta}}^b$ where the repeated indices
are summed over, we look for the identity representation in the product of the three representations on the right hand side of Eq.\ (\ref{eq:Dspinbilinear}).

The axial vector representation of ${\bf D}_{4h}$ is decomposed into
$A_{2g} \oplus E_g$. The $z$-component of the axial vector transforms according to the $A_{2g}$ representation, while  $\sigma^1$
and $\sigma^2$ transform as an $E_g$ doublet.
Recall that the low-energy states at the $\Gamma$-point transform as an $E_g$ doublet of ${\bf D}_{4h}$.
We can therefore form four independent bilinears composed out of these two states which transform according
to $E_g \otimes E_g = A_{1g} \oplus A_{2g} \oplus B_{1g} \oplus B_{2g}$. It follows that there is only one $\bk$-independent
invariant in $D^{AV}\otimes E_g \otimes E_g$ and that it only contains $\sigma^3$.

Similarly, for the low energy states at the $\bM$-point, we have
$E_{\bM_1}\otimes E_{\bM_1} = A_{1g}\oplus B_{2g} \oplus A_{2u} \oplus B_{1u}$ and $E_{\bM_3}\otimes E_{\bM_3}=A_{1g} \oplus B_{2g} \oplus A_{1u} \oplus B_{2u} $.
Therefore, there are no symmetry allowed, $\bk$-independent, spin-orbit terms with {\it both} $\psi^*$ {\it and} $\psi$ transforming
under $E_{\bM_1}$, or both transforming under $E_{\bM_3}$. On the other
hand, $E_{\bM_1} \otimes E_{\bM_3} = E_g \oplus E_u$. Therefore, there is
a $\bk$-independent spin-orbit term mixing $E_{\bM_1}$ and $E_{\bM_3}$ states, containing the Pauli spin matrices $\sigma^1$ and $\sigma^2$, but not $\sigma^3$.

Putting it all together, we find that
\be
  H_{so} = \sum_{\bk}\sum_{\sigma,\sigma'} \psi_{\sigma}^\dagger(\bk) \left ( \begin {array}{c c c} 0 & h^{so}_{\bM,\sigma\sigma'} & 0 \\
    \left ( {h^{so}_{\bM}}^\dagger  \right )_{\sigma\sigma'} & 0 & 0 \\ 0 & 0 & h^{so}_{\Gamma,\sigma\sigma'} \end {array} \right ) \psi_{\sigma'}(\bk),\nonumber\\ \label{eq:H0blockSO}
\ee
where
\be
  h^{so}_{\Gamma,\sigma\sigma'} &=& \frac 1 2 \lambda_{\Gamma}\left ( \begin {array}{c c}
    0
    & -i \\
    i & 0 \end {array} \right )\sigma_{\sigma\sigma'}^3, \label{eq:hGammaso}
\ee
and
\be
  h^{so}_{\bM,\sigma\sigma'} &=& \frac i2 \lambda_{\bM}\left[\left ( \begin {array}{c c}
    0 & 1 \\
    0 & 0 \end {array} \right )\sigma_{\sigma\sigma'}^1+\left ( \begin {array}{c c}
    0 & 0 \\
    1 & 0 \end {array} \right )\sigma_{\sigma\sigma'}^2\right]. \label{eq:hMso}
\ee
 Eq.\ (\ref{eq:hGammaso}) resembles the term introduced by Kane and Mele in the single layer graphene \cite {KaneMelePRL2005}. In fact,
at the $\Gamma$-point there is a close similarity with the effective Hamiltonian for the bilayer graphene, the difference
being the large particle-hole asymmetry term proportional to $\bk^2$ which is negligible in the bilayer graphene.

In addition to these, $\bk$-independent, spin-orbit terms, the symmetry of the lattice allows several other $\bk$-dependent terms
in the Hamiltonian. We may, however, neglect these because they originate from higher order hopping processes and are much smaller than the on-site
terms, Eqs.\ \eqref{eq:hGammaso} and \eqref{eq:hMso}.

\begin {figure}[h]
\begin {center}
\includegraphics[width=0.45\textwidth]{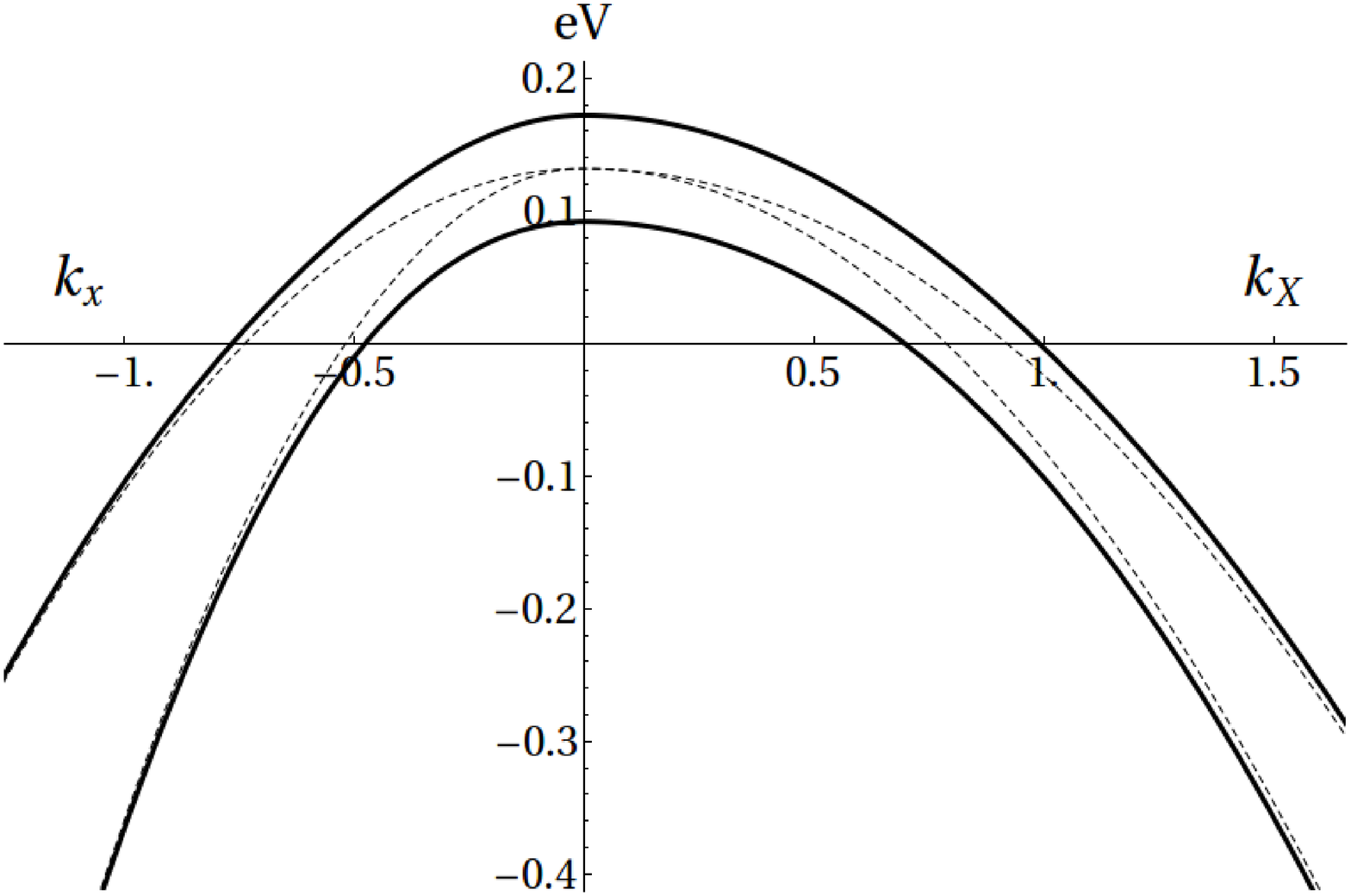}\\
a) \\
\includegraphics[width=0.45\textwidth]{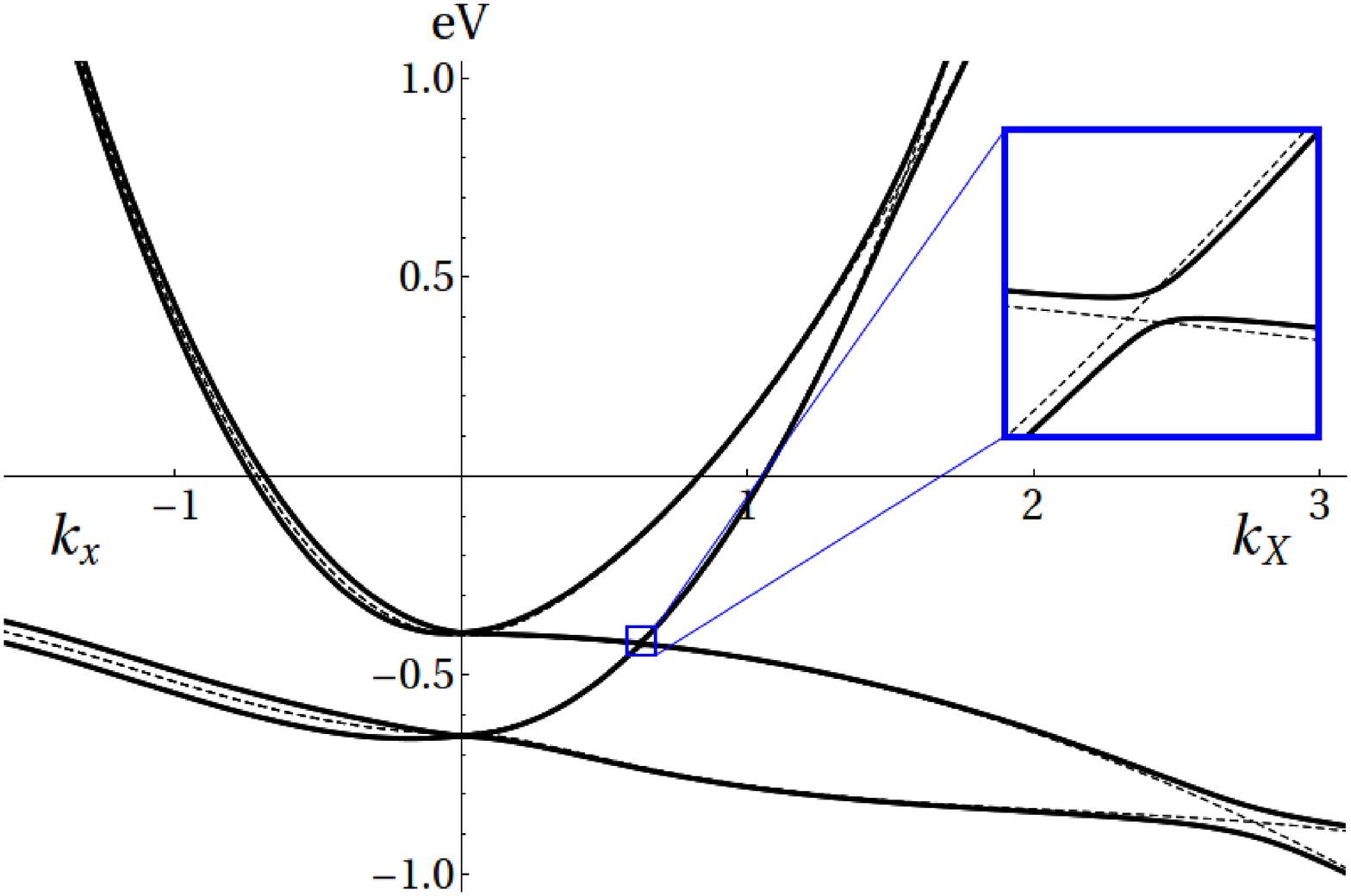} \\
b)
\end{center}
  \caption{The dispersion of the low-energy effective theory without (thin-dashed lines) and with (thick-solid line) spin-orbit effects included.
    The two panels correspond to the effective theory near the a) $\Gamma$-, and b) $\bM$-point. The strength of the
    spin-orbit coupling we use is $\lambda_\Gamma = \lambda_\bM = 80{\rm meV}$. Each band is doubly degenerate regardless the presence of
    the spin-orbit coupling. No two bands cross except at the $\bM$-point where bands meet in pairs and each state is four-fold degenerate.}
    \label{FigDispersionsWithSO}
\end{figure}

If Figs.\ \ref{FigDispersionsWithSO}a and \ref{FigDispersionsWithSO}b,
we compare the dispersion obtained using the low energy effective Hamiltonian $\mathcal{H}_0$ with and
without the spin-orbit term $H_{so}$ in Eq.\ (\ref{eq:H0blockSO}). We use the values $\lambda_\Gamma=\lambda_\bM=80{\rm meV}$
obtained from from Ref.\ \onlinecite {TiagoPRL2006}. Both with and without spin-orbit coupling interaction, we find that each band is doubly degenerate.
This degeneracy is an obvious consequence of the spin double degeneracy in the absence of spin orbit coupling. When the spin-orbit
coupling is present, the double degenerate states form Kramers doublets --- two states related to each other by a combined operation,
time reversal $\Theta$ followed by ``glide"-inversion $\{i|\frac{1}{2}\frac{1}{2}\}$, must be degenerate since this operation is a
symmetry of the Hamiltonian at any $\bk$-point in the Brillouin zone. Therefore, each band in Figs.\ \ref{FigDispersionsWithSO}a and \ref{FigDispersionsWithSO}b,
as well as generally in iron-pnictides, is doubly degenerate. We find that, in the presence of spin-orbit interaction, any two bands never cross
except at the $\bM$-point, where we find {\it only}  Dirac-like points (see Fig.\ \ref {FigDispersionsWithSO}b).
The Dirac points are guaranteed by the space group symmetry and cannot be removed without lowering this symmetry.

\subsection {Comparison  to other models for iron-pnictides}

After we have constructed our low-energy effective model for iron-pnictides, one may ask how does this
model compare to the existing models for iron pnictides and what are its advantages relative to other models.

The effective itinerant models for iron-pnictides can be broadly divided into two groups, those which  use
orbital or Wannier states as their basis, and those which use the exact eigenstates of the Hamiltonian, i.e.,
bands, as their eigenstates. In an earlier subsection, we have already discussed the advantages our effective model has in comparison to
the band-effective models . Among other issues, a model which uses the eigenstates of the Hamiltonian as its basis
suffers from the non-analyticity of the basis at the $\bM$-point. This may introduce problems
in dealing with the states in the vicinity of that point, as well as issues with the interaction terms.

The models which use orbital or Wannier states differ by the number of orbitals/states per iron atom in the model.
The simplest ones are two- and three-orbital models and we discuss these in the following subsubsections.
Speaking of models with four, five, or more orbitals, one can say that they reproduce the band structure of iron-pnictides with  a great
accuracy and preserving the proper symmetries. In fact, we rely on five orbital models for establishing the symmetry properties of the
states at the high symmetry points and also as a source for the numerical values of our model.  The problem with these models,
however, is that they carry too much information which is often irrelevant, but adds to the complexity of the problem.  Particularly
in weak-coupling problems, one should not have to worry about degrees of freedom which are far above or below from the Fermi level.
Within tight-binding models, removing these degrees of freedom, as we shall see, cannot be done without an impact on the qualitative features of the
spectrum.

\subsubsection {Two-orbital models}

The first and the simplest effective model for iron-pnictides, was introduced in Ref.\ \onlinecite{RaghuPRB2008}.
This two orbital model has been preferred due to its appealing simplicity.

The construction of the model is based on the observation that both hole and electron states at the Fermi surfaces
of iron-pnictides have the highest overlap with the $d_{xz}$ and $d_{yz}$ states. The model uses
only these two orbitals as the basis, and with only the nearest and next-nearest hopping processes between iron atoms, it
is able to reproduce the characteristic Fermi surfaces of iron based superconductors: two hole pockets around the $\Gamma$-point and
two electron pockets around the $\bM$-point, once the folding of the Brillouin zone, due to the doubling of the
proper unit cell, is taken into account.

\begin{figure}[h]
\begin{center}
\includegraphics[width=0.4\textwidth]{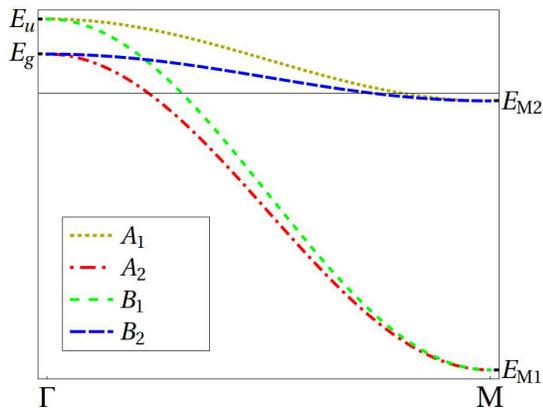}\\
\end{center}
  \caption{The band structure of the two-orbital effective model for iron-pnictides along the $\Sigma = \Gamma - \bM$ line based on Ref.\ \onlinecite{RaghuPRB2008}. Comparison with the Fig.\ \ref{FigHamBands} shows that the model
    does not reproduce the symmetry for all bands crossing the Fermi level.}\label{FigRaghu}
\end{figure}

The entire band structure of this minimal model for iron based superconductors is given in Ref.\ \onlinecite{RaghuPRB2008}. Using
the parameters given in that paper, we reconstructed the result and in Fig.\ \ref{FigRaghu}
plotted the band structure along the $\Sigma = \Gamma-\bM$ line
for comparison with the five band models, Fig.\ \ref{FigHamBands}. The two bands
crossing the Fermi level near the $\Gamma$-point correspond to the hole bands, the two bands that
cross the Fermi level near the $\bM$-point correspond to the electron bands. Clearly, this model can
reproduce the symmetry of neither hole nor electron bands. For example,  while one of the hole bands
originates from the $E_g$-doublet at the  $\Gamma$-point, and has $A_2$ symmetry properties along this line,
the other state originates from the wrong, $E_u$, doublet, and accordingly has $B_1$ and not $B_2$ symmetry,
as it should. Since $B_1$ states are even under the $\{ \sigma^z | \thalf \thalf \}$ mirror, any states on this
Fermi surface are even under this mirror, i.e., belong to $A'$ irreducible representation of $\bP_\bk$. Similarly,
both electron bands near the $\bM$-point originate in $E_{\bM 2}$-doublet. While one band, $B_2$, has the
proper symmetry along the $\Sigma$-line, the other band transforms as $A_1$, not $B_1$ as it should.

More recently, another two-orbital model was presented in Ref.\ \onlinecite{JHuPRX2012}.
This model uses the same states as Ref.\ \onlinecite{RaghuPRB2008}, but it includes hoppings up to third
nearest iron neighbor. We believe that this model violates mirror reflection symmetry about the $xz$- and $yz$- planes. Such reflection symmetries force $t_{1x}=-t_{1y}$ and $t_{3x}=t_{3y}$, implying $t_{1s}=0$ and $t_{3d}=0$ in Eq.\ (5) of the Ref.\ \onlinecite{JHuPRX2012}; such symmetry has not been reinforced as can be seen from the explicit values stated on page 6 of Ref.\ \onlinecite{JHuPRX2012}.

Can any modifications of the two-orbital model recover the correct symmetry
for its bands? We would argue that the answer is no --- this model does not have sufficient ingredients to match all the
bands parities correctly. The first problem is that it does not include the $E_{\bM 3}$-doublet states
at the $\bM$-point, inevitably missing states which in more accurate band structure calculations describe the electron bands. Further, the
parity of the bands under the $\{ \sigma^z | \thalf \thalf \}$ mirror does match: there are
two Fermi surfaces with odd, and two Fermi surfaces with even states under the mirror. A more accurate model must
reproduce one electron Fermi surface which is even under the $\{ \sigma^z | \thalf \thalf \}$ mirror, and the three
remaining Fermi surfaces (two hole and one electron) which are odd under this mirror.
We therefore conclude that any two-orbital model is unable to reproduce the symmetry properties of Fermi surface
states in iron-pnictides.

\subsubsection {Three-orbital models}

The next in complexity are the three-orbital models. The two representative three-orbital models we discuss here
were introduced in Refs.\ \onlinecite {LeeWenPRB2008} and \onlinecite {DaghoferPRB2010}. In both of these models
an additional iron orbital, $d_{x^2-y^2}$ is included in the basis of the effective model. This is motivated by the
ab initio calculations which demonstrate that the electron Fermi surface states have high overlap with
the Bloch states composed of the $d_{x^2-y^2}$ orbital.

\begin{figure}[h]
\begin{center}
\includegraphics[width=0.4\textwidth]{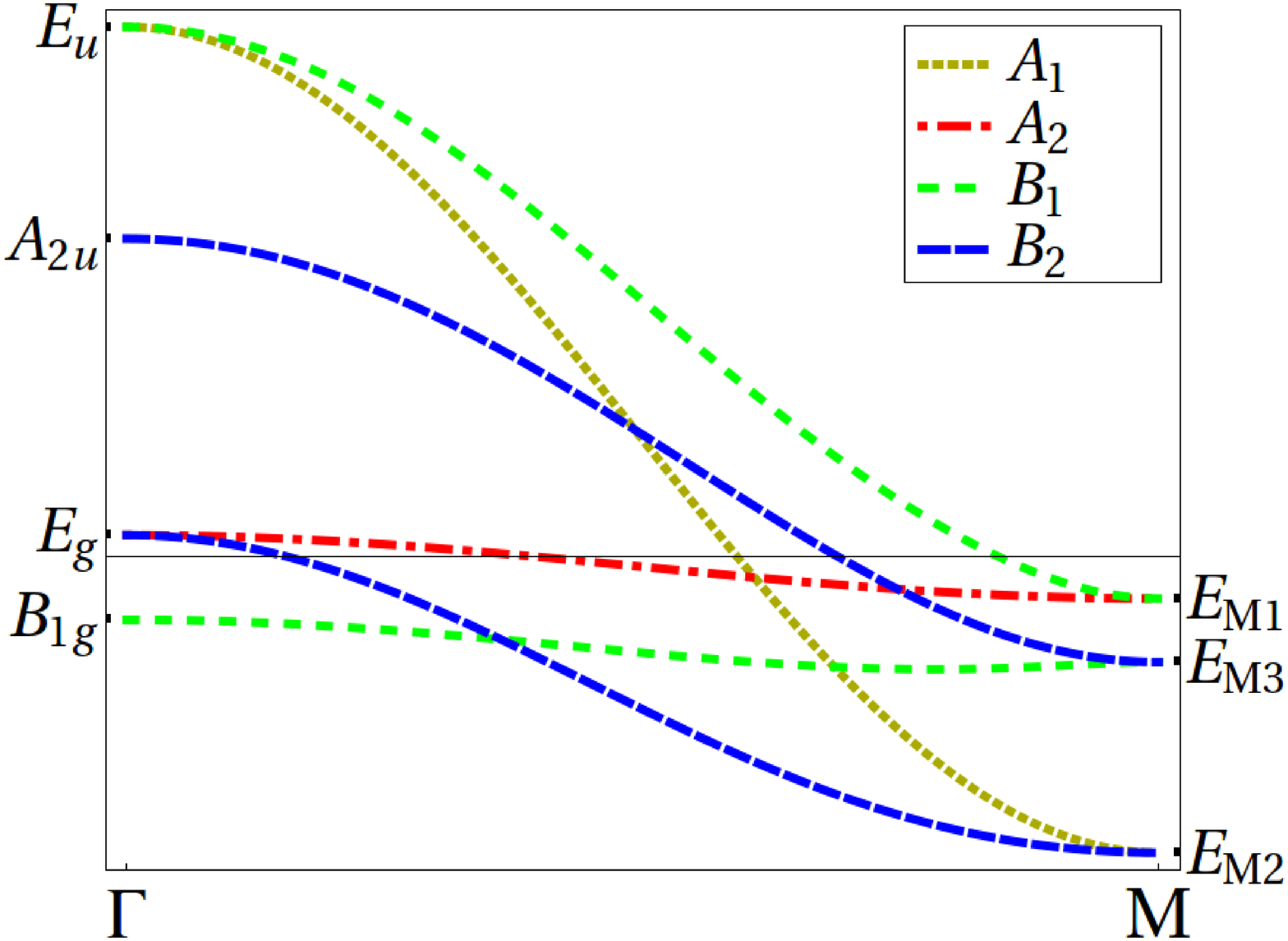}\\
a)\\
\includegraphics[width=0.4\textwidth]{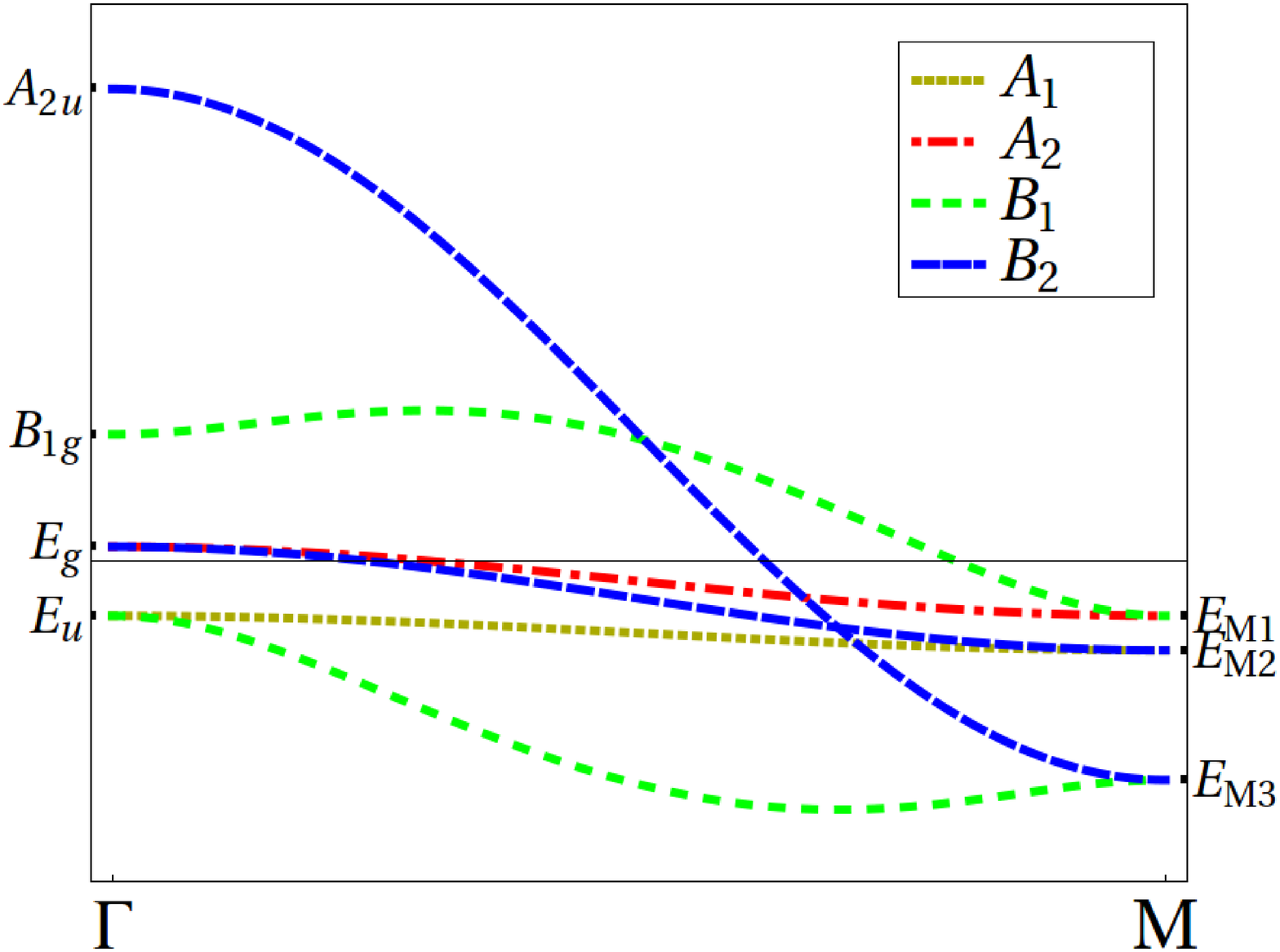} \\
b)
\end{center}
  \caption{The band structure in three-orbital effective models for iron-pnictides along the $\Sigma = \Gamma - \bM$ line, reconstructed
    from Refs.\ a) \onlinecite {LeeWenPRB2008} and b) \onlinecite {DaghoferPRB2010}. The symmetry properties of the hole and electron  Fermi surfaces
    states are correctly reproduced in both models, however, in a) a spurious Fermi surface is present; in b) there are no spurious
Fermi surfaces, but the ordering of the bands at the $\Gamma$- and $\bM$-points is incorrect. }\label{FigWenLee}
\end{figure}

In Fig.\ \ref {FigWenLee} we plot the dispersions for these two models along the $\Sigma$-line. We notice that
both models are able to properly reproduce the symmetry properties of the bands crossing the Fermi level: the
two hole bands originate in the $E_g$-doublet at the $\Gamma$-point; the electron bands near the $\bM$-point
originate from $E_{\bM 1}$- and $E_{\bM 3}$-doublets. Nevertheless, each of these two models is unable to
reproduce the qualitative aspects of the iron-pnictides band structure.

The band structure of the model from Ref.\ \onlinecite {LeeWenPRB2008} contains one additional band
which crosses the Fermi level, thus leading to one excess Fermi surface. The authors acknowledge this, and
conclude that, in order to remove this spurious Fermi surface, the model must include at least one additional
orbital. We find these statements to be in agreement with the symmetry properties of the bands. Since the band
which creates the spurious Fermi surface transforms according to $A_1$ along the $\Sigma$-line, it is
sufficient to include either $d_{xy}$ orbital in the model. One of the bands corresponding to this orbital
transforms according to the $A_1$ irreducible representation along the $\Sigma$-line. Since the band which produces
the spurious Fermi surface also transforms according to the $A_1$, the two bands must hybridize. At the
$\Gamma$- and $\bM$-points, where the hybridization is absent, the $d_{xy}$ state has a negative and
positive eigenvalue, respectively. Therefore, for large enough hybridization the spurious Fermi surface is removed,
but the price paid is the introduction of a new orbital.

By cleverly arranging the hopping amplitudes in the three-orbital model, Ref.\ \onlinecite {DaghoferPRB2010}
arrives at the band structure where the spurious Fermi level is absent. By looking at the dispersion in Fig.\ \ref {FigWenLee}
we find that this is achieved by moving the energy of the $E_u$-states at the $\Gamma$-point below the Fermi level,
while changing the energy of a single $B_{1g}$ state from negative to positive. In addition, for the parameters given in
Ref.\ \onlinecite {DaghoferPRB2010}, this model makes the energy of the $E_{\bM 2}$-doublet higher than that of the
$E_{\bM 3}$-doublet. Thus, this model violates the energy ordering of the states at both $\Gamma$- and $\bM$-points.
One can easily conclude, by looking at Fig.\ \ref {FigWenLee}, that any three-orbital model must either a) have a spurious
Fermi surface, or b) have the states at the $\Gamma$-point ordered incorrectly.

By the number of degrees of freedom, i.e., states per iron atom, the three-orbital models are equivalent to our low-energy effective theory.
Any model which would include additional orbitals in order to remedy the shortcomings of the three-orbital models is,
by construction, more complicated than the model we introduced.  Therefore, we  do not need to compare such models to the low-energy
effective model we introduce, and can safely assume that our model is the simplest model for iron-pnictides which respects the symmetries
of the states near the Fermi level.

\subsection{Electron-electron interactions without the spin-orbit interaction}

In this part of the section we present the term in the Hamiltonian which describes the contact electron-electron interaction. Such
interaction may lead to the symmetry breaking orders, such as spin-density waves or a
superconducting pairing. How the interaction terms lead to the broken symmetries is beyond the
scope of this paper and is left for the future. Nevertheless, for completeness, we present the corresponding  terms given that they  follow naturally
from the symmetry classification.
We focus only on the quartic term in this paper, but following similar steps, one can have a straightforward derivation of
higher order terms if these are desired.

Focusing on the quartic term, the translational invariance implies that it can be split into two parts,
\be
  {\mathcal H}_{\rm int} = H_{\rm int}^{(0)} + H_{\rm int}^{(\bM)}. \label{Hint_split}
\ee
The first term, $H_{\rm int}^{(0)}$, contains terms that are products of two charge neutral bilinears, each carrying no momentum,
therefore representing small momentum transfer processes. In the second term, $H_{\rm int}^{(\bM)}$, each
term is a product of two bilinears, both having momentum $\bM$, and it describes interaction processes
where large momentum $\bM$ is exchanged. The Umklapp processes are contained in $H_{\rm int}^{(\bM)}$.

A product of two bilinears has the symmetry properties given by the product of the individual representations. The product
table for the irreducible representations of ${\bf D}_{4h}$ implies that our quartic terms --- a product of two bilinears --- transforms under
the trivial representation only if both bilinears belong to the same irreducible representation. The time reversal
invariance similarly implies that bilinears must be either both even or both odd in order for the product to be
time reversal invariant.

Given a spin singlet bilinear $\sum_{\sigma} \psi^\dagger_\sigma (\br) \Gamma_{i, j}^{(m)} \psi_\sigma (\br)$,
the transformation of $\psi$'s under the space group operations and the time-reversal induce the transformations on $\Gamma_{i, j}^{(m)}$. The
36-dimensional representation, given by the $\Gamma$'s has been reduced and presented in Appendix \ref {AppGammas}.
For one-dimensional irreducible representations
$m=1$. For the two-dimensional ones, $m=1$ or $m=2$ denoting the component of the irreducible representation. Two
irreducible representations, $A_{1g}^+$ and $B_{2g}^+$, corresponding to $i=1$ and $i=7$ respectively, have multiplicity $M_{1} = M_{7} = 3$.
In other words,  there are three linearly
independent bilinears, enumerated by $j=1, 2, 3$, that transform according to $A_{1g}^+$ and $B_{2g}^+$ respectively. For
all the other irreducible representations, the multiplicity is $M_i=1$.

We divide $H_{\rm int}^{(0)}$ into a direct and mixed part,
\be
  H_{\rm int}^{(0)} = H_{\rm int, dir}^{(0)} +H_{\rm int, mix}^{(0)}. \label{Hint0_split}
\ee
The direct part contains products of two identical spin singlet bilinears. All the terms that are products of two different  bilinears belonging to a
same irreducible representation are contained in the mixed part. We can therefore write
\begin {align}
  H_{\rm int, dir}^{(0)} =& \frac 1 2 \int \rmd \br ~ \sum_{i=1}^{12} \sum_{j=1}^{M_i} g_{i}^{(j)} \times \nonumber \\
  & \sum_{m=1}^{dim(i)}  \left ( \sum_{\sigma = \uparrow, \downarrow} \psi_\sigma^\dagger (\br) \Gamma_{i, j}^{(m)} \psi_\sigma (\br) \right )^2,   \label{Hint0dir}
\end {align}
for the direct, and
\begin {align}
  H_{\rm int, mix}^{(0)} = \frac 1 2 \int \rmd \br ~ \sum_{i=1,4} \sum_{j=1}^{2} \sum_{j' = j+1}^3 \tilde g_{i}^{(j,j')} \times \qquad \quad \qquad \nonumber \\
    \left ( \sum_{\sigma = \uparrow, \downarrow} \psi_\sigma^\dagger (\br) \Gamma_{i, j}^{(1)} \psi_\sigma (\br) \right )
    \left ( \sum_{\sigma' = \uparrow, \downarrow} \psi_{\sigma'}^\dagger (\br) \Gamma_{i, j'}^{(1)} \psi_{\sigma'} (\br) \right ),  \label{Hint0mix}
\end {align}
for the mixed part.

Following the same argument, the quartic contact interaction term, $H_{\rm int}^{(\bM)}$, reads
\be
  H_{\rm int}^{(\bM)} &=& \frac 1 2 \int \rmd \br ~ \sum_{i=13}^{20} g_{i} \sum_{m=1}^{2}
    \left ( \sum_{\sigma = \uparrow, \downarrow} \psi_\sigma^\dagger (\br) \Gamma_{i, 1}^{(m)} \psi_\sigma (\br) \right )^2. \nonumber \\
  \label{HintM}
\ee
The multiplicity of each irreducible representation $E_{\bM i}^\pm$ is exactly one, hence we do not sum over $j$ in
this expression. So far, we have identified 30  independent coupling constants, 16 in Eq.\ \eqref {Hint0dir}, 6 in Eq.\ \eqref {Hint0mix},
and 8 in Eq.\ \eqref {HintM}.

In addition to the singlet-singlet quartic terms we have just presented, there are symmetry allowed triplet-triplet terms.
For each term in Eqs.\ \eqref {Hint0dir}-\eqref {HintM}, there is an
analogous triplet-triplet term, schematically obtained by the substitution
\be
  g_{mn} \left ( \psi^\dagger_\alpha \Gamma_m \psi_\alpha \right ) \left ( \psi^\dagger_\beta \Gamma_n \psi_\beta \right )
    \longrightarrow \qquad \qquad \nonumber \\
 g^{(t)}_{mn} \left (  \psi^\dagger_\alpha \Gamma_m \vec \sigma_{\alpha \alpha'} \psi_{\alpha'} \right )
    \cdot \left (\psi^\dagger_\beta \Gamma_n \vec \sigma_{\beta \beta'} \psi_{\beta'} \right ), \label{tripletfromsinglet}
\ee
where each $\Gamma$ stands for one of the $\Gamma_{i, j}^{(m)}$ matrices.
The summation is assumed only over the spin indices written in Greek letters. The terms obtained in this manner
obey the lattice, time-reversal and spin SU(2) symmetries. Seemingly, Eq.\ \eqref {tripletfromsinglet}
introduces additional 30 independent coupling constants, but this is not true for the contact interaction. The Pauli
matrix completeness relation,
\be
  \vec \sigma_{\alpha \alpha'} \cdot \vec \sigma_{\beta \beta'} = 2 \delta_{\alpha \beta'} \delta_{\alpha' \beta} -
    \delta_{\alpha \alpha'} \delta_{\beta \beta'}, \label{Paulicompleteness}
\ee
can be used in each triplet-triplet term. Rearranging
fermion operators in such an expression yields a quartic interaction term that is a product of two spin singlet
bilinears, Eqs.\ \eqref {Hint0dir}-\eqref {HintM}. Hence, none of the 30 triplet coupling constants, $g^{(t)}$, is independent.
We conclude that the terms in Eqs.\ \eqref {Hint0dir}-\eqref {HintM} represent the most general quartic contact interaction
terms, with spin SU(2) symmetry, in our low-energy effective model.

\section {Spin-density wave orders and some properties of the band structure in their presence}
\label{SecSDW}

The {\it only} spin-density wave order parameter at the wavevector $\bM$ with sublattice magnetization peaked on the iron sites corresponds to the irreducible representation $E_{\bM 4}$ (see Figs. \ref{FigLowestHarmonics} and \ref{FigNextHarmonics}).
The order parameters, corresponding to $E_{\bM 1}$, $E_{\bM 2}$, and $E_{\bM 3}$, vanish on the iron sites. Since they correspond to a different irreducible representation, they also must have different transition temperatures unless, of course, the system is fine tuned to a multicritical point. That's because there is no symmetry which guarantees the coefficients of the second order invariants to vanish at the same temperature.

When the spin SU(2) symmetry is present, the irreducible representations discussed in the Section \ref{SecP4nmm} can be used to classify the {\it orbital} component of the spin-density wave order parameters; the spin magnetization component is independent. If the spin-orbit coupling is included the spin SU(2) symmetry is absent. In such a case, under a symmetry operation, the direction of the spin must be transformed together with the spatial coordinates. The spin-density order parameters can be classified according to
the irreducible representations of the space group determined in Section \ref{SecP4nmm} as long as we are careful to include the sublattice magnetization when performing the symmetry operations. Any spin-density wave order parameter is odd under time-reversal.

With spin-orbit coupling, the spin part of the order parameter transforms
under the lattice space group operations together with the orbital part. Being an axial vector, the spin part of
an order parameter transforms under Eqs.\ (\ref{DAVsigmaX} - \ref{DAVsigmax}).
Therefore, when writing down the invariants, we must first determine the
overall symmetry properties of a spin-density order parameter, $\Delta_{E_{\bM i}^a}^b$. Here,
$b$ is the spin component index and $E_{\bM i}^a$ corresponds to the irreducible representation of the orbital part.
This is done by multiplying the corresponding representation, and the result is presented in Table \ref {TableSDWirreps}.

\begin{figure}[h]
\begin{center}
\includegraphics[width=0.2\textwidth]{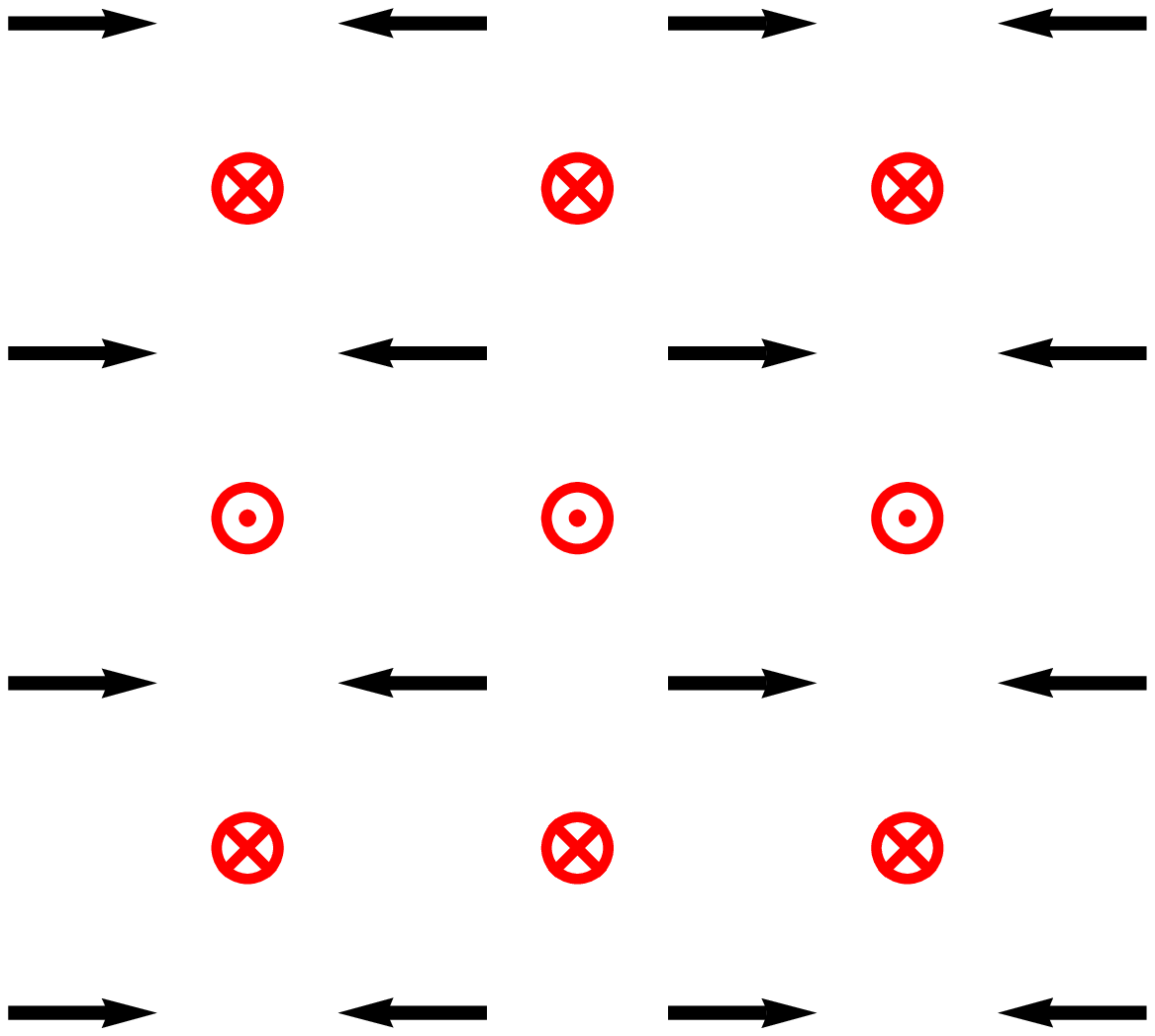}
\hspace {0.03\textwidth}
\includegraphics[width=0.2\textwidth]{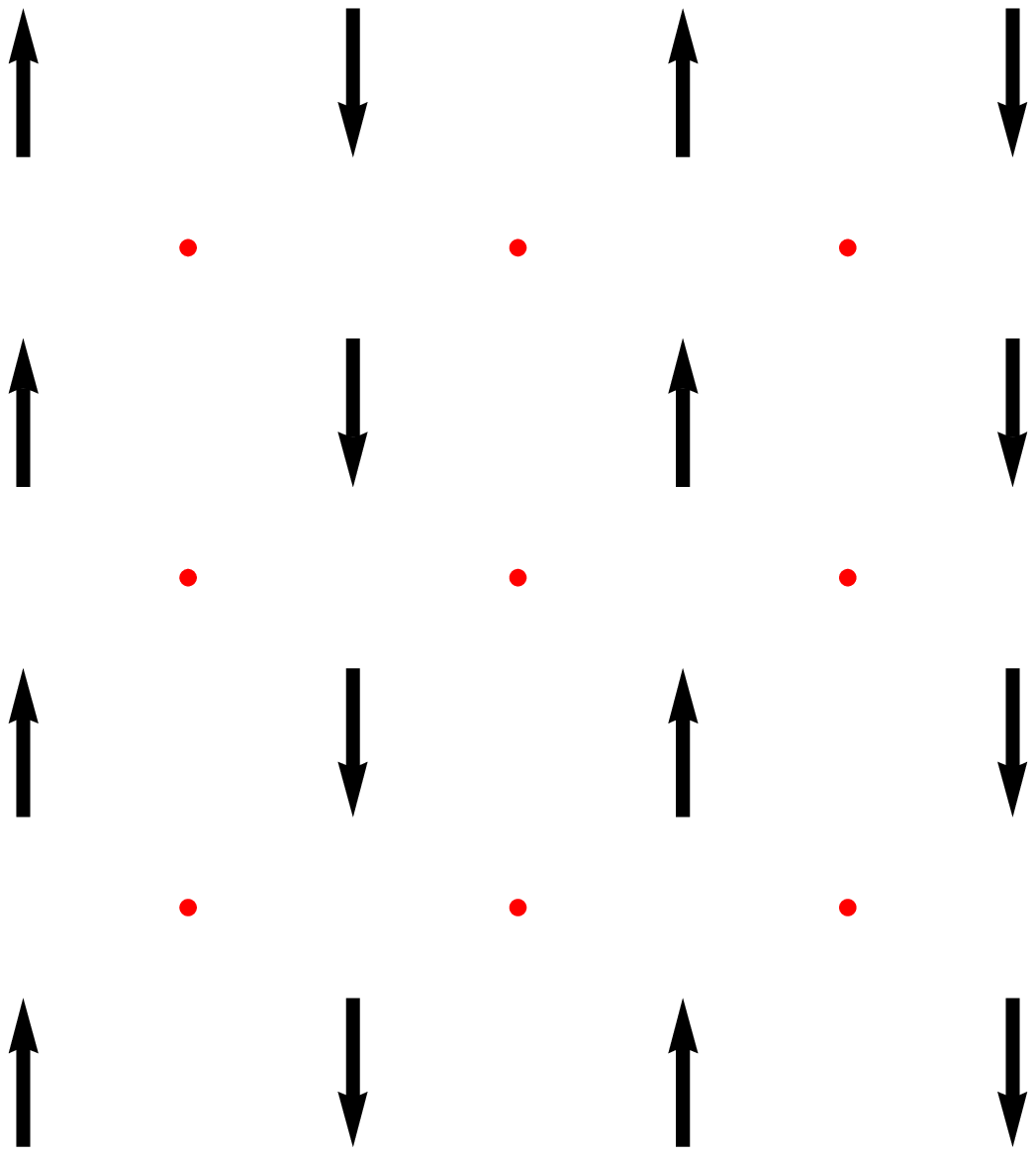} \\
$E_{\bM 1}^Y$ \hspace {0.2\textwidth} $E_{\bM 2}^Y$ \\
\vspace {0.2cm}
\includegraphics[width=0.2\textwidth]{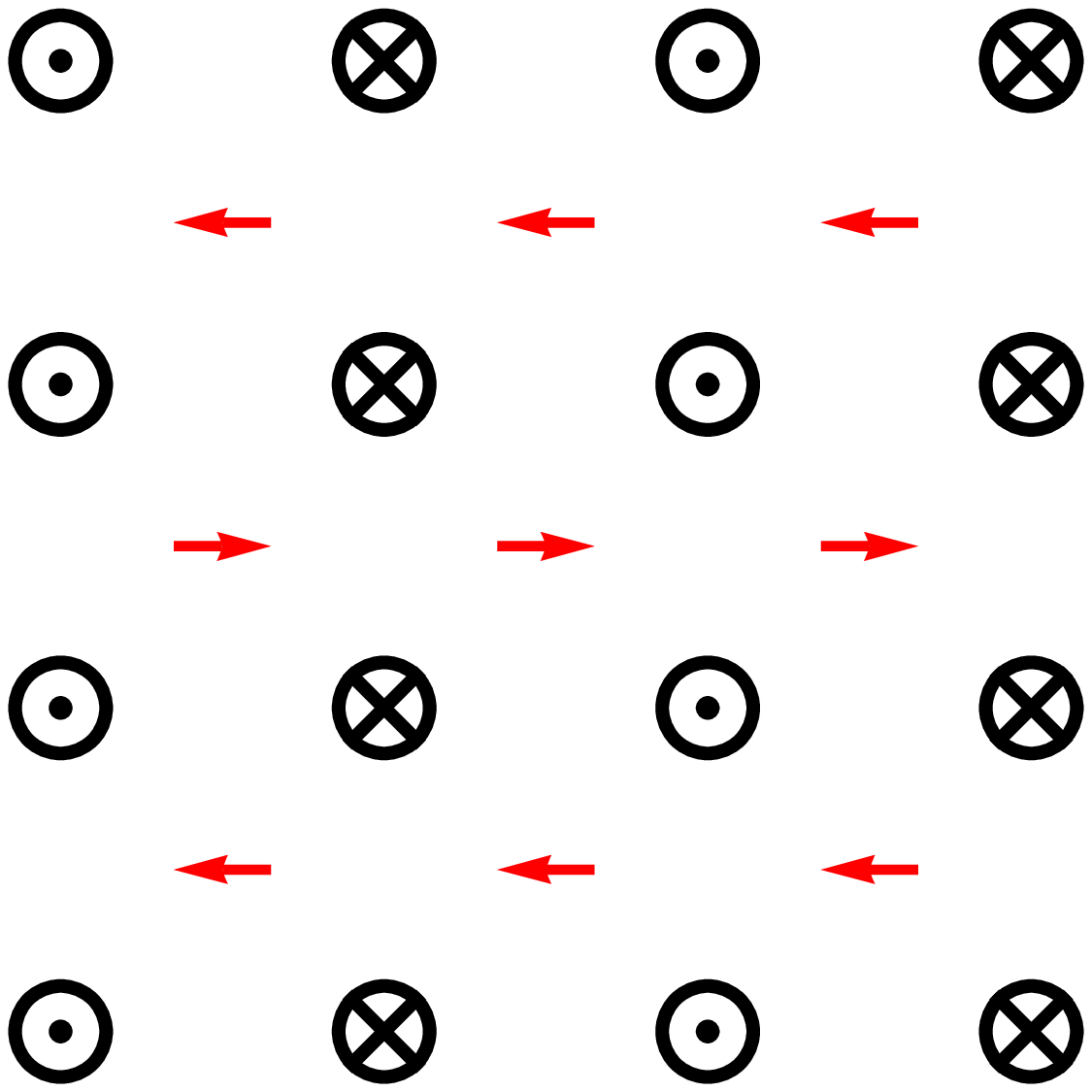}
\hspace {0.03\textwidth}
\includegraphics[width=0.2\textwidth]{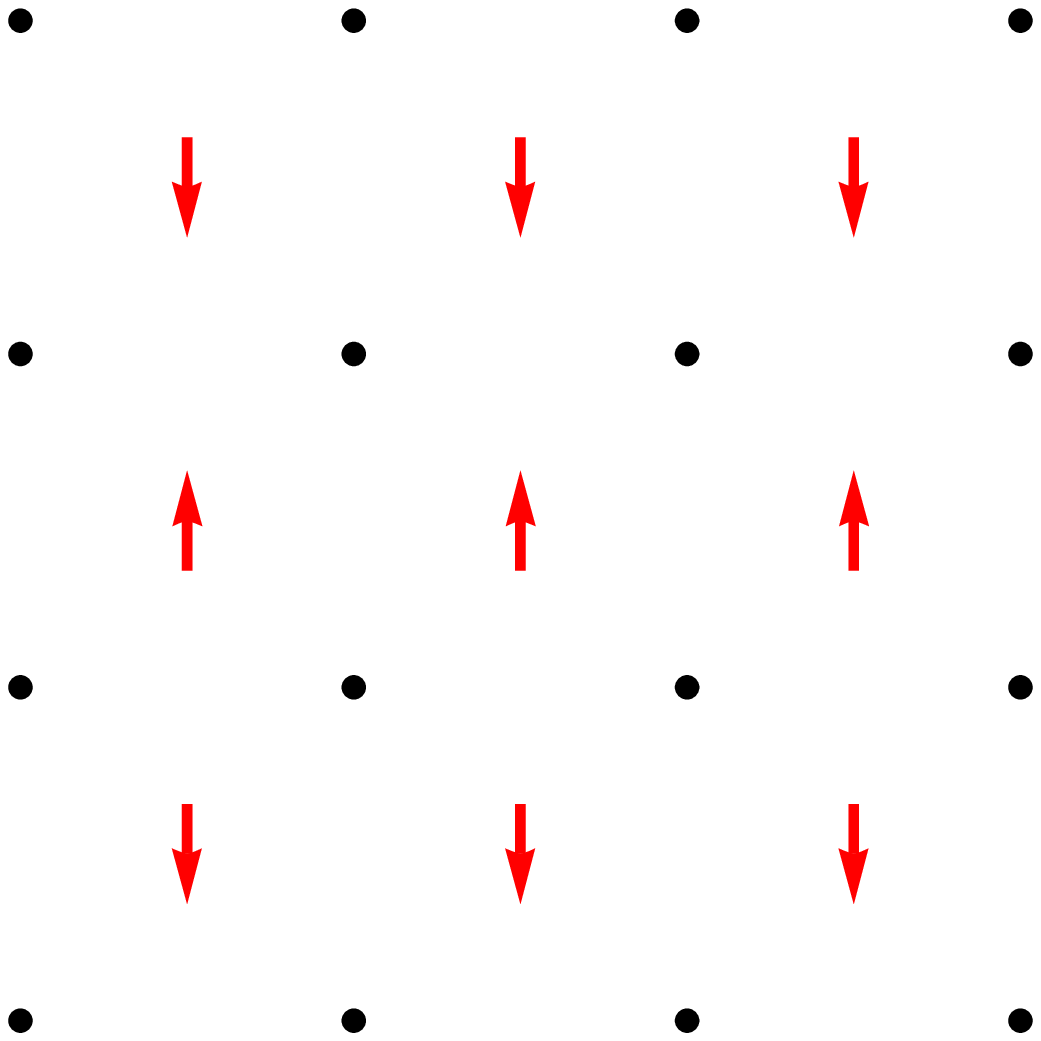}\\
$E_{\bM 3}^X$  \hspace {0.2\textwidth}$E_{\bM 4}^X$
\end {center}
  \caption {The magnetic moments on iron (black) and pnictide (red) atoms when only a single component of the overall order parameter
    condenses. The moments on both iron and pnictide sites are locked into $X$-, $Y$-, or $z$-direction.}
    \label{FigSDWirreps1}
\end{figure}

\begin{figure}[h]
\begin {center}
\includegraphics[width=0.2\textwidth]{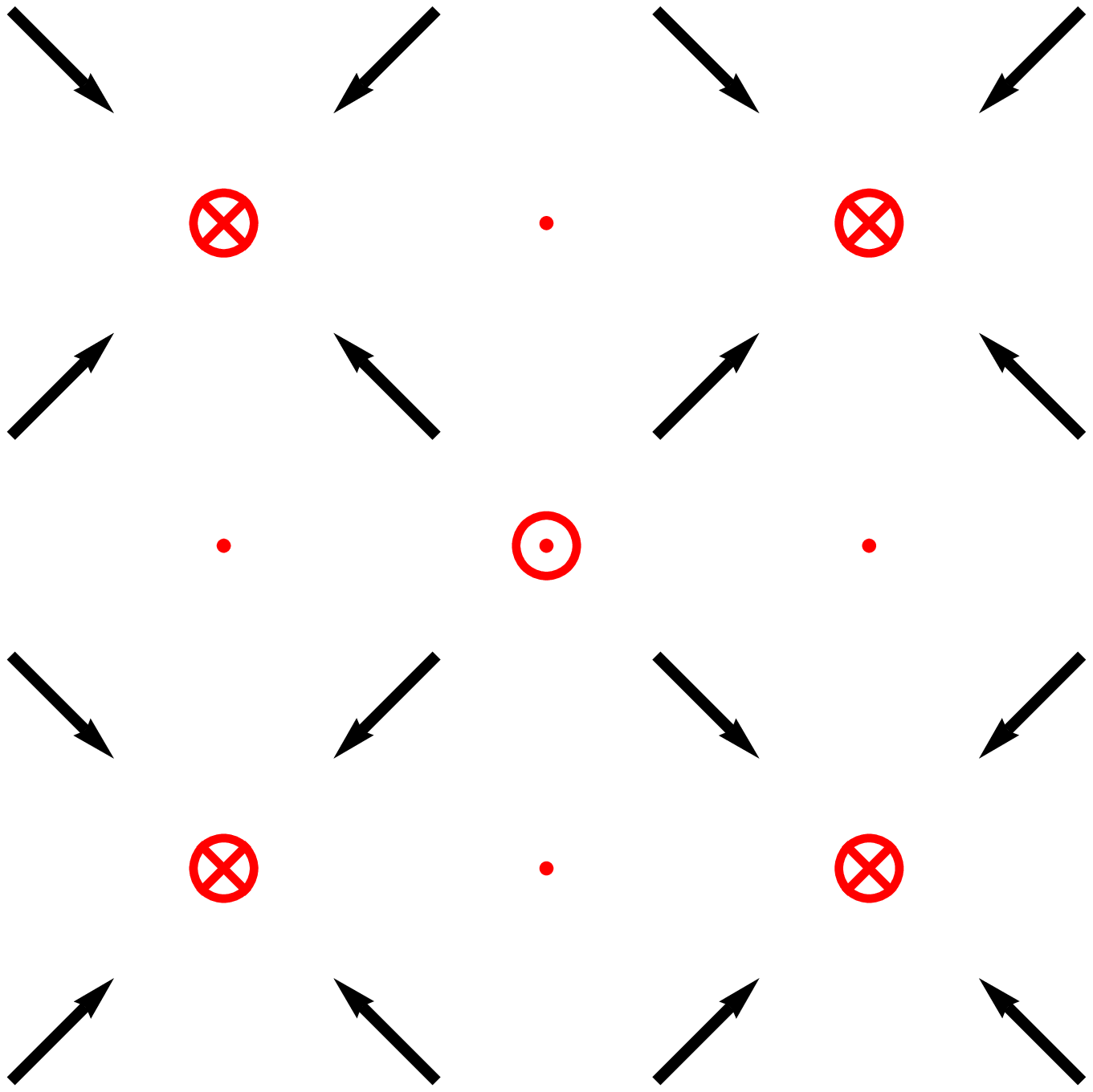}
\hspace {0.03\textwidth}
\includegraphics[width=0.2\textwidth]{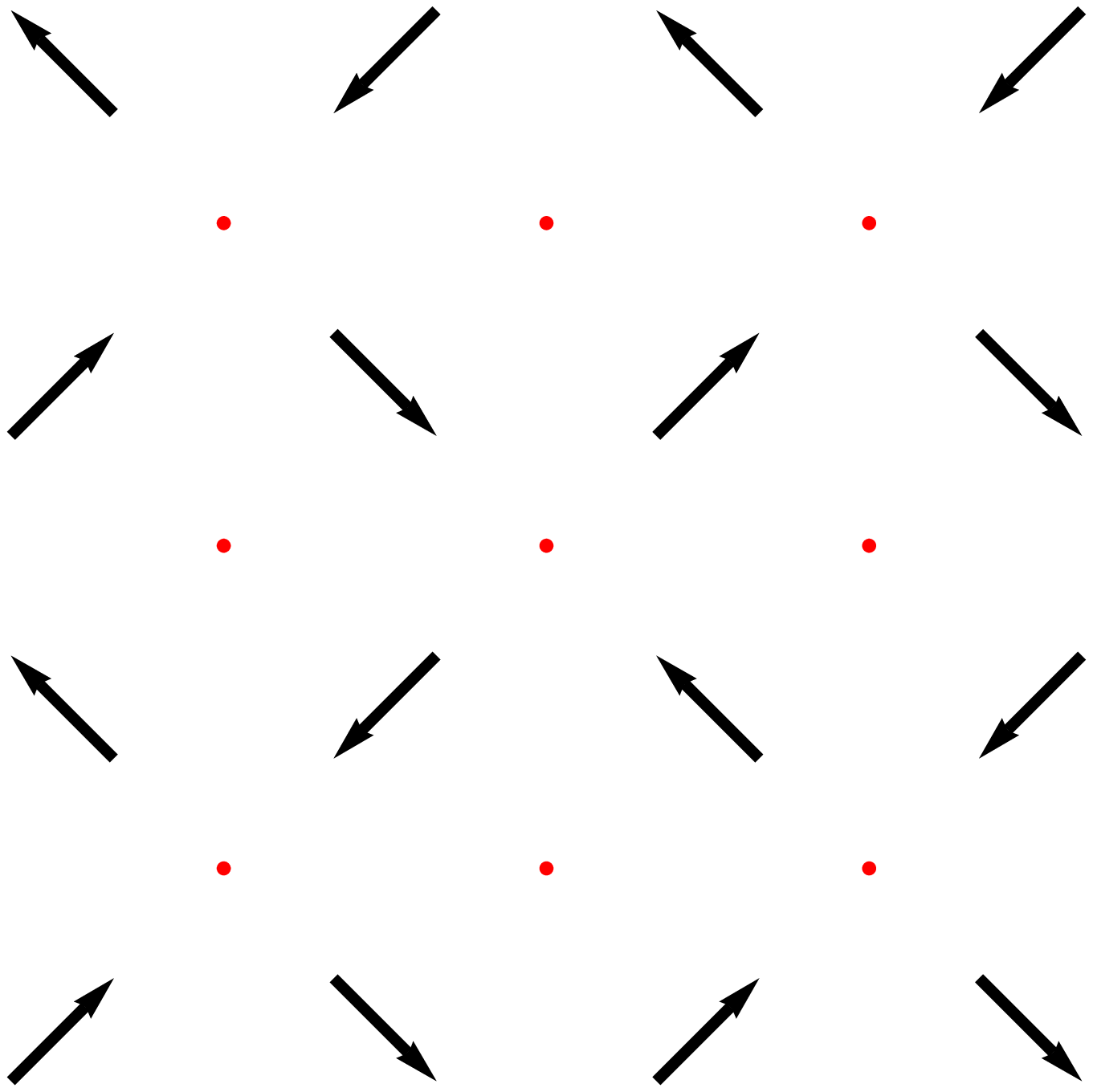} \\
$E_{\bM 1}^X+E_{\bM 1}^Y$ \hspace {0.14\textwidth} $E_{\bM 2}^X+E_{\bM 2}^Y$ \\
\vspace {0.2cm}
\includegraphics[width=0.2\textwidth]{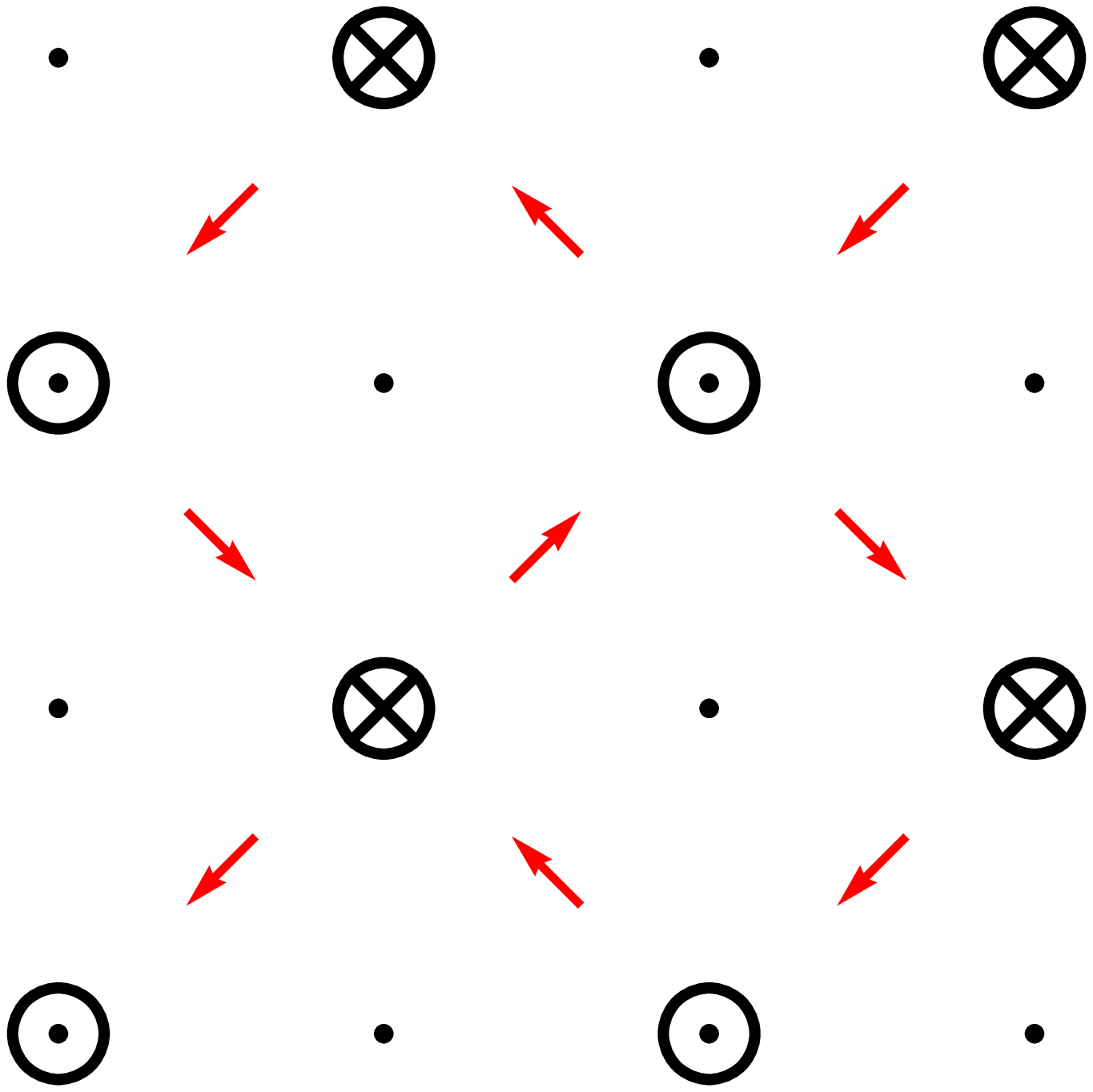}
\hspace {0.03\textwidth}
\includegraphics[width=0.2\textwidth]{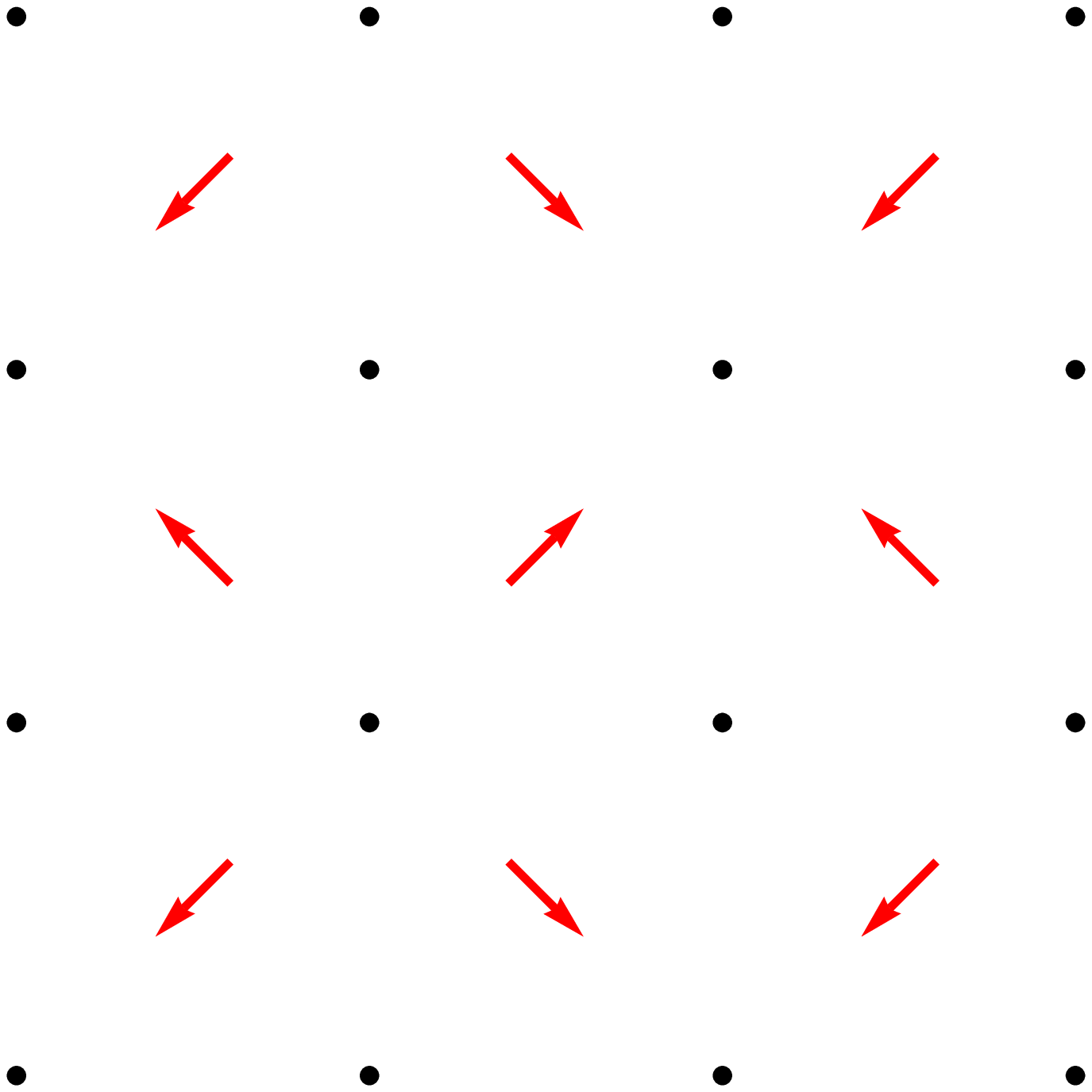} \\
$E_{\bM 3}^X+E_{\bM 3}^Y$  \hspace {0.14\textwidth}$E_{\bM 4}^X+E_{\bM 4}^Y$ \\
\end{center}
  \caption{The magnetic moments on iron (black) and pnictide (red) atoms when both order parameter components for a particular irreducible
    representation condense at the same time. The local moments are locked into $x$-, $y$-, or $z$-direction. }
    \label{FigSDWirreps2}
\end{figure}

To illustrate the connection between the orbital part and the overall symmetry of a spin-density wave order parameter, let
us consider the collinear spin-density wave \cite{delaCruzNature2008,GoldmanPRB2008} shown in the upper left corner of Fig.\ref{FigSDWirreps1}. The orbital part of this order
transforms as one of the $E_{\bM 4}$ components, let us set it to $E_{\bM 4}^X$. Then, the local moments on irons
are pointing in the $X$-direction too \cite{delaCruzNature2008,GoldmanPRB2008}, yielding the overall order parameter symmetry $E_{\bM 1}^Y$,
according to Table \ref {TableSDWirreps}. Had the spin pointed in $Y$- or $z$-direction, the overall order parameter symmetry
would have been $E_{\bM 2}^Y$ or $E_{\bM 3}^X$, respectively.

Table \ref {TableSDWirreps} may also be used in determining the symmetry properties of the orbital part of a
spin-density wave, once the symmetry properties of the overall order parameter have been set. In the physically
relevant case, where the overall order parameter is $E_{\bM 1}^Y$, we find that, next to $\Delta_{E_{\bM 4}^X}^X$
order parameter, there are two other spin-density wave order parameters
$\Delta_{E_{\bM 3}^X}^Y$ and $\Delta_{E_{\bM 2}^X}^z$ which belong to the same irreducible representation.
These order parameters are automatically induced by the presence of $\Delta_{E_{\bM 4}^X}^X$.  The latter
of the two has the orbital part transforming according to $E_{\bM 2}$, and as such, it is finite on pnictogen atoms.
This leads us to the prediction that the collinear spin-density wave of the kind reported in Refs.\ \onlinecite{delaCruzNature2008,GoldmanPRB2008}
must induce moments on pnictogen atoms in the presence of spin-orbit coupling.

For the overall order parameter $E_{\bM 2}^Y$, where the moment on iron sites points in the $Y$-direction (see Fig. \ref{FigSDWirreps1}),
there are no corresponding spin-density wave order parameters with an $E_{\bM 2}$ orbital part. Therefore,
such a collinear spin-density wave induces no moment on pnictogen atoms.
For the overall order parameter $E_{\bM 3}^X$, where the moment on iron sites points out of plane,
the induced spin-density wave results in the moment on pnictogen atoms pointing in-plane.
In both cases where a collinear density wave with a moment on pnictogen atoms is induced,
the orbital part of the induced order parameter is $E_{\bM 2}^Y$.

\begin{table}
  \begin {tabular}{| >$c<$ | >$c<$  >$c<$  >$c<$ |}
    \hline
    ~ & s^X & s^Y & s^z \\
    \hline
    E_{\bM 1}^X & E_{\bM 3}^Y & E_{\bM 4}^Y & E_{\bM 2}^X \\
    E_{\bM 1}^Y & E_{\bM 4}^X & E_{\bM 3}^X & E_{\bM 2}^Y \\
    E_{\bM 2}^X & E_{\bM 4}^Y & E_{\bM 3}^Y & E_{\bM 1}^X \\
    E_{\bM 2}^Y & E_{\bM 3}^X & E_{\bM 4}^X & E_{\bM 1}^Y \\
    E_{\bM 3}^X & E_{\bM 2}^Y & E_{\bM 1}^Y & E_{\bM 4}^X \\
    E_{\bM 3}^Y & E_{\bM 1}^X & E_{\bM 2}^X & E_{\bM 4}^Y \\
    E_{\bM 4}^X & E_{\bM 1}^Y & E_{\bM 2}^Y & E_{\bM 3}^X \\
    E_{\bM 4}^Y & E_{\bM 2}^X & E_{\bM 1}^X & E_{\bM 3}^Y \\
    \hline
  \end {tabular}
  \caption {This table gives the overall symmetry properties of a spin-density parameter $\Delta_{E_{\bM i}^a}^b$, whose orbital
    part transforms as $a$-component of the $E_{\bM i}$ irreducible representation (left), while the spin points
    in the $b$ direction (top). The table can also be used to determine the orbital part
    a spin-density wave order with an overall $E_{\bM i}^a$ symmetry when the spin points in a particular direction.
    Omitted from the table is the parity under the time reversal: the overall order parameter is
    odd under the time reversal thanks to its spin part; the orbital part is even.} \label{TableSDWirreps}
\end {table}

There is one additional  spin-density wave order parameter, namely the one
with the overall symmetry properties given by $E_{\bM 4}$. A spin-density wave order parameter with such an
overall symmetry cannot have an $E_{\bM 4}$ orbital part, as seen in Table \ref{TableSDWirreps}, therefore implying
that such a spin-density wave-order parameter carries no moment on the iron sites (see Fig.\ref{FigSDWirreps1}).

The spin-density wave orders which we consider is either a collinear spin-density wave or a co-planar four-fold symmetric density wave shown in the upper left corners of Figs.\ \ref{FigSDWirreps1} and \ref{FigSDWirreps2}.
The collinear
state, with the sublattice magnetization peaked on iron, occurs when only a single component of an $E_{\bM 1}$ order parameter condenses. This state is
ubiquitous in iron pnictides \cite{delaCruzNature2008,GoldmanPRB2008}. The four-fold
density wave may occur when both components of an $E_{\bM 1}$ order parameter condense with the same
magnitude. Such a state may have been observed recently\cite{AvciarXiv2013}.

From the symmetry properties of each component of the low-energy effective theory spinor in Eq.\ \eqref {eq:spinor},
we know how to construct symmetry breaking terms. Since the magnetic moment of these orders is situated on iron sites,
the orbital part in each symmetry breaking term is a bilinear which transforms as one of the $E_{\bM 4}$ components.
These are constructed and analyzed in the following subsections. But first, we address the issue of double degeneracy of
all bands, as found in the dispersions with the spin-orbit interaction absent, Fig.\ \ref {FigCompareTB2LoEff}, or present,
Fig.\ \ref {FigDispersionsWithSO}, in the model.

\subsection {Kramers degeneracy}
\label{SubsecKramers}

In a model with no spin-orbit interaction, each electron state is doubly, or even-, degenerate. This is
an automatic consequence of the spin SU(2) symmetry: the Hamiltonian is diagonal in the spin space, hence for each Bloch state
with spin up, $\psi_{\bk \uparrow} (\br) = e^{i \bk \cdot \br} u_\bk (\br) \left ( \begin {array}{c} 1 \\ 0 \end {array} \right )$, there exists a Bloch
state with the same orbital part, $u_\bk (\br)$, but with the opposite spin, $\psi_{\bk \downarrow}
(\br) = e^{i \bk \cdot \br} u_\bk (\br) \left ( \begin {array}{c} 0 \\ 1 \end {array} \right )$. The two states are orthogonal, and, since
they are related by a symmetry of the Hamiltonian, they have the same eigenvalue $\epsilon_\bk$.

Once the spin-orbit interaction is present in the problem, the SU(2) symmetry is lost and the double degeneracy
of each state does not follow automatically. It may or may not exist, depending on the remaining symmetry.
Generally, if the system is left invariant under an inversion  followed by
the time-reversal, for each single electron eigenstate of the Hamiltonian at $\bk$, there is another one which is
orthogonal to, and degenerate with it, at the same $\bk$. Note that, for such Kramers pairs to exist, it is required that only the product of the inversion and the time-reversal
is a symmetry of the system; the two operations individually need not be good symmetries.

To show that this holds, even when the inversion is followed by a translation, as is the case here, let us assume that one eigenstate of the Hamiltonian at $\bk$ is
\be
  \Psi_\bk (\br) = e^{i \bk \cdot \br} \left ( \begin {array}{c} u_\bk (\br) \\ v_\bk (\br) \end {array} \right ). \label{Kramers1}
\ee
Since the Hamiltonian is, in general, not spin SU(2) symmetric, an eigenstate will have both of its spin-up and spin-down components non-zero.
The Kramers partner of $\Psi_\bk (\br)$, obtained by the product of inversion and time-reversal (assumed to be a symmetry of the Hamiltonian),  is
\begin {align}
  \overline \Psi_\bk (\br) &= \Theta \{ i | \btau \} \Psi_\bk (\br) =  \Theta e^{i \bk \cdot (\btau - \br)} \left ( \begin {array}{c} u_\bk (\btau - \br) \\ v_\bk (\btau - \br) \end {array} \right ) \nonumber \\
  &= e^{i \bk \cdot \br} e^{- i \bk \cdot \btau} \left ( \begin {array}{c} \phantom - v^*_\bk (\btau - \br) \\ - u^*_\bk (\btau - \br) \end {array} \right ) \nonumber \\
  &= e^{- i \bk \cdot \btau} (i \sigma_2) K \Psi_\bk (\btau - \br). \label{Kramerspartner}
\end {align}
The inversion operation followed by a translation by vector $\btau$, $\{ i | \btau \}$, is equivalent to an inversion with respect to the point at
$\btau/2$. Any inversion keeps the spin part of the wave-function invariant, as it must, because spin is an axial vector.
The time-reversal, $\Theta$, complex conjugates the value of the wave-function ($K$),
while acting as $i \sigma_2$ in the spin space.  Since $\Psi_\bk (\br)$ is a Bloch state with momentum $\bk$,
from Eq.\ \eqref {Kramerspartner} it follows that $\overline \Psi_\bk (\br)$ is also a Bloch state with the same momentum.

Under the assumption that $\Theta \{ i | \btau \}$ is a symmetry of the Hamiltonian, it follows that the Kramers partner Eq.\ \eqref {Kramerspartner}
of the state $\Psi_\bk (\br)$ has the same eigen-energy. These two states are orthogonal,
\be
  \int \rmd \br ~ \overline \Psi^*_\bk (\br) \Psi_\bk (\br) = 0,
\ee
which follows directly from the definitions in Eqs.\ \eqref {Kramers1} and \eqref {Kramerspartner}.

In the absence of a symmetry breaking order, any operation $\{ i | \bt + \btau_0 \}$, where $\bt$ is an integer translation vector,
is the symmetry of the Hamiltonian. Unless there is an externally applied magnetic field, the time-reversal symmetry is also
automatically present, with or without the spin-orbit interaction.
Therefore, the product of these two operations is also a symmetry, implying the Kramers degeneracy.
This is in agreement with the dispersions shown in Fig.\ \ref {FigDispersionsWithSO}.

\subsection {Collinear spin-density wave order parameter in the low-energy effective model}

Since the magnetic moment rests on iron atoms, the orbital part of the collinear spin-density wave
order parameter transforms according to one of the components of the $E_{\bM 4}$.
Accordingly, this order parameter couples to any bilinears in the low-energy effective theory for which the orbital part has the
same symmetry properties. The only two bilinears in the low-energy effective theory forming an {\em orbital} $E_{\bM 4}$ doublet are
\be
  E_{\bM 4}^X:&& \sum_\bk \psi_{\Gamma, \alpha}^{\dagger}  (\bk) \left \lbrack \begin {array}{c c} 1 & 0 \\ 0 & 0  \end {array} \right \rbrack
    \vec{\sigma}_{\alpha\beta}\psi_{Y, \beta} (\bk) + h.c., \label{EM4Xbilinear} \\
  E_{\bM 4}^Y:&& \sum_\bk \psi_{\Gamma, \alpha}^{\dagger} (\bk)  \left \lbrack \begin {array}{c c} 0 & 0 \\ -1 & 0  \end {array} \right \rbrack
    \vec{\sigma}_{\alpha\beta}\psi_{X, \beta} (\bk) + h.c. \label{EM4Ybilinear}.
\ee

One of these two bilinears appears in the symmetry breaking term in the presence of the spin-density wave order.
We chose the collinear spin-density which is anti-periodic in the $X$-direction, thus the orbital part of the order parameter is
Eq.\ \eqref {EM4Xbilinear}. For such a choice of the orbital part of the order parameter, the experiments find the
magnetization to point in the $X$-direction \cite{delaCruzNature2008, GoldmanPRB2008}. The symmetry breaking term corresponding to such a
collinear spin-density wave is
\begin {align}
  H_{\rm SDW} =& \Delta_{\rm SDW} \sum_\bk \sum_{\alpha, \beta = \uparrow, \downarrow} \psi_{\Gamma, \alpha}^{\dagger}  (\bk) \nonumber \\
 & \quad \times   \left \lbrack \begin {array}{c c} 1 & 0 \\ 0 & 0  \end {array} \right \rbrack \sigma^1_{\alpha \beta} \psi_{Y, \beta} (\bk) + h.c. \label{HSDW}
\end {align}
In Eq.\ \eqref {HSDW}, we have written only the
$\bk$-independent symmetry breaking term of a collinear spin-density wave state. We neglect all $\bk$-dependent terms because the
radii of the hole and electron pockets are relatively small compared to the extent of the Brillouin zone.

\begin {figure}[h]
\begin {center}
\includegraphics[width=0.46\textwidth]{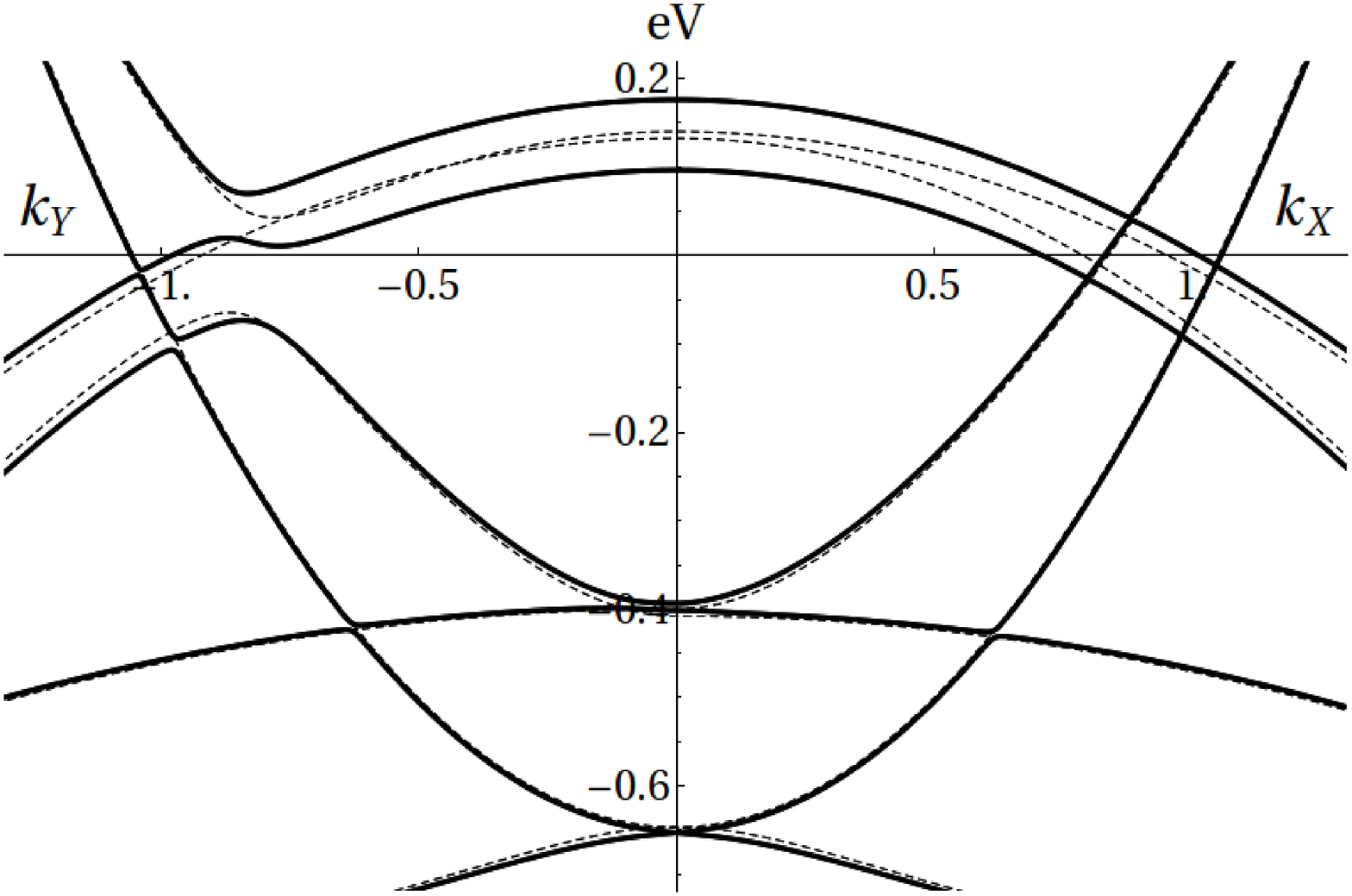}\\
a) \\
\includegraphics[width=0.24\textwidth]{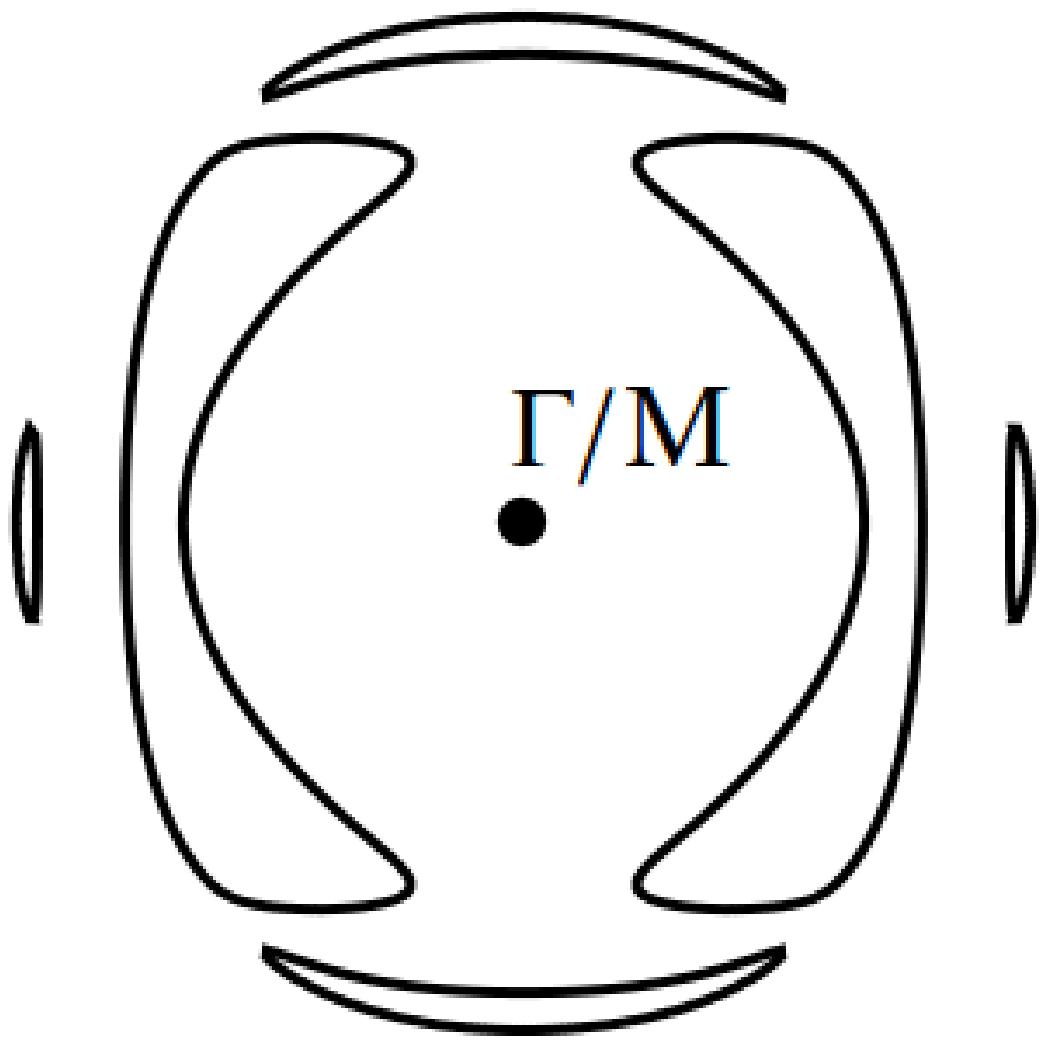}\\
b)
\end{center}
  \caption{The band structure from the low-energy effective model in the presence of a collinear  spin-density wave order.
    a) The dispersion of the low-energy effective model along the high symmetry lines. 
    The orbital part of the spin-density wave order parameter transforms as the $E_{\bM 4}^X$-component. The spin points in the $X$-direction. Without
    the spin-orbit coupling (thin dashed lines), both Dirac nodes and lines where bands cross are found in the dispersion. The spin-density wave order splits
    the four-fold degeneracy at the $\bM$-point. With the spin-orbit coupling, all the band crossings are avoided, although the splittings are too small to be visible
    in the plot. Each state is doubly degenerate, a consequence of the Kramers doublet existence in the collinear spin-density wave phase.
    b) The Fermi surfaces bear resemblance to the experimentally determined Fermi surfaces \cite{AnalytisPRB2009, TerashimaPRL2011}.}
    \label{FigSDWdispersion}
\end{figure}

In Fig.\ \ref{FigSDWdispersion}a we plot the  dispersion of the low-energy effective model  with the symmetry breaking term, Eq.\ \eqref {HSDW}.
The plot shows both the result in the absence of any spin-orbit interaction (dashed lines) and with the spin-orbit coupling terms, Eqs.\
\eqref {eq:hGammaso} and \eqref {eq:hMso}, included (solid line). The value of the spin-orbit coupling is set to $\lambda_\Gamma = \lambda_\bM = 80 {\rm meV}$
according to Ref.\ \onlinecite{TiagoPRL2006}. The value of the symmetry breaking parameter is set to $\Delta_{\rm SDW} = 65 {\rm meV}$ since this
value produces Fermi surfaces similar in shape to those found experimentally \cite {AnalytisPRB2009, TerashimaPRL2011}.

All bands in the dispersion in Fig.\ \ref {FigSDWdispersion} are doubly degenerate, and find this to be true even away from the
high symmetry lines. This is the Kramers degeneracy, described in the previous subsection. The collinear spin-density wave
state is odd under the time reversal; it is also odd under any inversion $\{ i | \bt + \thalf \thalf \}$, where $\bt$ is an even translation.
Therefore, the collinear spin-density wave is even under the product of these two operations and all the states come in Kramers pairs.

In the absence of any spin-orbit coupling, the dispersion of one of the electron bands is not affected by a presence of the collinear
spin-density wave. In our example, it is the two $\psi_{X, \sigma}$ components of the spinor, Eq.\ \eqref {eq:spinor},  that have are decoupled
from the remaining degrees of freedom, even with the spin-density wave term, Eq.\ \eqref {HSDW}, present. Had we considered the
collinear state where the other component of the $E_{\bM}$-doublet condenses, it would be the two components in $\psi_{Y, \sigma}$
that are decoupled and unaffected by the presence of the spin-density wave order. Either way, there are two bands decoupled from the others and
the crossings of these two bands with the others results in line degeneracies. These crossings are guaranteed by the $\{ \sigma^z | \thalf \thalf \}$
mirror reflection symmetry. Even if one considered a spin-density wave order parameter which is $k_x$- and $k_y$-dependent, these lines would
remain. However, if the spin-density wave order parameter has a $k_z$-dependence, such that it has a piece which is a $k_z$ odd function,
then, based on the symmetry, additional terms would appear in Eq.\ \eqref {HSDW} which would lead to avoided crossings between the
bands. When the moments are localized on iron sites only, as it is in the absence of spin-orbit coupling, the spin-density wave order is
always an even $k_z$ function, hence the Dirac lines are unavoidable. This can also be shown following the Wigner-von Neumann type
of arguments for the presence of degeneracies in the system.

In our example in Fig.\ \ref {FigSDWdispersion}, the Fermi surface crossing
band which is decoupled from the hole bands is the inner electron band in the right hand side of the plot ($B_1$ in Fig.\ \ref {FigHamBands}a, and Fig.\ \ref {FigFermiSurfaces}),
and the outer electron band (also $B_1$ on Fig.\ \ref {FigFermiSurfaces}) on the left hand side of the plot.  For a $\bk$ not on a
high symmetry line, the electron band not affected by the spin-density wave term Eq.\ \eqref {HSDW} is the $A'$ band in
Fig.\ \ref {FigFermiSurfaces}. On the other hand, the electron band coupled by Eq.\ \eqref {HSDW} to the hole bands is the $A''$ band
in Fig.\ \ref {FigFermiSurfaces}, which also has $B_2$ symmetry properties along the high symmetry lines plotted in Fig.\ \ref {FigSDWdispersion}.

Putting the two decoupled ``spinor'' components, $\psi_{X, \sigma}$, aside, we note that the collinear spin-density wave term in Eq.\ \eqref {HSDW} couples the first
components of $\psi_{\Gamma, \sigma}$, and the first component of $\psi_{Y, \sigma}$. This leads to a gap opening between
the hole and electron bands for $\bk$'s which do not lie on the high symmetry lines. For $\bk$'s on the high symmetry lines, plotted
in Fig.\ \ref {FigSDWdispersion}, we notice that only one band crossing, on the left hand side of the plot ($k_X=0$) between the
lower hole band and one electron band, is avoided; all the other band crossings remain in the absence
of spin-orbit interaction regardless of the strength of the spin-density wave order parameter $\Delta_{\rm SDW}$ in Eq.\ \eqref {HSDW}.
Had we considered a $\bk$-dependent spin-density wave order parameter $\Delta_{\rm SDW} (\bk )$, a gap would open at another band
crossing, the one on the right hand side of the plot ($k_Y=0$) between the lower hole band and one electron band.

Apparently, the spin-density wave order does not open a gap between the upper hole band and the electron band along these two directions, thus leading to Dirac cones in the
dispersion of the spin-density wave state in iron-pnictides \cite {RanPRB2009}. These crossings are protected by the vertical mirror reflection
$\{ \sigma^Y | \thalf \thalf \}$ on the $k_Y=0$ line (and similarly by $\{ \sigma^X | \thalf \thalf \}$ on the $k_X=0$ line). When we
classified the bands along the $\Sigma$-line, the two hole bands, originating from the $E_g$-doublet at the $\Gamma$-point,
transformed according to $A_2$ and $B_2$ irreducible representations, respectively. Therefore, one hole band is even, and one is odd,
under the mirror reflection $\{ \sigma^Y | \thalf \thalf \}$. The symmetry breaking term Eq.\ \eqref {HSDW} couples the electron $B_2$ electron
band along this line to only one of the two hole bands. In our example, it is the $B_2$ hole band. If there is any coupling between the electron band
and the $A_2$ hole band, it must be odd in $k_Y$ so that the overall symmetry breaking term has the same parity under the
$\{ \sigma^Y | \thalf \thalf \}$ mirror reflection as the $k_Y$ even term which couples the $B_2$ hole band and the electron band. The
presence of these nodes can alternatively be deduced from Wigner-von Neumann type of arguments as we do in Subsection \ref{SubsecWvN}.

These Dirac nodes are protected only as long as no other bands enter the effective model, i.e., as long as the $\Delta_{\rm SDW}$ energy
scale is lower than the energy of the bands excluded in the low-energy effective theory. We studied the evolution of the full tight-binding
model \cite{CvetkovicTesanovicEPL2009} dispersion in the presence of the spin-density wave symmetry breaking term on each iron orbital,
and determined that the nodes vanish once the $A_{1u}$ band at the $\Gamma$-point gets close to the Fermi level. This happens
for the values of $\Delta_{\rm SDW} \approx 700{meV}$.

Alternatively, we found the Dirac nodes to be instantaneously removed once the spin-orbit coupling is included in the Hamiltonian.
This is a consequence of the coupling of the two hole bands ($A_2$ and $B_2$ along the $\Sigma$-line) in Eq.\ \eqref {eq:hGammaso}.
Since, in the presence of the spin-orbit interaction, each hole band on the $\Sigma$-line is an admixture of both $E_g$ doublet states,
in a collinear spin-density wave state a gap opens at each crossing of the hole bands with the $B_2$ electron band. Similarly,
the spin-orbit coupling also couples the two electron bands, thus it leads to the removal of the all the degeneracy lines where the
dispersion of $\psi_{X, \sigma}$ states  used to intersect with the other dispersions in the absence of the spin-orbit interaction.
Therefore, we find that the dispersion of a collinear spin-density wave state has no four-fold degeneracies anywhere once the spin-orbit coupling
is present in the model. The four-fold degeneracy at the $\bM$-point, which was protected even with the spin-orbit coupling Eq.\ \eqref {eq:hMso},
is  also split due to the collinear spin-density wave order.

\subsection {The four-fold symmetric (coplanar) spin-density wave in the low-energy effective model}
In a four-fold symmetric coplanar state, in addition to one component of the order parameter,
here $E_{\bM 1}^X$, the other component of the total order parameter, $E_{\bM 1}^Y$, also acquires an expectation value, which is
the same in magnitude.
The symmetry breaking term, therefore, contains a term equivalent to Eq.\ \eqref{HSDW}, and its $E_{\bM}$ doublet partner, i.e., the term obtained
by the action of $\{ \sigma^x | 00 \}$ mirror reflection,
\begin {align}
  H_{C_4} =& \Delta_{C_4} \sum_\bk \sum_{\alpha, \beta = \uparrow, \downarrow} \bigg \lbrack  \psi_{\Gamma, \alpha}^{\dagger}  (\bk)
    \left \lbrack \begin {array}{c c} 1 & 0 \\ 0 & 0  \end {array} \right \rbrack \sigma^1_{\alpha \beta} \psi_{Y, \beta} (\bk) \nonumber \\
  &  + \psi_{\Gamma, \alpha}^{\dagger}  (\bk)
    \left \lbrack \begin {array}{c c} 0 & 0 \\ -1 & 0  \end {array} \right \rbrack \sigma^2_{\alpha \beta} \psi_{X, \beta} (\bk) \bigg \rbrack + h.c. \label{HC4}
\end {align}

\begin {figure}[h]
\begin {center}
\includegraphics[width=0.46\textwidth]{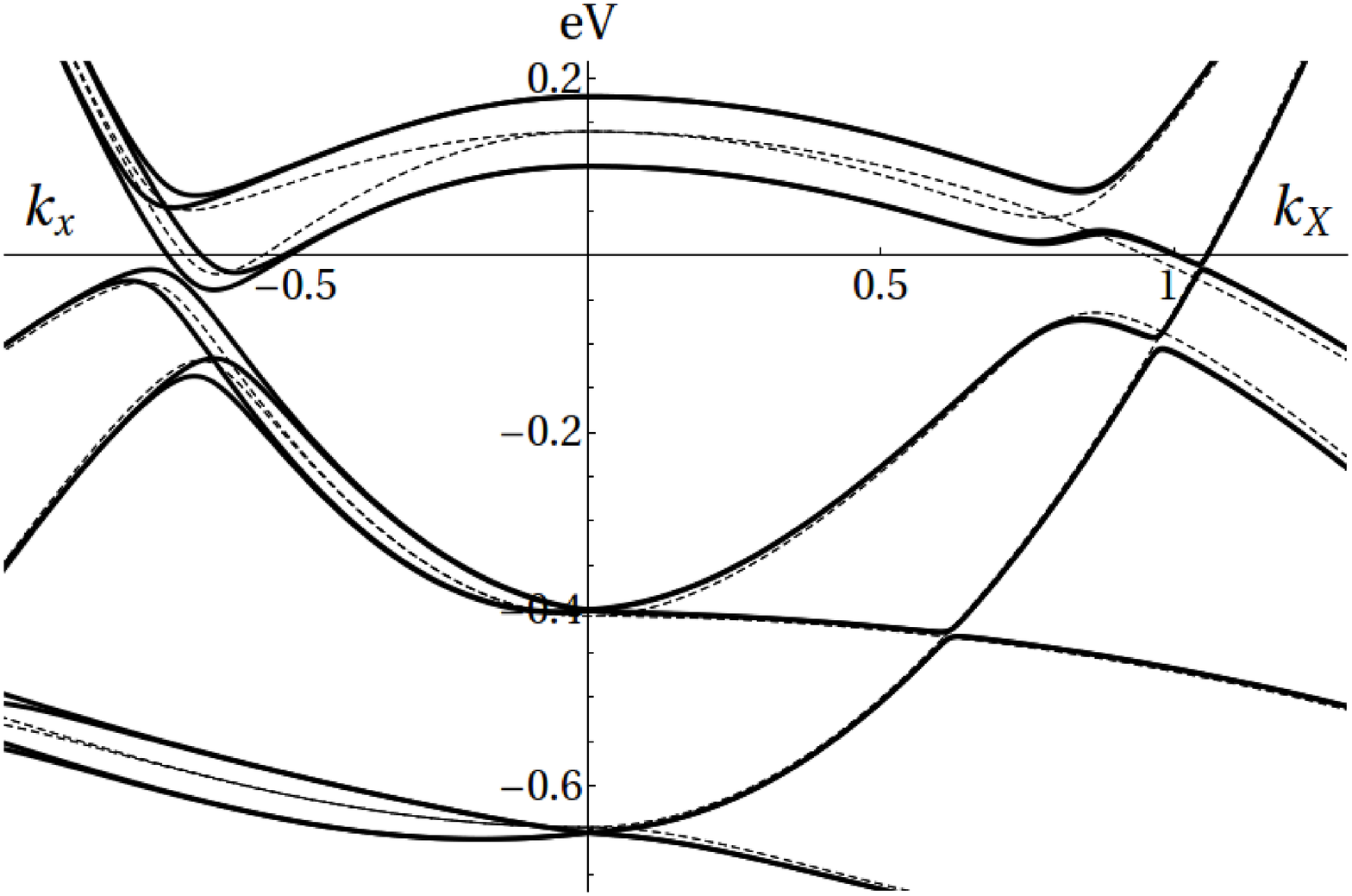}\\
a)\\
\includegraphics[width=0.24\textwidth]{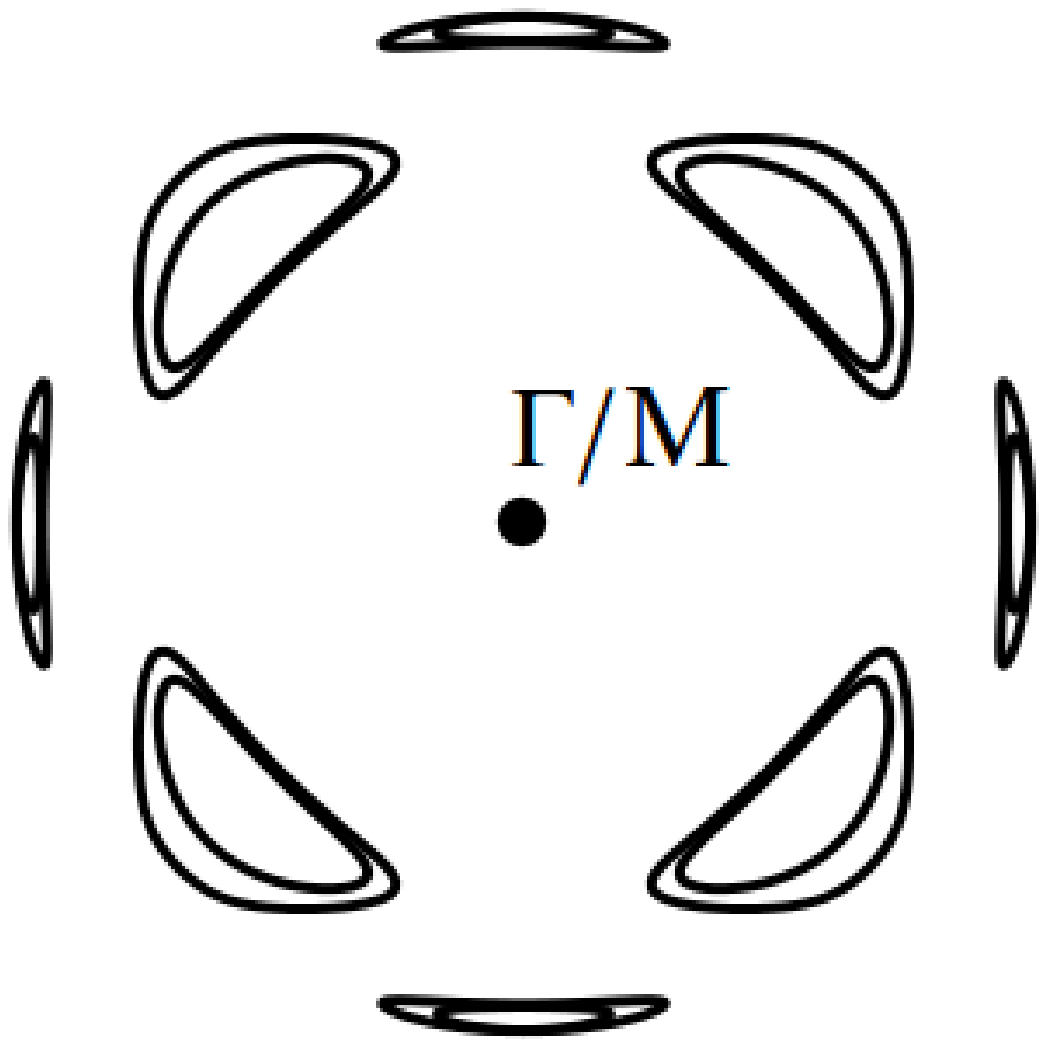}\\
b)
\end{center}
  \caption{The band structure of the low-energy effective model in the presence of the four-fold symmetric spin-density wave order.
    The local moments are plotted in Fig.\ \ref {FigSDWirreps2}.
    a) The dispersion of the low-energy effective model along the high symmetry lines 
    without  the spin-orbit coupling (thin dashed lines), and with the spin-orbit coupling (solid lines). The Kramers degeneracy is broken
   at every $\bk$-point. On the right hand side of the plot ($k_X=0$ or $k_Y=0$) there are no band crossings. On the left hand side of
    the plot ($k_x=0$ or $k_y=0$), the bands crossing is allowed leading to Dirac points in the spectrum of this phase.
    b) Thanks to the absence of the Kramers degeneracy, the Fermi surfaces appear in weakly split pairs.}
    \label{FigC4dispersion}
\end{figure}

In Fig.\ \ref {FigC4dispersion} we plot the dispersion in the low-energy effective model with this symmetry breaking term present. We choose the
strength of the order parameter to be the same as in the previous subsection, $\Delta_{\rm C4} = 65 {\rm meV}$. The first notable property of
the dispersion is that the bands are not doubly degenerate, i.e., the Kramers degeneracy is broken. One can understand this by noticing that
there are two symmetry breaking terms in Eq.\ \eqref {HC4}. For any lattice inversion, $\{ i | \bt + \btau_0 \}$, where $\bt$ is an even lattice
translation vector, the first term in Eq.\ \eqref {HC4} is odd under the action of this lattice symmetry; the second
term is even. If we consider the inversions where $\bt$ is an odd lattice translation vector, then the first term is even and the second term is odd.
Since both terms are odd under the time reversal symmetry operation, the entire symmetry breaking term is neither even nor odd
under the combined transformation, therefore, for an arbitrary $\bk$ the Kramers degeneracy is lifted as seen in Fig.\ \ref {FigC4dispersion}.

Since the Kramers degeneracy is broken in the four-fold coplanar spin-density wave, for an arbitrary $\bk$ we find each band to be non-degenerate.
However, at $\Gamma$ and $\bM$ we find each state to be doubly degenerate. The degeneracy is exact only at these points, while for
arbitrary small momenta, the dispersion is linear, i.e., we find six Dirac cones. We also find other Dirac points in the spectrum along
the two special directions, $k_x=0$ or $k_y=0$.

\subsection {Wigner-von Neummann analysis of the degeneracies: Dirac points and lines }
\label{SubsecWvN}

Until now, we have analyzed the excitation spectrum in the non-superconducting states of iron based superconductors by
diagonalizing particular model Hamiltonians. Such an approach is useful as it provides us with an intuition about the
presence or absence of degeneracies in the spectrum. More generally, the degeneracies, even the accidental ones,
are deeply tied to the symmetry of the problem. In this section we use a variant of the
Wigner-von Neumann argument to prove when the degeneracies, such as Dirac nodes
or lines, may  appear in the spectrum, and when the bands must avoid each other.

The basic idea is to consider two variational states $| \Psi_\bk^{(1)} \rangle$ and $| \Psi_\bk^{(2)} \rangle$
for each $\bk$. Each of these states is assumed to be symmetry adapted and chosen to
transform according to one of the irreducible representations of the group of the wave-vector at $\bk$, i.e.,  $\bP_\bk$. Then,
the Hamiltonian is projected onto these states for each $\bk$. The symmetry properties of these states
dictate the form of the projected Hamiltonian which can, in turn, be used to determine the feasibility of the band
touching, or whether additional fine tuning is necessary.

We are mainly interested in the ``high-temperature'' state, which breaks no symmetries, and the
collinear spin-density wave state. For both of these states, each band  is doubly degenerate due to the Kramers
theorem, as proven in the previous subsection. Therefore, an assumption that the two bands cross requires us to consider at least four states:
$| \Psi_\bk^{(1)} \rangle$, $| \Psi_\bk^{(2)} \rangle$, and their Kramers partners $| \overline \Psi_\bk^{(1)} \rangle$, and
$| \overline \Psi_\bk^{(2)} \rangle$. The coordinate and spin
representation for these states are given in Eqs.\ \eqref {Kramers1} and  \eqref {Kramerspartner}.
In the basis formed by the four states,
\be
  \left ( | \Psi_\bk^{(1)} \rangle, | \overline \Psi_\bk^{(1)} \rangle,  | \Psi_\bk^{(2)} \rangle, | \overline \Psi_\bk^{(2)} \rangle \right ), \label{WvNbasis}
\ee
the projected Hamiltonian is a $4\times 4$ matrix
\be
  H^{\rm eff}_\bk = a (\bk) \tau_3 \otimes \bbone + b (\bk) \tau_1 \otimes \bbone + \sum_{i=1}^3 c_i (\bk) \tau_2 \otimes \sigma_i. \label{H4eff}
\ee
Seemingly, there are sixteen independent coefficients, but the fact that  $| \Psi^{(1)}_{\bk} \rangle$
and $| \Psi^{(2)}_{\bk} \rangle$ come with their Kramers partners, reduces the number to five. This can be seen
by observing that under the inversion and the time reversal,
\be
  \Theta \{ i | \btau \} H_\bk^{\rm eff} = (\bbone \otimes \sigma_2) \left \lbrack K H_{\bk}^{\rm eff} \right \rbrack (\bbone \otimes \sigma_2).
\ee

When the spin-orbit interaction is absent, the Hamiltonian has the spin SU(2) symmetry, and any spin rotation is
independent of the lattice symmetry operations. Both $| \Psi^{(1)}_{\bk} \rangle$
and $| \Psi^{(2)}_{\bk} \rangle$ can be chosen to be spin up, and their Kramers partners must be spin down. With such a
choice, all three coefficients $c_i (\bk)$  must vanish for every $\bk$.
One is then left with only two independent coefficients, $a (\bk)$ and $b(\bk)$, in Eq.\ \eqref {H4eff}. Therefore,
in order for the degeneracy to appear, it is enough for the two functions to vanish at the same $\bk$. Since,
for $k_z = 0$, each is a function of two variables, they vanish along one-dimensional curves. Two such curves
may generically intersect without any further fine tuning. This would lead to Dirac points, provided that $| \Psi_\bk^{(1)} \rangle$
and $| \Psi_\bk^{(2)} \rangle$ have the same parity under the in-plane mirror reflection followed by a `half-translation'.
If the two states have an opposite parity, then, the Hamiltonian must commute with $\tau_3 \otimes \bbone$,
and this sets $b(\bk) = 0$ for all $\bk$'s. We see that in such a case, only a single function of two variables $a(k_x, k_y)$ must vanish
in order for accidental degeneracies to appear. Since this may happen along one-dimensional curves without any further
fine tuning, the two bands may intersect along lines in $\bk$-space. In fact, we may already see this in Eq.\ \eqref {eq:H0block},
where $h_\bM^+ (\bk)$ and $h_\bM^- (\bk)$ are not coupled; this holds to any order in $\bk$ because, one block
corresponds to the states even and the other block to the states odd under the in-plane mirror. This, in principle, allows
one to `unfold' the two-iron Brillouin zone to one-iron Brillouin zone; as soon as spin-orbit is taken into account, such
unfolding can never be done. The above argument goes through in the collinear spin-density wave state
because one can always choose a `half-translation' so that such a state is invariant under the combined
operation with the in-plane mirror. Additionally, vertical plane mirror operations can be used to locate the
positions of the Dirac nodes along the  Brillouin zone diagonal, thus making connection with the results of Ref.\ \onlinecite {RanPRB2009}.
However, as we will now show, including the effects of the spin-orbit coupling in the collinear spin-density wave removes all the accidental
degeneracies without additional fine tuning.

\subsubsection {Absence of degeneracies at a low symmetry $\bk$}

For an arbitrary $\bk = (k_x, k_y, 0)$, away from the high symmetry points
and lines, there is only a single symmetry operation in $\bP_\bk$.
In the high temperature state, where no symmetries are broken, the mirror reflection, followed by a `half-translation'  $\{ \sigma^z | \thalf \thalf \}$
corresponds to such an operation. On the other hand, in the collinear spin-density wave, the unit cell is doubled. The symmetry breaking term in the
Hamiltonian may be even or odd under  $\{ \sigma^z | \thalf \thalf \}$, depending on whether the overall spin-density wave
order parameter transforms according to $E_{\bM 1}^Y$ or $E_{\bM 1}^X$, respectively. If the symmetry breaking term is even under
$\{ \sigma^z | \thalf \thalf \}$, then it is odd under $\{ \sigma^z | \bar \thalf \thalf \}$,
and vice versa. Therefore, when analyzing the degeneracies in the collinear spin-density wave, only the one
which leaves the Hamiltonian invariant, i.e., the `even' one must be considered. To
keep the discussion general, therefore we assume that $\{ \sigma^z | \btau_z \}$ leaves the Hamiltonian invariant, where $\btau_z$ will be specified according
to the state we analyze: $E_{\bM 1}^Y$, $E_{\bM 1}^X$, or the high temperature symmetric state.

$| \Psi_\bk^{(1)} \rangle$ is either even or odd under the $\{ \sigma^z | \btau_z \}$; simultaneously, the
parity of its Kramers partner must be  opposite. Similarly for $| \Psi_\bk^{(2)} \rangle$. Therefore, the
Hamiltonian in Eq.\ \eqref {H4eff} must commute with either $\bbone \otimes \sigma_2$ or with $\tau_3 \otimes \sigma_2$,
depending on whether the parities of $| \Psi_\bk^{(1)} \rangle$ and $| \Psi_\bk^{(2)} \rangle$ are the
same or the opposite. In the first case,  $c_1 (\bk)$ and $c_2 (\bk )$ must vanish for every $\bk$; in the second case,
$b (\bk)$ and $c_3 (\bk )$ must vanish. In either case this requires three functions of two variables to
vanish simultaneously, which does not happen without fine tuning.

\subsubsection {Absence of degeneracies on the $\Gamma-\bM$ line (excluding the high symmetry points)}

If we consider the effect of additional symmetries at a generic $\bk$-point along the high symmetry line $\Sigma = \Gamma-\bM$
excluding the points $\Gamma$ and $\bM$, then there are additional constraints
on the functions in Eq.\ \eqref {H4eff}. Note, however, that the constraints also remove one independent variable, because,
unlike in the above discussion, we must move along the high symmetry line.

The corresponding symmetry operation is $\{ \sigma^Y | \btau_Y \}$. Here, analogous to the above,  $\btau_Y$ is the
`half-integer' translation which, combined with the (vertical) mirror along the face diagonal leaves the Hamiltonian invariant,
even in the collinear spin-density wave state. When acting on our basis Eq.\ \eqref{WvNbasis}, up to an overall phase factor,
it can be represented either by $\bbone \otimes \sigma_2$ or by $\tau_3 \otimes \sigma_2$, depending on the relative parities of $u^{(1)}_\bk (\br)$ and $u^{(2)}_\bk (\br)$
under $\{ \sigma^Y | \btau_Y \}$. If $c_1 (\bk) = c_2 (\bk) =0$, as required by the in-plane mirror,
then, either $c_3 (\bk) =0$ or $b(\bk) = 0$, respectively. In either case, we are left with two functions of {\em one} variable,
which do not vanish at the same point without additional fine tuning. If $b(\bk) = c_3 (\bk) = 0$, then, either $c_1 (\bk) = 0$
or $c_2 (\bk) = 0$, respectively. Again, in either case we are left with two functions of one variable which do not vanish  simultaneously unless
fine tuned. Therefore, whether in the normal state or in the collinear spin-density wave state, the bands generically
avoid each other. This effect is captured within our low-energy effective Hamiltonian.

\subsubsection {The high symmetry points $\Gamma$ and $\bM$}

Looking at the Table \ref{tab:Gamma generators irreps} we see that, at the $\Gamma$-point, if the symmetry adapted functions come from one dimensional representations,
then there is no constraint on $a(\bk)$. Since, there is no longer any freedom in choosing $\bk$, it being at $\Gamma$,
the bands avoid each other without additional fine tuning.

If the symmetry adapted functions come from a two-dimensional irreducible representations,  we should note
that, because the two sates must have equal parity under the in-plane mirror, only the case $c_1 (\bk) = c_2 (\bk) =0$
needs to be considered. Similarly, the parities under the vertical mirror $\{ \sigma^Y | \btau_Y\}$ are opposite, therefore
$b (\bk) =0$. The remaining generator $\{ \sigma^x | 00 \}$ requires the effective Hamiltonian to commute with
$\tau_1 \otimes (\sigma_1 - \sigma_2)$. While this forces $a(\bk)=0$, it puts no constraints on $c_3 (\bk)$ at $\Gamma$.
Again, since there is no longer any freedom in choosing $\bk$, the bands avoid each other without the fine tuning. This effect
can be clearly seen in Fig.\ \ref{FigDispersionsWithSO}a. Note that the remaining term has a form analogous to the Kane and Mele
term in graphene \cite {KaneMelePRL2005}.

Because, as long as no symmetries are broken, at the $\bM$-point there are only two-dimensional representations,
we should consider $e^{i \bM \cdot \br} u_\bM^{(1)} (\br)$, and $e^{i \bM \cdot \br} u_\bM^{(2)} (\br)$
transforming as a doublet of the same irreducible representation. Note that this is not the case in the
collinear spin-density wave. The two components of any $E_{\bM i}$ doublet have the opposite parity under the
in-plane mirror, hence $b (\bk) = c_3 (\bk) = 0$. Under the vertical mirror $\{ \sigma^Y | \btau_Y \}$,
the two components of $E_{\bM 1}$ doublet have the same parity. The same is true for the $E_{\bM 2}$ doublet.
Therefore, for $E_{\bM 1}$ and $E_{\bM 2}$, $c_1 (\bk) =0$. Conversely, for $E_{\bM 3}$ and $E_{\bM 4}$, the two
components of the doublet have the opposite parities under $\{ \sigma^Y | \btau_Y \}$, giving $c_2 (\bk) = 0$.
The vertical mirror $\{ \sigma^x | 00 \}$ requires the effective Hamiltonian to commute with $\tau_1 \otimes (\sigma_1 - \sigma_2)$.
Similarly, $\{ \sigma^y | 00 \}$ requires the effective Hamiltonian to commute with $\tau_1 \otimes (\sigma_1 + \sigma_2)$.
Therefore, it must commute with both $\tau_1 \otimes \sigma_1$ and $\tau_1 \otimes \sigma_2$. This
automatically sets $a(\bk)=0$, as well as, the remaining $c_2 (\bk)=0$ or $c_1(\bk)=0$. Therefore, the two Kramers
degenerate bands touch at the $\bM$-point. Moving away from it, splits it linearly in momentum as seen in Fig.\ \ref {FigDispersionsWithSO}b.

In the presence of the collinear spin-density wave, the symmetry is broken such that the unit cell doubles and the
$\bM$-point folds to $\Gamma$. Our discussion on the absence of the degeneracies at the $\Gamma$ is applicable;
therefore, the two Kramers bands avoid each other in the absence of fine tuning. For the value of the collinear spin-density wave
order parameter used in Fig.\ \ref{FigSDWdispersion}, the splitting of the lowest, $E_{\bM 3}$, doublet is $\sim 0.13{\rm meV}$, and
for the other, $E_{\bM 1}$, doublet is $\sim 7.8{\rm meV}$.

\section{Superconductivity}
\label{SecSuperconductivity}

In this section we describe the superconducting state using our low-energy effective theory introduced earlier.
We will only consider pairing with no overall momentum. This is consistent with the current experimental findings \cite{MazinPhysicaC2009,WangScience2011,BasovChubukovNatPhys2011}; adding finite
momentum pairing is straightforward. The pairing
term therefore involves either two states at $\Gamma$ or two states at $\bM$, allowing us to analyze them separately.

In the first part of the section we consider only spin singlet pairing. All such pairing terms in the low-energy effective theory can be
classified according to the symmetry of the lattice, $P4/nmm$. We show that the $A_{1g}$-superconductivity can be well described by
three parameters. Depending on the relative sign
and size of these parameters, the ground state of our model is $s_{++}$-, $s_{+-}$-, or a nodal-$s$-wave superconductor. This effective
model successfully describes the gap anisotropy for the electron Fermi surfaces near the $\bM$-point. In the absence of spin-orbit
coupling, the gap is isotropic on the hole Fermi surfaces near $\Gamma$. In the second part
of this section, we show that spin-orbit coupling necessarily admixes a spin triplet component which can give rise to an anisotropy
of the gap near the $\Gamma$-point.

We begin by constructing spin singlet pairing bilinears:
\be
  H_{\rm singlet} = \sum_{\bk}\psi_{\alpha}^a (-\bk) (i \sigma_2)_{\alpha \beta} M_{ab} \psi_{\beta}^b (\bk) + h.c.,  \label{DeltaSCM}
\ee
where $M_{ab}$ must be a symmetric matrix, and $\psi_{\bk, \alpha}^a$, with $a=1, \ldots 6$ are the components
of spinor Eq.\ \eqref {eq:spinor} with spin label $\alpha$. The form of the matrix $M_{ab}$ determines the symmetry properties
of the pairing term, Eq.\ \eqref {DeltaSCM}. Since, for any spin projection, the individual components of the spinor Eq.\ \eqref {eq:spinor} transform according
to $E_g$, $E_{\bM 1}$, or $E_{\bM 3}$, the symmetry properties of the pairing term Eq.\ \eqref {DeltaSCM}
are easily deduced from the multiplication tables for these irreducible representations, Tables \ref{TableExE} and \ref{TableExD4h}.
Under a space group operation $R$ we have
\begin {align}
  &\sum_{c d}D_{ac, bd} (R) \left \lbrack \sum_{\bk} \psi_{\alpha}^c (-\bk) (i \sigma_2)_{\alpha \beta} \psi_{\beta}^d (\bk) \right \rbrack = \nonumber\\
  & \sum_\bk \left ( \sum_{c} D_{ac} (R) \psi_{\alpha}^{c} (-\bk ) \right) (i \sigma_2)_{\alpha \beta} \left(\sum_{d} D_{bd} (R) \psi_{\beta}^d (\bk ) \right ).  \label{eq:Dppbilinear}
\end {align}
Note that, because this term is a spin singlet, the spin structure is left invariant under $R$. This is not the case when we consider spin triplet pairing terms,
later in this section.

In order to find the symmetry properties of the pairing terms in $H_{\rm singlet}$, we decompose the outer product $D^T (R) \otimes D (R)$ into
the irreducible representations of the group of the wave vector at $\Gamma$- and $\bM$-points, $\bP_\Gamma$ and $\bP_\bM$. The coefficients
in the decomposition are unaffected by the transpose due to the orthogonality of the characters, which are, of course, invariant under  the transpose.

At the $\Gamma$-point such a decomposition reads
\be
  E_g \otimes E_g = A_{1g} \oplus B_{1g} \oplus B_{2g} \oplus A_{2g}. \label{EgxEg}
\ee
Because the $A_{2g}$ behaves as a $z$-component of an axial vector, $L_z$, it corresponds to an odd angular momentum pairing.
This makes it odd under the exchange, and incompatible with the spin singlet; it would require $M_{ab}$ to be antisymmetric,
contrary to Eq.\ \eqref {DeltaSCM}. Therefore, the pairing terms at the $\Gamma$-point may
transform only according to one of the first three terms in Eq.\ \eqref {EgxEg}. The $A_{1g}$ corresponds to the $s$-wave symmetry;
the $B_{1g}$ and $B_{2g}$ correspond to $d_{x^2-y^2}$- and $d_{xy}$-wave symmetry, respectively.
We focus on  the $s$-wave symmetry. The corresponding $A_{1g}$ pairing bilinear reads
\be
  \sum_\bk \psi_{\Gamma, \downarrow}^T (-\bk) \psi_{\Gamma, \uparrow} (\bk ). \label{DeltaGammaA1g}
\ee

The symmetry properties of the singlet pairing terms at the $\bM$-point can, similarly, be determined from the products
\begin {align}
  &E_{\bM 1} \otimes E_{\bM 1} = E_{\bM 3} \otimes E_{\bM 3} = A_{1g} \oplus B_{2g} \oplus A_{2u} \oplus B_{1u}, \label{EMxEM} \\
  &E_{\bM 1} \otimes E_{\bM 3} = E_g \oplus E_u. \label{EM1xEM3}
\end {align}
In the first equation, the $B_{1u}$ can be ignored, because it corresponds to an antisymmetric combination
of fermionic operators at the $\bM$-point, and is incompatible with the spin singlet. The two independent $s$-wave pairing terms are
\be
  \sum_\bk \sum_{a=X, Y} \psi_{a, \downarrow}^T (-\bk) \left ( \begin {array}{c c} 1 & 0 \\ 0 & 0 \end {array} \right ) \psi_{a, \uparrow} (\bk ), \label{DeltaM1A1g} \\
  \sum_\bk \sum_{a=X, Y} \psi_{a, \downarrow}^T (-\bk) \left ( \begin {array}{c c} 0 & 0 \\ 0 & 1 \end {array} \right ) \psi_{a, \uparrow} (\bk ). \label{DeltaM3A1g}
\ee
The remaining pairing terms, i.e., the ones not transforming according to $A_{1g}$, as well as those
with the $(\pi, \pi)$ pairing momentum, are spelled out in detail in Appendix \ref {AppSCpairs}.

\subsection {Singlet spectral gaps near $\Gamma$ and $\bM$}

Following Eq.\ \eqref {DeltaGammaA1g}, the only $s$-wave pairing term at $\Gamma$ is
\be
  H_{\Gamma, {\rm SC}} &=& \Delta_\Gamma \sum_\bk \psi_{\Gamma, \downarrow}^T (-\bk) \psi_{\Gamma, \uparrow} (\bk ) + h.c., \label{HSCGamma}
\ee
where we choose $\Delta_\Gamma$ to be real. In order to determine the quasi-particle energy spectrum near $\Gamma$,
we  define the Nambu spinor $\Psi_\Gamma^\dagger (\bk) = (\psi_{\Gamma \uparrow}^\dagger (\bk), \psi^T_{\Gamma, \downarrow} (-\bk) )$.
The resulting Bogoliubov-de Gennes Hamiltonian is
\be
  H_{\Gamma, {\rm BdG}} = \sum_\bk ~ \Psi_\Gamma (\bk )^\dagger \left ( \begin {array}{c c} h_\Gamma (\bk) & \Delta_\Gamma \bbone_2 \\
    \Delta_\Gamma \bbone_2 & -h_\Gamma (\bk) \end {array} \right ) \Psi_\Gamma (\bk), \label{HBdGGamma}
\ee
where we used $h_\Gamma (-\bk)^* = h_\Gamma (\bk)$, which follows from Eq.\ \eqref {eq:hGamma}.

Notice that diagonalizing $h_\Gamma (\bk)$ leaves the pairing term invariant. Performing a unitary transformation,
${\mathcal U} = \left ( \begin {array}{c c} U & 0 \\ 0 & U \end {array} \right )$, which brings us to such
``band'' basis, we therefore find
\be
  {\mathcal U}^\dagger
  \left ( \begin {array}{c c} h_\Gamma (\bk) & \Delta_\Gamma \bbone_2 \\
    \Delta_\Gamma \bbone_2 & -h_\Gamma (\bk) \end {array} \right ) {\mathcal U} =
  \left ( \begin {array}{c c c c} \epsilon_\bk^{(1)} & 0 & \Delta_\Gamma & 0 \\
    0 & \epsilon_\bk^{(2)} & 0 & \Delta_\Gamma \\
    \Delta_\Gamma & 0 & -\epsilon_\bk^{(1)} & 0 \\
    0 &\Delta_\Gamma & 0 & -\epsilon_\bk^{(2)} \end {array} \right ), \label{HbandGamma}
\ee
where $U^\dagger h_\Gamma (\bk) U= \left ( \begin {array}{c c} \epsilon_\bk^{(1)} & 0 \\ 0 & \epsilon_\bk^{(2)} \end {array} \right )$.
The $\epsilon_\bk^{(1, 2)}$ are, of course, the  energy dispersions of  the two hole bands. From Eq.\ \eqref {HbandGamma}
the dispersions for the quasi-particles follow,
\be
  E_\bk^{(i)} = \sqrt {\left ( \epsilon_\bk^{(i)}\right )^2 + \Delta_\Gamma ^2}. \label{EkGammaSCsinglet}
\ee
We conclude that the spin singlet $A_{1g}$ pairing leads to an isotropic superconducting gap, which is identical for the two hole bands.
Any gap anisotropy on the hole Fermi surfaces results either from higher order momentum dependent $A_{1g}$ pairing terms, or from spin-orbit
interaction induced $A_{1g}$ spin triplet admixture. The latter will be considered in the following subsection.

For the low-energy  states near the $\bM$-point, the spin singlet pairing term with the $A_{1g}$ symmetry is
\begin{align}
  H_{\bM, {\rm SC}} =& \sum_\bk \sum_{a=X, Y} \psi_{a, \downarrow}^T (-\bk)
    \left ( \begin {array}{c c} \Delta_{\bM 1} & 0 \\ 0 & \Delta_{\bM 3} \end {array} \right ) \psi_{a, \uparrow} (\bk ) + h.c.. \label{HSCM}
\end {align}
We define  the Nambu spinor at the $\bM$-point as $\Psi^\dagger_\bM (\bk) = ( \psi^\dagger_{X, \uparrow} (\bk), \psi^\dagger_{Y, \uparrow} (\bk),
\psi^T_{X, \downarrow} (-\bk), \psi^T_{Y, \downarrow} (-\bk) )$. The resulting Bogoliubov-de Gennes
Hamiltonian is
\begin {align}
  H_{\bM, {\rm BdG}} = \sum_\bk \Psi_\bM (\bk )^\dagger \left ( \begin {array}{c c} h_\bM (\bk) & \DDelta_\bM^* \\
    \DDelta_\bM & -h_\bM (\bk) \end {array} \right ) \Psi_\bM (\bk), \label{HBdGM}
\end {align}
where
\be
  h_{\bM} (\bk) = \left ( \begin {array}{c c} h_{\bM}^+ (\bk)  & 0 \\ 0 & h_\bM^- (\bk) \end {array} \right ).
\ee
To obtain Eq.\ \eqref {HBdGM}, we have used $h_\bM (-\bk)^* = h_\bM (\bk)$, which follows from Eq.\ \eqref {eq:hMpm}; the diagonal pairing matrix is
$\DDelta_\bM = \diag (\Delta_{\bM 1}, \Delta_{\bM 3}, \Delta_{\bM 1}, \Delta_{\bM 3})$.

The Hamiltonian in Eq.\ \eqref {HBdGM} may be brought into a block diagonal form such that the first  block describes the $X$-components of the $E_{\bM 1}$
and $E_{\bM 3}$ doublets, and their pairing, and the other block describes the $Y$-components and their pairing. The
two blocks are related by the mirror reflection about the $yz$-plane, $\{ \sigma^x | 00 \}$, and therefore it is sufficient to consider just
the first block. This reads
\begin {align}
  &H_{\bM^X, {\rm BdG}} = \nonumber \\
  & \sum_\bk \left ( \psi_{X, \uparrow }^\dagger (\bk), \psi_{X, \downarrow}^T (-\bk) \right ) \left ( \begin {array}{c c}
    h_\bM^+ (\bk) & \delta \\ \delta & - h_\bM^+ (\bk) \end {array} \right ) \left ( \begin {array}{c} \psi_{X, \uparrow } (\bk) \\ \psi_{X, \downarrow}^* (-\bk) \end {array} \right ) \label{HMXBdG}
\end {align}
where
\be
  \delta &=& \left ( \begin {array}{c c} \Delta_{\bM 1} & 0 \\ 0 & \Delta_{\bM 3} \end {array} \right ). \label{SCdelta}
\ee
We note that the Bogolyubov-de Gennes matrix in Eq.\ \eqref {HMXBdG} can be written as $\tau_3 \otimes h_{\bM}^+ (\bk) + \tau_1 \otimes \delta$.
Therefore, it anticommutes with $\tau_2 \bbone$, and, as a result, for every eigenstate at $\bk$ with energy $E_\bk$, there is another one
at $\bk$ with energy $-E_\bk$. Therefore, the solution of the fourth order secular equation reduces to the solution of a quadratic equation only.
The explicit form of the solutions is presented in Appendix \ref {AppBdGeigenvalues}.

Before we present the full solution, let us illustrate the behavior of the pairing gap on the Fermi surface in the limit where $\Delta_{\bM 1}$ and $\Delta_{\bM 3}$ are
much smaller than the energy difference between the two eigenvalues of $h_\bM^+ (\bk)$, $\tilde \epsilon_\bk^{(1+)}$ and $\tilde \epsilon_\bk^{(2 +)}$.
Performing a unitary transformation which diagonalizes $h_{\bM}^+ (\bk) \equiv u_+ (\bk) \bbone + u_- (\bk) \sigma^z + v_+ (\bk) \sigma^y$, where
\begin {align}
  u_{\pm} (\bk) = \frac 12 (\epsilon_1 \pm \epsilon_3 ) + \frac {\bk^2}4 \left ( \frac 1 {m_1} \pm \frac 1{m_3} \right ) + \frac 1 2 (a_1 \pm a_3) k_x k_y, \label{upm}
\end {align}
and $v_+ (\bk)$ was given in Eq.\ \eqref {vpm}, we find that the matrix in Eq.\ \eqref {HMXBdG} becomes
\begin {widetext}
\begin {align}
  \left ( \begin {array}{c c c c} \tilde \epsilon_\bk^{(1+)} & 0 & \Delta_{\bM 1}^* \cos^2 \frac {\chi_\bk} 2 + \Delta_{\bM 3}^* \sin^2 \frac {\chi_\bk} 2 &
    \frac 1 2 (\Delta_{\bM 1}^* - \Delta_{\bM 3}^*) \sin \chi_\bk \\
    0 & \tilde \epsilon_\bk^{(2 +)} & \frac 1 2 (\Delta_{\bM 3}^* - \Delta_{\bM 1}^*) \sin \chi_\bk & \Delta_{\bM 1}^* \sin^2 \frac {\chi_\bk} 2 + \Delta_{\bM 3}^* \cos^2 \frac {\chi_\bk} 2 \\
    \Delta_{\bM 1} \cos^2 \frac {\chi_\bk} 2 + \Delta_{\bM 3} \sin^2 \frac {\chi_\bk} 2 &  \frac 12 (\Delta_{\bM 1} - \Delta_{\bM 3}) \sin \chi_\bk
    & - \tilde \epsilon_\bk^{(1+)} & 0 \\
    \frac 12 (\Delta_{\bM 3} - \Delta_{\bM 1}) \sin \chi_\bk & \Delta_{\bM 1} \sin^2 \frac {\chi_\bk} 2 + \Delta_{\bM 3} \cos^2 \frac {\chi_\bk} 2
    & 0 & -\tilde \epsilon_\bk^{(2 +)} \end {array} \right ).  \label{HbandM}
\end {align}
\end {widetext}
where $\cos \chi_\bk = u_-(\bk) / \sqrt {u_-(\bk)^2 + v_+(\bk)^2}$, $\sin \chi_\bk = v_+(\bk) / \sqrt {u_-(\bk)^2 + v_+(\bk)^2}$,
and $\tilde \epsilon_\bk^{(1+,2+)} = u_+ (\bk) \pm \sqrt {u_- (\bk)^2 + v_+ (\bk)^2}$.

If the band with the dispersion $\tilde \epsilon_\bk^{(1+)}$ crosses the Fermi level, while $\tilde \epsilon_\bk^{(2 +)}$ remains
well separated from it, then, to leading order, we can neglect everything but the paring term which couples $\tilde \epsilon_\bk^{(1+)}$
with $-\tilde \epsilon_\bk^{(1+)}$, which become degenerate at the Fermi level.
Therefore, the gap on the Fermi surface --- described by the zeros of $\tilde \epsilon_\bk^{(1+)}$ ---  can be approximated by
\begin{align}
  \Delta^{(1+)} =& \Delta_{\bM 1} \cos^2 \frac {\chi_\bk} 2 + \Delta_{\bM 3} \sin^2 \frac {\chi_\bk} 2 + \ldots \nonumber \\
  =& \frac 12 ( \Delta_{\bM 1} + \Delta_{\bM 3}) + \frac 12 ( \Delta_{\bM 1} - \Delta_{\bM 3})
    \frac {u_-(\bk)}{\tilde \epsilon_\bk^{(1+)} - u_+(\bk)}, \label{DeltaSCband1}
\end{align}
where the $\ldots$ corresponds to terms of order $O \left \lbrack { \left ( \Delta_{\bM 1} - \Delta_{\bM 3} \right )^2}/ |
{ \tilde \epsilon_\bk^{(2+)} } - { \tilde \epsilon_\bk^{(1+)} } | \right \rbrack$,
dropped in the second line. Similarly, the gap on the other band $\tilde \epsilon_\bk^{(2+)}$
can be  directly read off from Eq.\ \eqref {HbandM}, however, unless the system is heavily hole doped, this band is always far below the
Fermi level and need not be considered.

\begin{figure}[h]
\begin{center}
\includegraphics[width=0.44\textwidth]{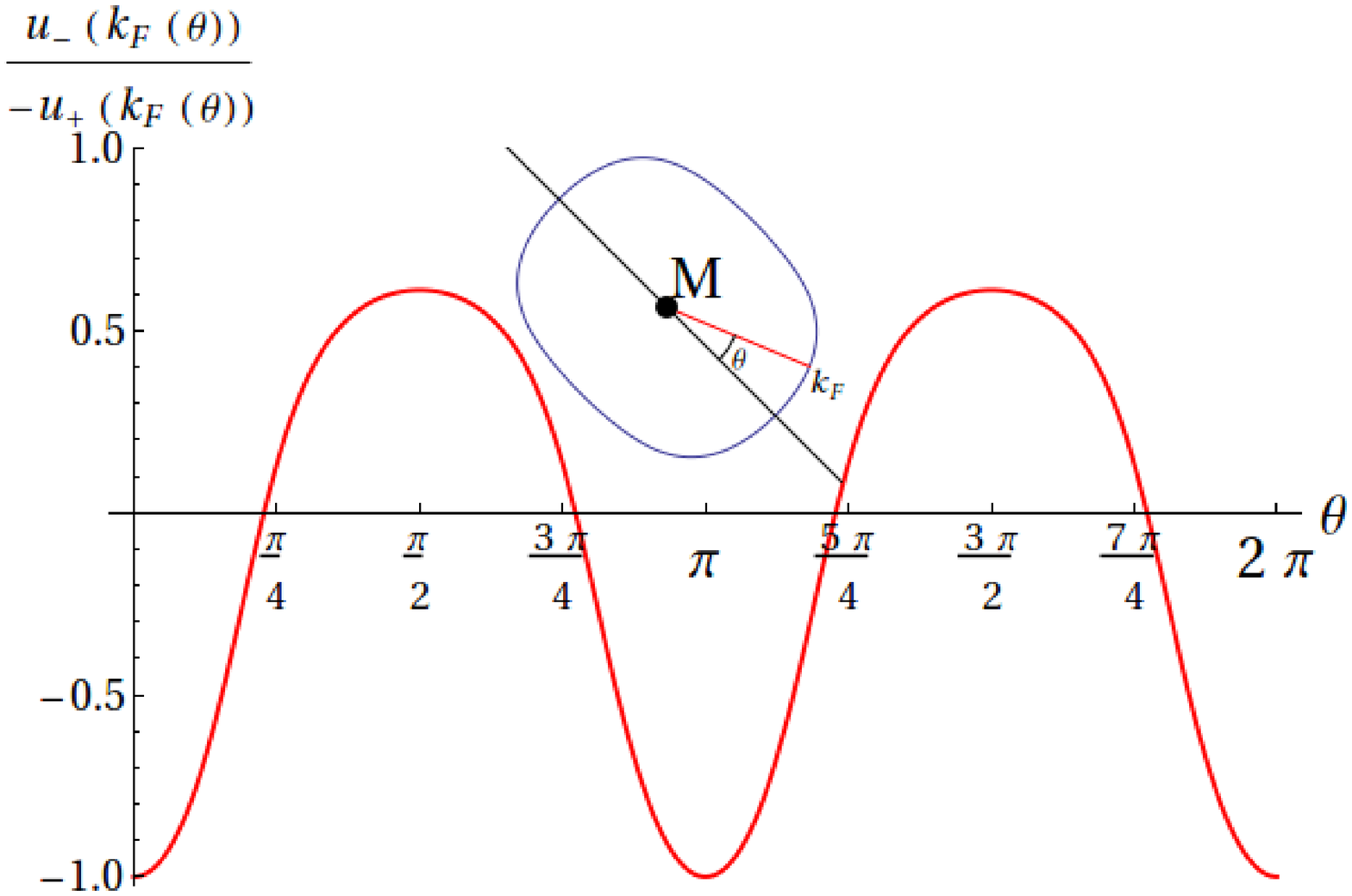} \\
a) \\
\includegraphics[width=0.44\textwidth]{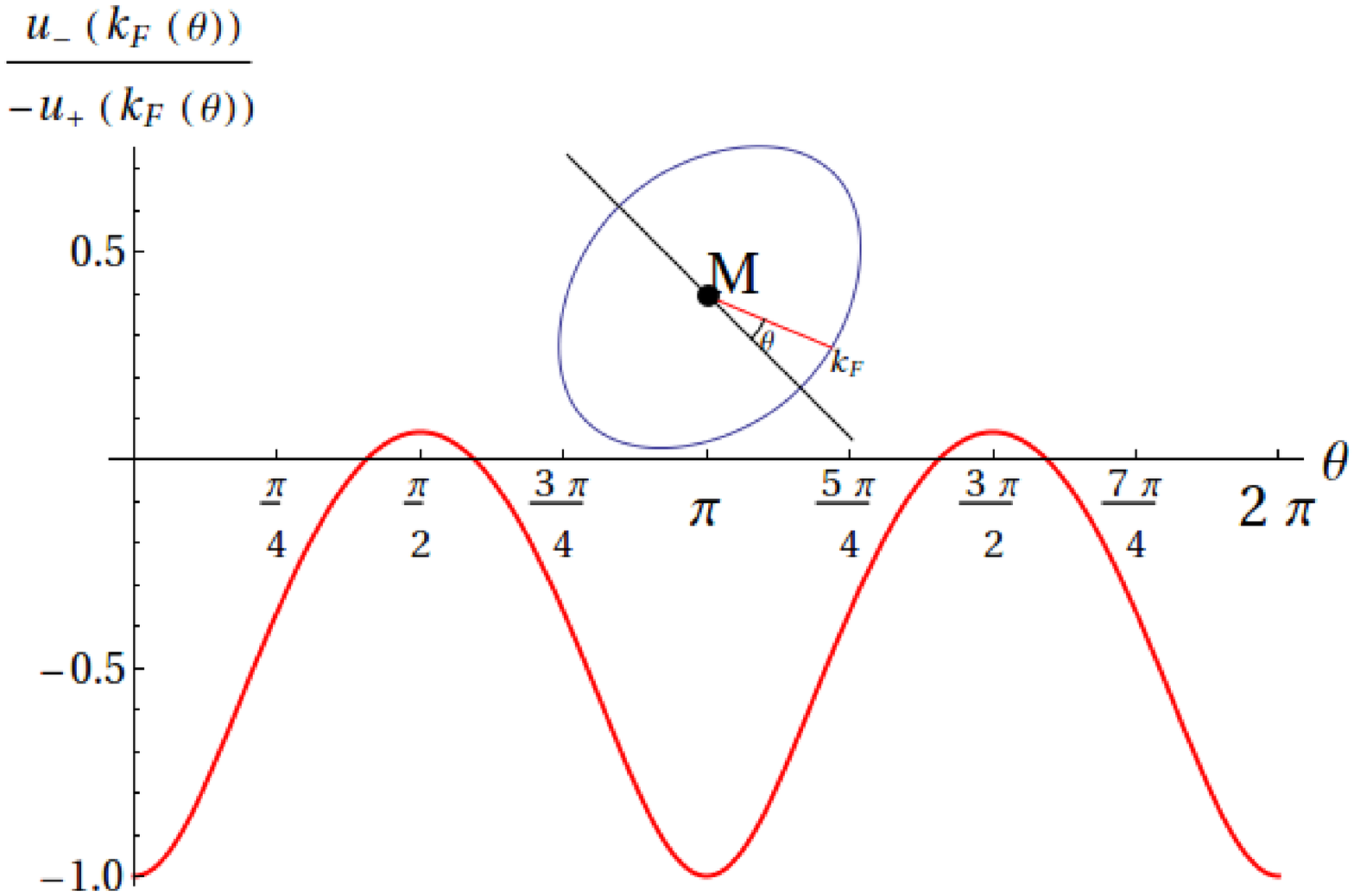} \\
b)
\end{center}
  \caption{The gap on the electron Fermi surface is given by Eq.\ \eqref {DeltaSCband1}. Here we plot its angular dependent part, i.e., $-u_- (\bk) / u_+ (\bk)$,
     at the Fermi surface, where $\tilde \epsilon_\bk^{(1+)} = 0$.}\label{FigAngularGap}
\end{figure}

Note that, unless $\Delta_{\bM 1} = \Delta_{\bM 3}$, the pairing gap at the $\bM$-point is anisotropic. This follows directly from
Eq.\ \eqref {HbandM}, and can also be seen in the approximate expression Eq.\ \eqref {DeltaSCband1}. Along one of the zone diagonals where
$\bk = (k, -k)$,  $v_+ (\bk)$ vanishes by symmetry. This implies that $\tilde \epsilon_\bk^{(1+)} - u_+(\bk) = | u_- (\bk)|$, and the pairing gap  in
Eq.\ \eqref {DeltaSCband1} is either  $\Delta_{\bM 1}$ or $\Delta_{\bM 3}$.  This result does not rely on the approximation made in Eq.\ \eqref {DeltaSCband1},
because along this line $\chi_\bk = 0$ or $\pi$, and, as is clear from
the structure of the matrix Eq.\ \eqref {HMXBdG}, the value of the gap along this line is indeed $\Delta_{\bM 1}$ or $\Delta_{\bM 3}$.

\begin{figure}[h]
\begin{center}
\includegraphics[width=0.44\textwidth]{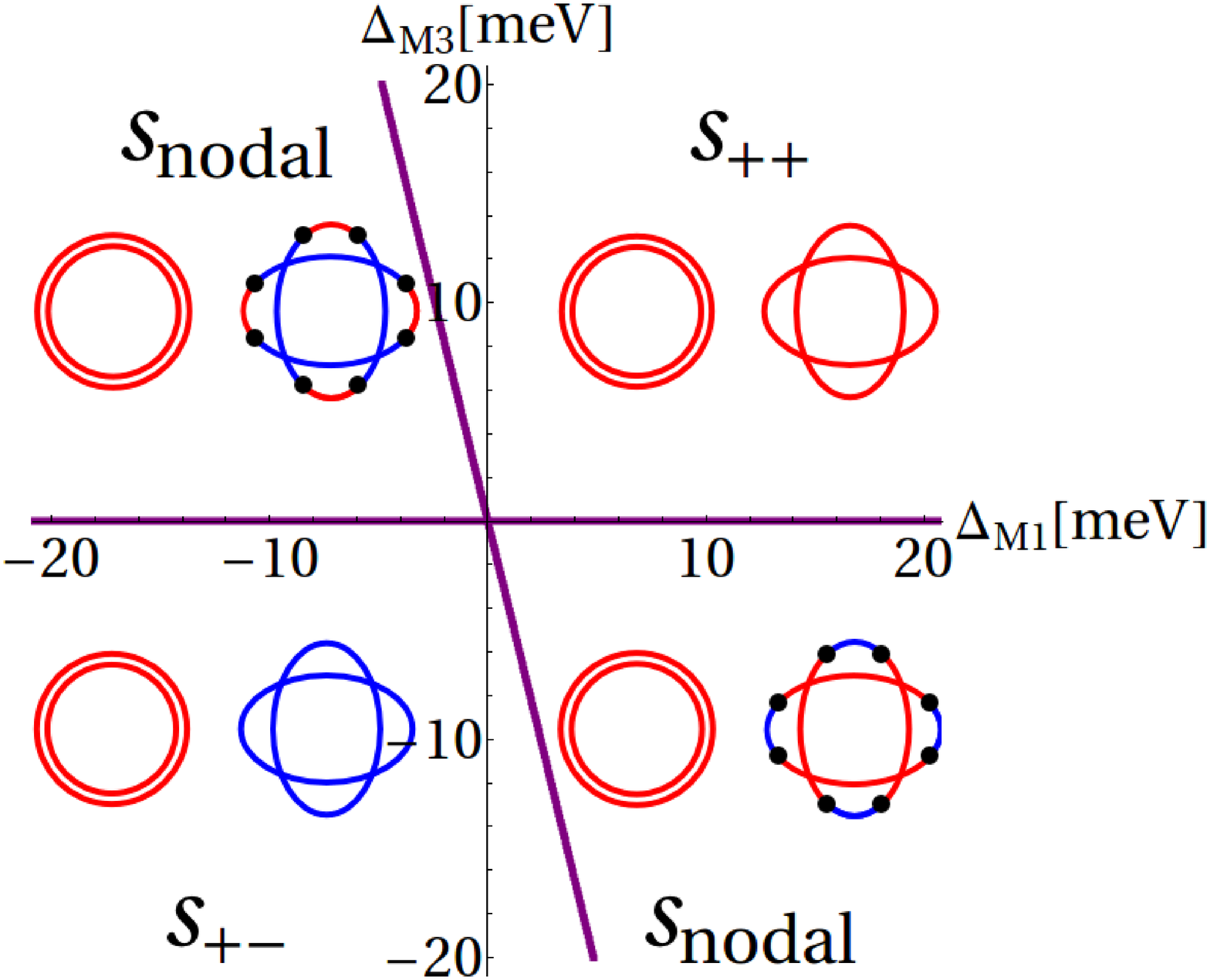}
\end{center}
  \caption{The zero-temperature phase diagram for the superconducting phase of iron-pnictides. Two parameters, $\Delta_{\bM 1}$
    and $\Delta_{\bM 3}$ determine the superconducting gap. The horizontal transition line is given by $\Delta_{\bM 3} =0$; the
    vertical line is determined numerically, and for $\Delta_{\bM 3} \lesssim 10{\rm meV}$ it coincides with the line where
    the zeroth order expression for the gap in Eq.\ \eqref {HbandM} vanishes. In this figure we chose $\Delta_\Gamma > 0$. When $\Delta_\Gamma < 0$,
    only the sign of the gap on the hole Fermi surfaces changes, hence regions with $s_{++}$ and $s_{+-}$ gap exchange places in the phase diagram.}\label{FigNodalPlot}
\end{figure}

We plot the variation of $f(\theta) = - u_-(\bk) / u_+(\bk)$ along the electron Fermi surface in Fig.\ \ref {FigAngularGap} for the fitting parameters given
in Table \ref {TableLoEff}. Along the zone diagonal this expression indeed equals $-1$ in both cases; this means that
the value of the gap is $\Delta_{\bM 3}$.

Fig.\ \ref {FigAngularGap} show that $f(\theta)$ is monotonically increasing as the angle $\theta$ is varied from $0$ to $\pi/2$. The extremal
values of the superconducting gap along the Fermi surface are therefore at $\theta = 0$ and $\theta = \pi/2$. As mentioned,
$f (0) = -1$, therefore one extremal value for the gap is $\Delta_{\bM 3}$. The other extremal  gap value, at $\theta = \pi/2$,  is
\begin {align}
  \Delta (\pi / 2) = \frac 12 (\Delta_{\bM 1} + \Delta_{\bM 3}) + \frac 12 (\Delta_{\bM 1} - \Delta_{\bM 3}) f(\pi/2). \label{Deltapi2}
\end{align}
From these two gap values we can determine the nature of the $s$-wave superconducting state.
Let us assume, for concreteness, that the gap on the hole Fermi surfaces is positive, $\Delta_\Gamma >0$. If both $\Delta_{\bM 3}$ and $\Delta (\pi/2)$
are positive, then the gap on the entire electron Fermi surface is positive. This corresponds to the $s_{++}$-wave
state. Analogously, if both $\Delta_{\bM 3}$ and $\Delta (\pi/2)$ are negative, then the gap is negative on the entire electron Fermi surface, and this is the $s_{+-}$-wave state.
Lastly, when the gaps at their extremal values have different signs, the gap value, being real, must vanish somewhere along the Fermi surface ---
the spectrum of such an $s$-wave superconductor has nodes.

The transitions between the nodal and nodeless $s$-wave states occur when either $\Delta_{\bM 3}$ or $\Delta (\pi / 2)$ vanish.
The latter occurs if
\be
  \Delta_{\bM 3} / \Delta_{\bM 1} = \frac {f (\pi / 2) + 1}{f (\pi / 2) - 1}. \label{Deltaratio}
\ee
For $\theta \neq 0$ and $\theta \neq \pi$, the value of $f(\theta)$ is model dependent. We
find $f (\pi/2) = 0.611$ (Ref.\ \onlinecite {CvetkovicTesanovicEPL2009}) or $f (\pi/2) = 0.067$ (Ref.\ \onlinecite {KurokiPRL2008}), which implies the
that the ratio in Eq.\ \eqref {Deltaratio} is $-4.14$ and $-1.14$, respectively.

We compare these results to the ones obtained without the approximation leading to Eq.\ \eqref {DeltaSCband1}. As shown in Fig.\ \ref {FigNodalPlot},
the transition between the nodal and the nodeless $s$-wave superconductor is indeed well captured by a straight line with a slope
of $-4.14$ in agreement with the approximation in Eq.\ \eqref {DeltaSCband1}.
The four superconducting phases in the plot do not break any space group symmetries because they all correspond to $A_{1g}$
pairing.  While at any finite temperature there is no phase boundary separating them, at $T=0$ the ground state energy is non-analytic
at the transition lines, implying that these are distinct quantum phases.

\subsection {Superconductivity and spin-orbit coupling}

In the previous subsection  we considered only spin singlet pairing terms with a
focus on the $s$-wave superconductivity. In this section we turn our attention to spin-orbit
interaction which necessarily induces  spin triplet pairing terms. In the absence of
any spin-orbit interaction,  there is, of course, spin SU(2) symmetry, and therefore the spin triplet
pairing terms are not induced by the spin singlet pairing. Once the spin-orbit interaction is
included, the spin SU(2) symmetry is lost, and the spin part of any spin triplet pairing term transforms
together with its orbital part under the operations of the space group. Therefore, some spin singlet and some spin triplet
pairing terms may belong to a same irreducible representation of the space group. When this is the case, the
presence of a spin singlet pairing, necessarily induces all spin triplet pairings which belong to the same irreducible representation
as the spin singlet. This is conceptually similar to the effect of the spin-orbit interaction on a spin-density wave order discussed in the section \ref{SecSDW}; in the absence of the
spin-orbit interaction, two spin-density wave orders with mutually perpendicular magnetizations are never related to each other, however,
once the spin-orbit interaction is present, the two spin-density waves may belong to the same irreducible representation and
necessarily induce each other.

We begin by constructing spin triplet pairing bilinears
\be
  H_{\rm triplet} = \sum_{\bk} \psi_{\alpha}^a (-\bk) (i \sigma_2 \vec \sigma)_{\alpha \beta} \cdot \vec M_{ab} \psi_{\beta}^b (\bk) + h.c.,  \label{DeltaSCMtriplet}
\ee
where the three components of $\vec M_{ab}$ are antisymmetric matrices, and $\psi_{\alpha}^a (\bk)$, with $a=1, \ldots 6$ are the components
of the ``spinor'' in  Eq.\ \eqref {eq:spinor} with the spin label $\alpha$. Unlike in the previous subsection, where the spin singlet was invariant under
all space group operations, here we have to ensure that the spin part of the bilinears in Eq.\ \eqref {DeltaSCMtriplet} also transforms
under space group operations. The equation analogous  to Eq.\ \eqref {eq:Dppbilinear} for spin triplet pairing terms is
\begin {align}
  &\sum_{c d j}D_{ac, bd}^{ij} (R) \left \lbrack \sum_{\bk}\psi_{\alpha}^c (-\bk) (i \sigma_2 \sigma_j)_{\alpha \beta} \psi_{\beta}^d (\bk) \right \rbrack = \nonumber\\
    & \sum_{cd} D_{ac}^{\alpha \gamma} (R) D_{bd}^{\beta \delta} (R) \left \lbrack \sum_{\bk}
     \psi_{\gamma}^c (-\bk) (i \sigma_2 \sigma_i)_{\alpha \beta} \psi_{\delta}^d (\bk) \right \rbrack .  \label{eq:Dppbilineartriplet}
\end {align}
The spin part of the triplet pairing term in Eq.\ \eqref {eq:Dppbilineartriplet} transforms as an axial vector, i.e.,
\begin {align}
  &\sum_{cd} D_{ac}^{\alpha \gamma} (R) D_{bd}^{\beta \delta} (R) \left \lbrack \sum_{\bk}
     \psi_{\gamma}^c (-\bk) (i \sigma_2 \sigma_j)_{\alpha \beta} \psi_{\delta}^d (\bk) \right \rbrack = \nonumber \\
  & \quad \sum_\bk \left ( \sum_{c} D_{ac} (R) \psi_{\alpha}^{c} (- \bk) \right) \left ( i \sigma_2 \sum_{j'} D^{\rm AV}_{j j'} (R) \sigma_{j'} \right  )_{\alpha \beta} \nonumber \\
  & \quad \times   \left(\sum_{d} D_{bd} (R) \psi_{\beta}^d (\bk) \right ).  \label{eq:Dppbilineartriplet2}
\end{align}
To show this, we use the   transformation property of the spinor under a mirror reflection through the plane perpendicular to a unit vector ${\bf m}$:
\be
  D^{\alpha \beta}_{ab} ( \sigma^{\bf m} ) \psi^b_{\beta} (\bk) = D_{ab} (\sigma^{\bf m}) (i {\bf m} \cdot \vec \sigma)_{\alpha \beta} \psi^b_{\beta} (\bk'),
\ee
where $\bk'$ is the result of the mirror operation $\sigma^{\bf m}$ on $\bk$.
Substituting this into the first line of Eq.\ \eqref {eq:Dppbilineartriplet2}, we find that
the spin part of the bilinear in Eq.\ \eqref {eq:Dppbilineartriplet} transforms as
\be
   (i \sigma_2 \sigma_j) \longrightarrow (i {\bf m} \cdot \vec \sigma )^T (i \sigma_2 \sigma_j) (i {\bf m} \cdot \vec \sigma) \label{pptripletspin}
\ee
under the mirror $\sigma^{\bf m}$. Because $(\sigma_l)^T \sigma_2 = -\sigma_2 \sigma_l$ for any $l=1,2,3$,
we have
\be
  (i \sigma_2 \sigma_j) \longrightarrow i \sigma_2 ({\bf m} \cdot \vec \sigma) \sigma_i ({\bf m} \cdot \vec \sigma).
\ee
Hence, the component of $\vec M_{ab}$ which is parallel to the ${\bf m}$ is invariant under the mirror reflection $\sigma^{\bf m}$;
the other two  components of $\vec M^{ab}$, which are parallel to the mirror plane, change sign. This is
a property of an axial vector. Similarly, by replacing $\sigma_j$ with $\bbone$ in Eq.\ \eqref {eq:Dppbilineartriplet},
it is readily seen that the bilinears in Eq.\ \eqref {eq:Dppbilinear} are spin singlets.

The second line in Eq.\ \eqref {eq:Dppbilineartriplet2} tells us that the symmetry properties of a bilinear are given by the
product of the irreducible representations of the orbital part of the bilinear and
the axial vector representation of the spin triplet. The symmetry
properties of the orbital part of pairing bilinears have been presented in the previous subsection. The symmetry properties
of the spin part,  an axial vector, are given by
the $A_{2g}$, when the spin points in the $z$-direction, and by the $E_g$, when the spin is in $X$- or $Y$-direction.

\subsubsection {Superconductivity and spin-orbit coupling near $\Gamma$}

At the $\Gamma$-point only the orbital $A_{2g}$ bilinear is antisymmetric in the (orbital) components. When $\vec M_{ab}$
points in the $z$-direction, the entire spin triplet pairing term at $\Gamma$  transforms according to $A_{2g} \otimes A_{2g} = A_{1g}$.
When $\vec M_{ab}$ is perpendicular to the $z$-axis, the spin triplet pairing term  transforms according to $E_g \otimes A_{2g} = E_g$.
We are interested in the former, i.e., the $A_{1g}$,
since such a pairing term has the same overall symmetry properties as the spin singlet $s$-wave  considered
earlier. The resulting pairing Hamiltonian reads
\begin {align}
  H_{\Gamma, {\rm triplet}} = \Delta_{\Gamma t} \sum_{k} \psi_{\Gamma, \downarrow}^T (-\bk) \left ( \begin {array}{c c} 0 & -i \\ i & 0 \end {array}  \right )
    \psi_{\Gamma, \uparrow} (\bk ) + h.c. \label{HSCGammatriplet}
\end {align}
where by time reversal symmetry, which we assume to be present, $\Delta_{\Gamma t}$ is real.

We can use the same Nambu spinor for the states near the $\Gamma$-point as in Eq.\ \eqref {HBdGGamma}.  The Bogolyubov-de Gennes
Hamiltonian is
\begin {align}
  &H_{\Gamma, {\rm BdG}} = \nonumber \\
  & \sum_\bk \Psi_\Gamma (\bk)^\dagger \left (
    \begin {array}{c c} h_\Gamma (\bk) + \thalf \lambda_\Gamma \tau_2 & \Delta_\Gamma \bbone +  \Delta_{\Gamma t} \tau_2 \\
    \Delta_\Gamma \bbone + \Delta_{\Gamma t} \tau_2 & - h_\Gamma (\bk) - \thalf \lambda_\Gamma \tau_2 \end {array} \right ) \Psi_\Gamma (\bk), \label{HGammaBdGtriplet}
\end {align}
where $\tau_2$ acts in the same space as $h_\Gamma (\bk)$ defined in Eq.\ \eqref {eq:hGamma}.

\begin{figure}[h]
\begin{center}
\includegraphics[width=0.44\textwidth]{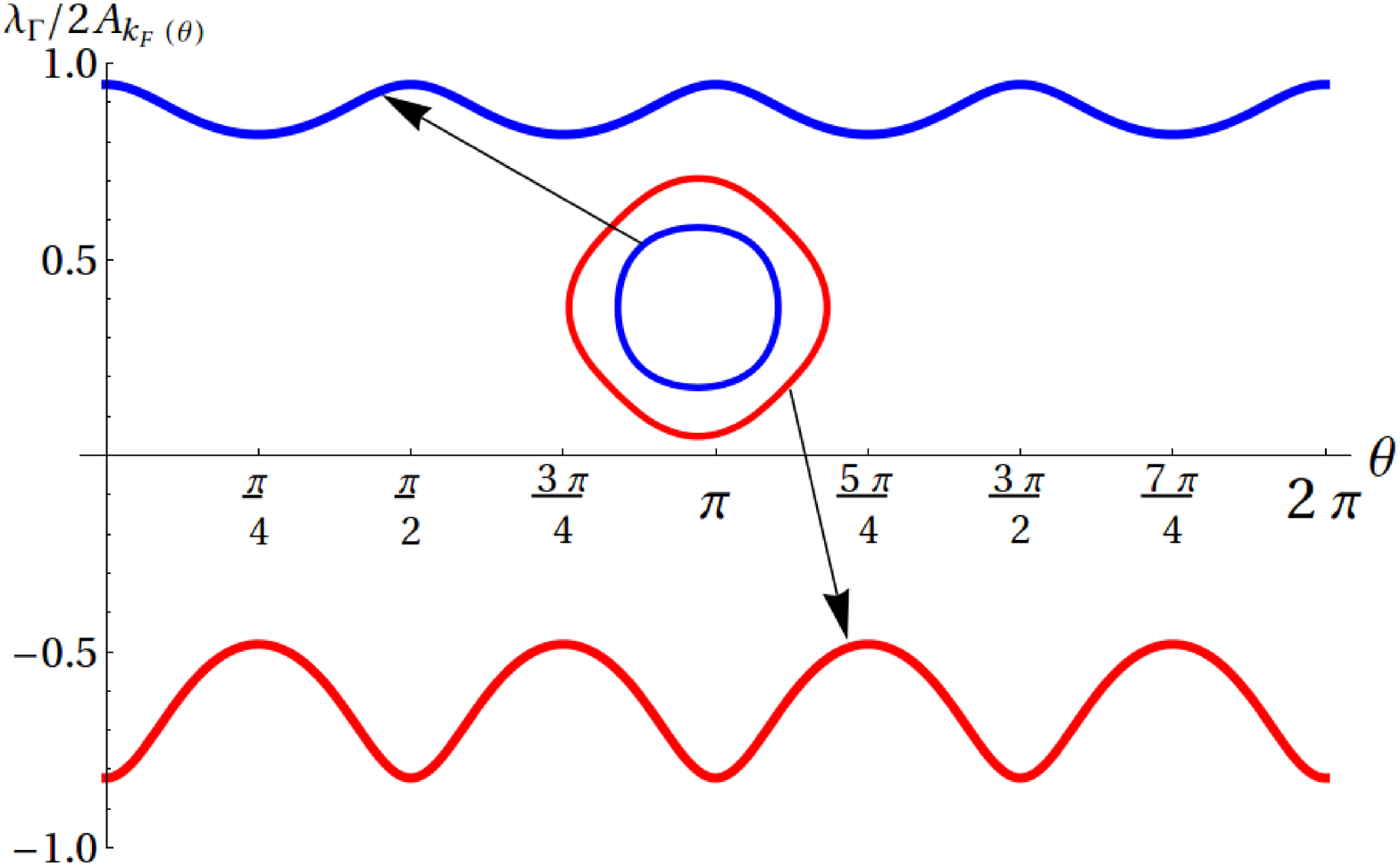}
\end{center}
  \caption{The angular dependence of $\lambda_\Gamma / 2A_{\bk}$ along the inner (top) and the outer (bottom) hole Fermi surfaces, calculated for the parameters
    from Ref.\ \onlinecite {CvetkovicTesanovicEPL2009}, Table \ref {TableLoEff}. The variation
    of this ratio enters into the formula for the anisotropy of the superconducting gap on the hole Fermi surfaces, Eq.\ \eqref {DeltaGammaFS}.}\label{FigGammaGapAnisotropy}
\end{figure}

Using the shorthand notation
\be
  h_\Gamma (\bk) + \thalf \lambda_\Gamma \tau_2 &=& A_\bk \bbone + \vec \tau \cdot \vec B_\bk, \nonumber \\
  \Delta_\Gamma \bbone + \Delta_{\Gamma t} \tau_2 &=& C_\bk + \vec \tau \cdot \vec D_\bk,
\ee
we use the Eq.\ \eqref {HBdGappendix}
in the appendix. The eigenvalues of the Bogolyubov-deGennes Hamiltonian Eq.\ \eqref {HGammaBdGtriplet} are given by Eq.\ \eqref {HBdGspectrumappendix},
\begin {align}
  E_\bk^2 =& A_\bk^2 + \vec B_\bk ^2 + C_\bk^2 + \vec D_\bk^2  \nonumber \\
    & \pm 2\sqrt{ \left(A_\bk \vec B_\bk + C_\bk\vec D_\bk\right)^2+ {\vec B}^2_\bk \vec D^2_\bk -\left(\vec B_\bk\cdot \vec D_\bk\right)^2}.  \label{eq:E2BdGGamma}
\end {align}
The normal state dispersions for the two hole bands in the presence of the spin-orbit coupling are
\be
  \epsilon_\bk^{(\pm)} = A_\bk \pm | \vec B_\bk | = A_\bk \pm B_\bk, \label{eq:epsilonGamma}
\ee
where we write $B_\bk = | \vec B_\bk |$. We can therefore eliminate $B_\bk$ in the Eq.\ \eqref {eq:E2BdGGamma} in favor of $\pm (\epsilon_\bk^{(\pm)} - A_\bk )$,
while noting that the scalar product $\vec{B}_\bk\cdot\vec{D}_{\bk} = \frac{1}{2}\lambda_\Gamma \lambda_{\Gamma t}$  is $\bk$-independent.
At the Fermi surface, one of the $\epsilon_\bk^{(\pm)}$'s vanishes, and the gap is given by
\begin {align}
  &\Delta_\Gamma^{\rm FS} = \bigg \lbrack 2 A_\bk^2 + \Delta_\Gamma^2 + \Delta_{\Gamma t}^2 \nonumber \\
  &- 2 \sqrt {A_\bk^4 + \Delta_{\Gamma t}^2 (A_\bk^2 + \Delta_\Gamma^2 - \tfrac 1 4 {\lambda_\Gamma^2} ) +
    A_\bk \lambda_\Gamma \Delta_{\Gamma} \Delta_{\Gamma t} } \bigg \rbrack^{1/2}. \label{DeltaGammaanisotropic}
\end {align}
In the above expression, $\bk$ lies either on the smaller or on the larger hole Fermi surface, which are not circular. Because $A_\bk = \epsilon_\Gamma
+ \bk^2 / (2 m_\Gamma)$, is circularly symmetric, the gap on either of the Fermi surfaces is anisotropic. In the limit when $\Delta_{\Gamma t}$
vanishes, however, the gap is isotropic and given by $\Delta_{\Gamma}$, even for a non-zero spin-orbit coupling constant $\lambda_\Gamma$.

To determine the size of the anisotropy,  we note that, for the parameters of Ref.\ \onlinecite {CvetkovicTesanovicEPL2009}, and for $\lambda_\Gamma = 80{\rm meV}$,
the magnitude of $A_\bk$ is at least $40 {\rm meV}$ on both hole Fermi surfaces. Therefore, for $\Delta$'s small compared to this energy
scale, we can expand Eq.\ \eqref {DeltaGammaanisotropic} in powers of $\Delta_\Gamma / A _\bk$, and $\Delta_{\Gamma t} / A_\bk$. In this limit we find that
the gap on the Fermi surfaces is given by
\be
  \Delta_\Gamma^{\rm FS} = \left | \Delta_\Gamma - \frac {\lambda_\Gamma}{2 A_\bk} \Delta_{\Gamma t} \right | + \ldots . \label{DeltaGammaFS}
\ee
In Fig.\ \eqref {FigGammaGapAnisotropy} we plot ${\lambda_\Gamma} /2{A_\bk}$ on the inner and the outer hole Fermi surfaces for the parameters of the model in
Ref.\ \onlinecite {CvetkovicTesanovicEPL2009} given in Table \ref {TableLoEff}. This approximate expression vanishes if $\Delta_{\Gamma} / \Delta_{\Gamma t}$
intersects either the upper blue curve or the lower red curve in Fig.\ \ref {FigGammaGapAnisotropy}. Therefore, to this order we would
conclude that either the inner or the outer hole Fermi surface has gap nodes. Closer analysis of the expression Eq.\ \eqref {DeltaGammaanisotropic}
reveals that the true gap nodes appear only under much more stringent conditions: $\Delta_{\Gamma} = \Delta_{\Gamma t}$, and
$A_\bk = \lambda_\Gamma / 2$ at some $\bk$ on a Fermi surface. Therefore, the gap nodes determined from the approximate expression Eq.\ \eqref {DeltaGammaFS}
are only ``near gap nodes'', in that ``$\ldots$'' terms give the gap a finite but small value (of order $\Delta_\Gamma^3 / A_\bk^2$).
Nevertheless,  because ${\lambda_\Gamma} / {A_\bk}$
has opposite sign on the inner and the outer hole Fermi surface, we see that, if the ``near gap nodes'' appear, then they only appear on one of the Fermi surfaces;
the gap on the other Fermi surface remains relatively isotropic.

\subsubsection {Superconductivity and spin-orbit coupling near $\bM$}

At the $\bM$-point we have to consider the direct products: $E_{\bM 1} \otimes E_{\bM 1}$, $E_{\bM 3} \otimes E_{\bM 3}$,
and $E_{\bM 1} \otimes E_{\bM 3}$. These are given in Table \ref {TableExE}. We see that there the only product which contains $E_g$
is $E_{\bM 1} \otimes E_{\bM 3}$. Hence we can combine it with the planar, $E_g$, components of the axial vector $\vec M_{ab}$
and form an $A_{1g}$ pairing bilinear. Since it is also antisymmetric in its orbital indices, it satisfies the proper anticommutation
relations, and such term will necessarily get induced by the spin-orbit coupling in the presence of the $A_{1g}$ singlet.
Therefore, the spin triplet pairing Hamiltonian at the $\bM$-point reads
\begin {align}
  H_{\bM, {\rm triplet}} =& \Delta_{\bM t} \sum_{\bk} \bigg ( \psi_{X \downarrow}^T (-\bk) \left \lbrack \begin {array}{c c} 0 & -i \\ -1 & 0 \end {array} \right \rbrack
     \psi_{Y \downarrow} (\bk) \nonumber \\
    & \quad + \psi_{Y \uparrow}^T (-\bk) \left \lbrack \begin {array}{c c} 0 & 1 \\ -i & 0  \end {array} \right \rbrack \psi_{X \uparrow} (\bk) \bigg  ) + h.c. \label{HSCMtriplet}
\end {align}
The triplet pairing parameter $\Delta_{\bM t}$ is real by time reversal symmetry, which we assume to be present.

We now construct the eight component Nambu spinor
\be
  \Psi_\bM (\bk) = \left ( \begin {array}{c} \psi_{X \uparrow} (\bk) \\ \psi_{Y \downarrow} (\bk) \\ \psi^*_{X \downarrow} (-\bk) \\ -\psi^*_{Y \uparrow} (-\bk) \end {array} \right ),
\ee
which we use to set up the Bogolyubov-deGennes Hamiltonian
\be
  H_{{\rm BdG}, \bM} = \sum_\bk \Psi_\bM^\dagger (\bk)
  \left \lbrack
  \begin {array}{c c}
    {\mathcal H} (\bk) & \DDelta \\
    \DDelta & - {\mathcal H} (\bk)
  \end {array}
  \right \rbrack
  \Psi_\bM (\bk), \label{HBdGMtriplet}
\ee
where
\be
  \DDelta &=& \left ( \begin {array} {c c c c} \Delta_{\bM 1} & 0 & 0 & - i \Delta_{\bM t} \\ 0 & \Delta_{\bM 3} & - \Delta_{\bM t} & 0 \\
    0 & - \Delta_{\bM t} & \Delta_{\bM 1} & 0 \\ i \Delta_{\bM t} & 0 & 0 & \Delta_{\bM 3} \end {array} \right ), \\
  {\mathcal H} (\bk) &=& \left ( \begin {array}{c c} h_{\bM}^+ (\bk) & \Lambda \\ \Lambda^\dagger & h_{\bM}^- (\bk) \end {array} \right ), \\
  \Lambda &=& \frac 1 2 \lambda_\bM \left ( \begin {array}{c c} 0 & i \\ 1 & 0 \end {array} \right ).
\ee
Because the matrix in Eq.\ \eqref {HBdGMtriplet} can be writen as $\tau_3 \otimes {\mathcal H} (\bk) + \tau_1 \otimes \DDelta$, it manifestly
anticommutes with $\tau_2 \otimes \bbone$. Therefore,  for any
eigenstate at $\bk$ with energy $E_{\bk}$, there is an eigenstate at $\bk$ with energy $-E_{\bk}$.

To proceed with the approximate solution to the spectrum, we first perform a unitary transformation which diagonalizes
$h_{\bM}^+ (\bk)$ and $h_{\bM}^- (\bk)$; for the parameters of our model, the two eigenvalues for each are split by an energy which is
larger than both the spin-orbit coupling energy scale and the pairing scale. Therefore, we can safely project onto the
states which are near the Fermi level. This will result in a $4 \times 4$ Hamiltonian with the particle hole symmetry, meaning
that the secular equation will reduce to quadratic. The corresponding basis is
\be
  \left ( \begin {array}{c} | + \rangle \\ 0 \\ 0 \\ 0 \end {array} \right ), \quad
  \left ( \begin {array}{c} 0 \\ | - \rangle \\ 0 \\ 0 \end {array} \right ), \quad
  \left ( \begin {array}{c} 0 \\ 0 \\ | + \rangle \\ 0 \end {array} \right ), \quad
  \left ( \begin {array}{c} 0 \\ 0 \\ 0 \\ | - \rangle \end {array} \right ), \quad
\ee
where $h_{\bM}^\pm (\bk) | \pm \rangle = \tilde \epsilon^{(1 \pm)}_\bk | \pm \rangle$. In this basis, the effective Bogolyubov-de Gennes
Hamiltonian reads
\be
  {\mathcal H}_{\bM, {\rm BdG}} = \left ( \begin {array}{c c} \tilde {\mathcal H} (\bk) & \tilde \DDelta (\bk) \\ \tilde \DDelta (\bk) & -\tilde {\mathcal H} (\bk) \end {array} \right ),
    \label{HBdGMtriplet2}
\ee
where
\be
  \tilde {\mathcal H} (\bk) &=& \left ( \begin {array}{c c} \tilde \epsilon_\bk^{(1+)} & \thalf \lambda_\bM \kappa \\
    \thalf \lambda_\bM \kappa^* & \tilde \epsilon_\bk^{(1-)}  \end {array} \right ), \\
  \tilde \DDelta &=& \left ( \begin {array}{c c} \Delta^{(1+)} & - \Delta_{\bM t} \kappa \\
    - \Delta_{\bM t} \kappa^* & \Delta^{(1-)}  \end {array} \right ).
\ee
where $\Delta^{(1+)}$ was given in Eq.\ \eqref {DeltaSCband1}, and analogously,
\begin {align}
  \Delta^{(1-)} =& \frac 12 ( \Delta_{\bM 1} + \Delta_{\bM 3}) + \frac 12 ( \Delta_{\bM 1} - \Delta_{\bM 3})
    \frac {w_-(\bk)}{\tilde \epsilon_\bk^{(1-)} - w_+(\bk)}, \label{DeltaSCband1m}
\end {align}
where $w$'s are defined as $h^- (\bk ) = w_+ (\bk) \bbone + w_- (\bk) \sigma^z + v_- (\bk ) \sigma^y$;
\be
  \kappa =& \langle + | \left ( \begin {array}{c c} 0 & i \\1 & 0 \end {array} \right ) | - \rangle.
\ee

The phase of this complex number may be eliminated by a unitary transformation generated by $\bbone \otimes \sigma_3$. This transformation
leaves invariant all terms which do not contain $\kappa$; in all other terms $\kappa$ is transformed into
\be
  | \kappa | = \left \lbrack \frac 12 \left ( 1 -  \frac {u_-(\bk)}{\tilde \epsilon_\bk^{(1+)} - u_+(\bk)}  \frac {w_-(\bk)}{\tilde \epsilon_\bk^{(1-)} - w_+(\bk)} \right ) \right \rbrack ^{1/2}
\ee

The Bogolyubov-de Gennes Hamiltonian in Eq.\ \eqref {HBdGMtriplet} can be written in the form
$\tau_3 \otimes \tilde {\mathcal H} (\bk) + \tau_1 \otimes \tilde \DDelta (\bk)$.
Eigenvalues of such matrices are derived in the Appendix \ref {AppBdGeigenvalues}. The spectrum of Eq.\ \eqref {HBdGMtriplet} can be obtained
exactly by substituting
\begin {align}
  A_\bk =& \frac{1}{2}\left ( \tilde \epsilon_\bk^{(1+)} + \tilde \epsilon_\bk^{(1-)} \right ), \\
  {\vec B}_\bk =& \frac 12 \left ( \lambda_\bM |\kappa|, 0,  \tilde \epsilon_\bk^{(1+)} - \tilde \epsilon_\bk^{(1-)} \right ), \\
  C_\bk =& \frac{1}{2}\left ( \Delta^{(1+)} + \Delta^{(1-)} \right ), \\
  {\vec D}_\bk =& \left ( -\Delta_{\bM t} | \kappa |, 0, \frac{1}{2}\left ( \Delta^{(1+)} - \Delta^{(1-)} \right ) \right ),
\end {align}
into Eq.\ \eqref {HBdGspectrumappendix},
\begin {align}
  E_\bk^2 =& A_\bk^2 + \vec B_\bk ^2 + C_\bk^2 + \vec D_\bk^2  \nonumber \\
    & \pm 2\sqrt{ \left(A_\bk \vec B_\bk + C_\bk\vec D_\bk\right)^2+ {\vec B}^2_\bk \vec D^2_\bk -\left(\vec B_\bk\cdot \vec D_\bk\right)^2}
\end {align}

The dispersion in the normal state is $\tilde \epsilon^\pm_\bk = A_\bk \pm | \vec B_\bk | = A_\bk \pm B_\bk$, where we, just like in the previous subsection,
define $B_\bk \ = | \vec B_\bk |$. Because along
the zone diagonal $\tilde \epsilon_\bk^{(1+)} = \tilde \epsilon_\bk^{(1-)}$, it is the non-zero spin-orbit coupling $\lambda_\bM$
that lifts the crossings between the two "elliptical" electron Fermi surfaces. The ``$+$'' sign therefore corresponds to the inner electron
Fermi surface, and the ``$-$'' sign to the outer Fermi surface. Eliminating $A_\bk$ in favor
of $\tilde \epsilon^\pm_\bk \mp B_\bk $ and noting that at an electron Fermi surface, one of $\tilde \epsilon_\bk^\pm$ vanishes, we find that the gap is given by
\begin {align}
  &\Delta_{\bM \pm}^{\rm FS} = \bigg \lbrack 2 {B}_\bk ^2 + C_\bk^2 + \vec D_\bk^2- \nonumber \\
  & 2 \sqrt {  {B}_\bk ^4 + {B}_\bk ^2 \Big ( \vec D_\bk^2 - (\hat {\bf b}_\bk \cdot \vec D_\bk)^2\ \mp
    2 C_\bk \hat {\bf b}_\bk \cdot \vec D_\bk  \Big ) + C_\bk^2 D_\bk^2 } \bigg \rbrack^{1/2}, \label{DeltaMFS}
\end {align}
where we defined the unit vector $\hat {\bf b}_\bk = \vec B_\bk / B+\bk$.
In this expression, the momentum $\bk$ lies on one of the electron Fermi surfaces.

The magnitude of $B_\bk$ is at least $28{\rm meV}$ on either of the Fermi surfaces for the parameters in Table \ref {TableLoEff}
corresponding to Ref.\ \onlinecite {CvetkovicTesanovicEPL2009} and $\lambda_\bM = 80{\rm meV}$. For $\Delta$'s small
compared to this energy scale we can expand the Eq.\ \eqref {DeltaMFS} in powers of $\Delta_{\bM 1} /  B_\bk$, $\Delta_{\bM 3} / B_\bk $,
and $\Delta_{\bM t} /  B_\bk$. In this limit we find that the gap on Fermi surfaces is
\begin {align}
  \Delta_{\bM \pm}^{\rm FS} =& | C_\bk  \pm \hat {\bf b}_{\bk} \cdot \vec D_\bk | + \ldots \nonumber \\
  =& | \alpha_{\bM 1} (\bk) \Delta_{\bM 1} + \alpha_{\bM 3} (\bk) \Delta_{\bM 3} + \alpha_{\bM t} (\bk) \Delta_{\bM t} | + \ldots  \label{DeltaMFSseries}
\end {align}
Here ``$\ldots$'' represents terms of order $\Delta^2 / B_\bk $.

\begin{figure}[h]
\begin{center}
\includegraphics[width=0.44\textwidth]{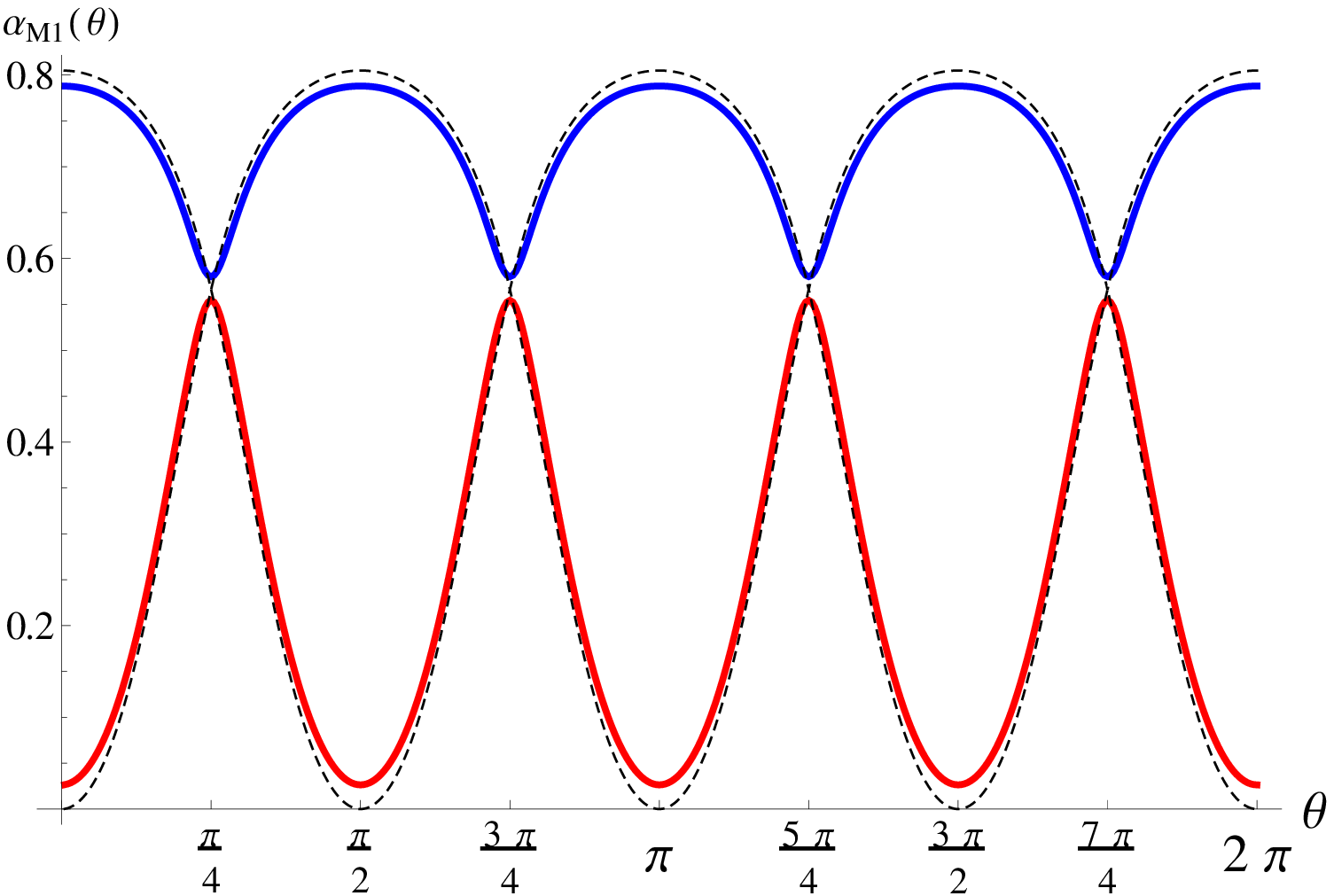} \\
a)\\
\includegraphics[width=0.44\textwidth]{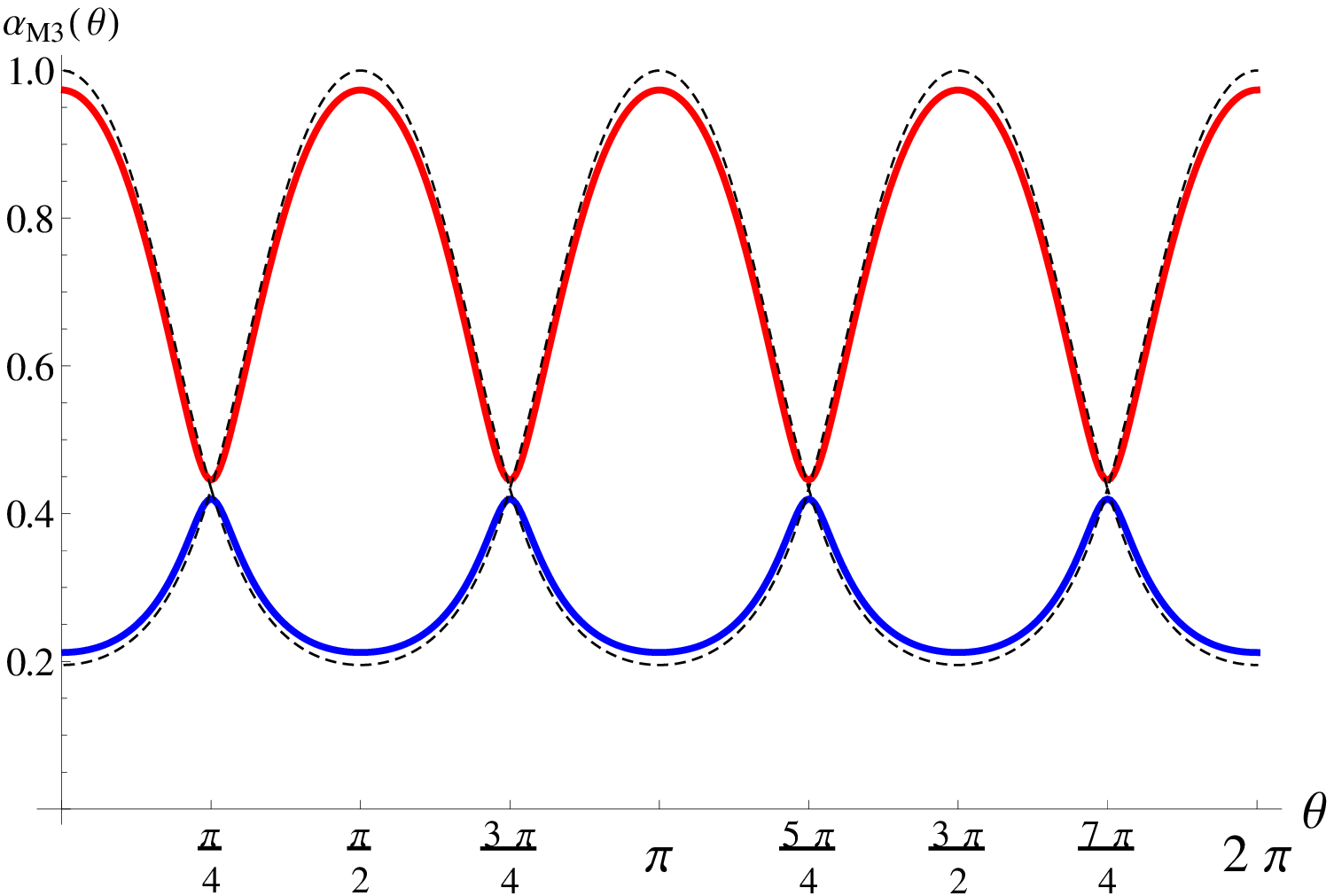} \\
b)\\
\includegraphics[width=0.44\textwidth]{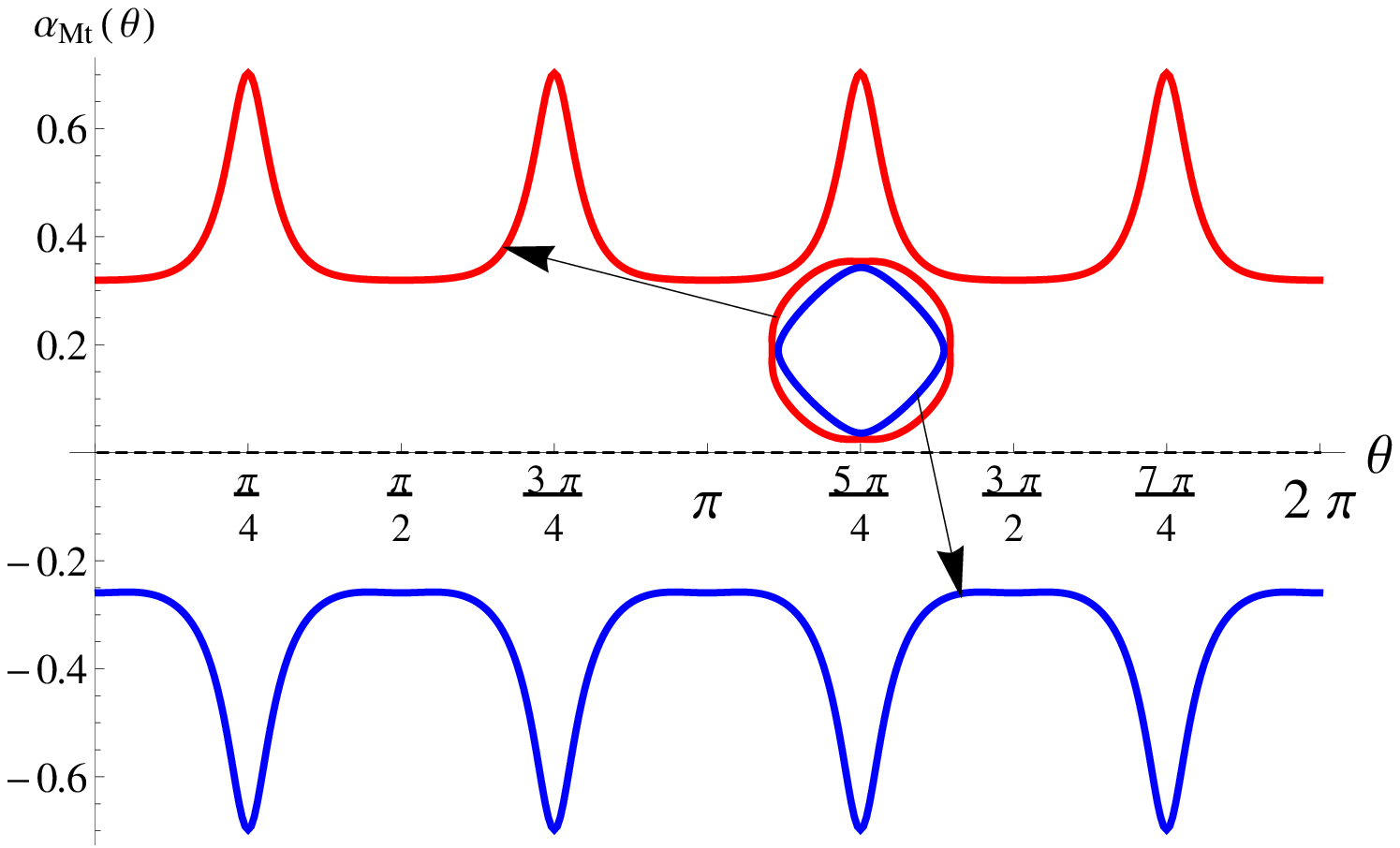} \\
c)
\end{center}
  \caption{The coefficients $\alpha (\bk)$ in the electron Fermi surfaces gap estimate, Eq.\ \eqref {DeltaMFSseries}, as a function of angle $\theta$. The
    coefficients imply a four-fold symmetric gap on each electron Fermi surface. The red
    and blue line correspond to the outer and inner Fermi surfaces respectively which do not touch in the presence of spin-orbit coupling.
    The dashed lines in a) and b) correspond to no spin-orbit interaction, in which case $\alpha_{\bM t}$ is absent.}\label{FigTriplealpha}
\end{figure}

The second line of Eq.\ \eqref{DeltaMFSseries} defines the coefficients $\alpha_j (\bk)$, which are plotted in Fig.\ \ref{FigTriplealpha}
as a function of angle $\theta$ on the two electron Fermi surfaces: inner (blue) and outer (red). For comparison, in Figs.\ \ref{FigTriplealpha}a and \ref{FigTriplealpha}b,
we also plot $\alpha_j$'s in the absence of spin-orbit interaction (dashed curves). The most important consequences of the spin-orbit coupling on the superconducting quasiparticle spectrum
are: a) coupling of the two electron Fermi surfaces via
lifting of the degeneracy which occurred for $\lambda_\bM=0$ at the Brillouin zone edge --- the Fermi surfaces reconstruct and each Fermi surface is four-fold symmetric; b)
the admixture of the spin triplet pairing, $\Delta_{\bM t}$, which contributes to the overall size of
gap with a coefficient $\alpha_{\bM t}$,  and which is comparable to the two
spin singlet coefficients $\alpha_{\bM 1}$, $\alpha_{\bM 3}$ in the Eq.\ \eqref {DeltaMFSseries}.

\section{Discussion}

We applied the method of invariants \cite{Luttinger1956, BirPikus} to construct an effective low-energy effective continuum model for iron based superconductors.
Instead of repeating the conclusions which were stated in the introduction, in this section we discuss directions along which this line of reasoning
can be extended. As we have mentioned, among the advantages of our approach is that it readily allows one to analyze the effects of an external magnetic field.
Therefore, unlike previous continuum models, ours has the advantage of being suitable for a calculation of cyclotron resonances.

While for the spin-density wave order we have mapped out all of the symmetry distinguishable order parameters with wave-vector $\bM$,
we focused primarily on the $A_{1g}$ superconducting order which preserves the time-reversal symmetry. In the Appendix \ref{AppSCpairs} we list the
remaining symmetry distinguishable
pairing terms without analyzing their effect on the quasi-particle spectrum. Since in some of the iron based materials there is a possibility of
having a non-$A_{1g}$ superconducting order, such analysis can be readily performed using our low-energy effective theory.

The method of invariants used here is generic \cite{BirPikus}. It relies on identifying the symmetry of the {\em exact} Bloch states at isolated
points in the Brillouin zone, and an expansion in powers of momentum away from such points. As long as the Fermi momenta are
a small fraction of the Brillouin zone, low order expansions are sufficient.
We demonstrated the  applicability of the method by fitting to  two tight-binding band structure calculations reported in Refs.\
\onlinecite{CvetkovicTesanovicEPL2009} (8 bands) and \onlinecite{KurokiPRL2008} (5 bands).
The method can be applied to construct a low-energy effective theory based on a microscopic band structure for any
other iron based  material with the same lattice symmetry, but possibly different
spectrum with small pockets. To do so, one simply has to select the isolated points in the Brillouin zone and the corresponding
states near the Fermi level as the basis states. Then project  the Hamiltonian  onto this basis, and perform the expansion in powers of momentum away from such points.
If some states do not cross the Fermi level, but lie close to it, one could include those as well by extending the number of
components of our ``spinor'' $\psi_{\bk, \sigma}$, Eq.\ \eqref{eq:spinor}.

The space group we study, $P4/nmm$, is the group of symmetries for the parent compounds with a single iron layer per unit cell, such as those within the
1111, 111, and 11 families \cite {JohrendtZKrist2011}.  Since this group is non-symmorphic, the conclusions we draw  for these compounds
apply directly. For the 122 family the space group of symmetries is $I4/mmm$. The puckered iron-pnictide layer is still the
basic structural unit, however, the puckering alternates along the $z$-direction from layer to layer. As a result,
this group is symmorphic, and therefore our results must be modified when applied to such materials. In the limit when the
inter-plane coupling can be ignored,  our results apply without modifications. While we have implicitly  assumed absence of
$k_z$-dispersion, our low-energy effective theory has the same form, with possibly modified parameters at $k_z=0$ and $k_z = \pi/ c$.
For weak inter-layer coupling, as applicable to these systems, the $k_z=0$ and $k_z= \pi / c$ planes in the Brillouin zone should be folded.
Our results can then be used as an appropriate starting point, defining the basis which is weakly mixed.

In the paper we present the irreducible representations of $P4/nmm$ the way we obtained them. We followed the physically motivated method of
C.\ Herring \cite {Herring1942}, by constructing the factor groups for non-symmorphic space groups, and the method of induced representations
outlined in  Ref.\ \onlinecite {InuiTanabeOnodera}. Alternatively, using different methods, generators, product tables and irreducible
representation for all 230 space groups --- including $P4/nmm$ --- along the high symmetry lines,
or at high symmetry points, appear in the Ref.\ \onlinecite{BradleyCracknell}. Our results are fully consistent.

The Landau free energy function, built based on the irreducible representations at the $\bM$-point, displays two distinct
isolated extrema below the ordering temperature $T_{\bM}$. One corresponds to the collinear spin-density wave, in our notation transforming
as $E_{\bM 1}^X$ or $E_{\bM 1}^Y$, whose presence in iron-pnictides
have been well established \cite{delaCruzNature2008, GoldmanPRB2008}. The other corresponds to the coplanar, four-fold symmetric, spin-density wave order,
$E_{\bM 1}^X \pm E_{\bM 1}^Y$,  which may have been observed recently \cite{AvciarXiv2013}. When the former order parameter
corresponds to a minimum, the latter is a saddle point of the Landau free energy, and vice versa. Which is preferred depends on the sign of the
coefficient $\lambda'$ in the quartic term, $\lambda'\left(\Delta^2_{E_{\bM 1}^X}-\Delta^2_{E_{\bM 1}^Y}\right)^2$. We speculate that the transition observed in Ref.\ \onlinecite{AvciarXiv2013}
is due to the temperature dependence of $\lambda'$ which is negative above the coplanar ordering temperature and positive below.
Such transition is necessarily first order, in agreement with Refs.\ \onlinecite{AvciarXiv2013} and \onlinecite{FernandesPRB2012},
although without invoking any spin-nematic model.
Our method allows us to study the difference in the electronic spectrum for the collinear and the coplanar spin-density waves.
Interestingly, while the former breaks the time-reversal symmetry, the electronic spectrum is still Kramers degenerate. That is
because, the product of time-reversal and inversion is preserved. On the other hand, the coplanar spin-density wave
breaks both the time reversal and inversion, as well as their product. The resulting electronic spectrum has no Kramers degeneracies.
Therefore, in this phase all Fermi surfaces split as seen in Fig.\ \ref{FigC4dispersion}. We expect this to have interesting ramifications
for the nature of the superconducting state if it microscopically coexists with the coplanar spin-density wave.

Another advantage of the low-energy effective model is that it is able to capture the anisotropy of the $s$-wave superconducting
state with a minimal number of  $\bk${\em-independent} parameters. The anisotropy of the quasi-particle gap on the electron pockets, and consequently their
nodal structure, depends on the ratio of $\Delta_{\bM 1}$ and $\Delta_{\bM 3}$, as seen in Fig.\ \ref {FigNodalPlot}. This
holds even when the
spin-orbit coupling is ignored. The spin-orbit coupling introduces an additional parameter, $\Delta_{\bM t}$, corresponding to
the pairing strength in the spin triplet channel which is necessarily induced. The spin-orbit also removes the degeneracies
at the Fermi surface crossings, thereby splitting them. The resulting quasi-particle gap is therefore unambiguously
four-fold periodic, see Fig.\ \ref {FigTriplealpha}. A similar conclusion was obtained in Ref.\ \onlinecite{KhodasPRB2012}
where a phenomenological {\em spin-independent} hybridization parameter, $\lambda (\varphi)$, was included. We
see that, in the absence of spin-orbit coupling, the two Fermi surfaces are hybridized only if the $P4/nmm$ lattice symmetry is broken.
In the presence of such symmetry, the hybridization parameter necessarily arises from the spin-orbit coupling,
and its detailed spin-structure is captured within our approach.

On the hole pockets, a single $A_{1g}$ spin singlet pairing parameter, $\Delta_\Gamma$, yields isotropic quasi-particle gap.
In the presence of the spin-orbit coupling, the spin triplet, $\bk$-{\em independent}, pairing, parameterized by $\Delta_{\Gamma t}$, is induced.
The addition of this term leads to the gap anisotropy on the hole Fermi surfaces. In this case, we find the anisotropy to have a different
strength on the two hole pockets. This result may be tested using spin polarized ARPES.

\section {Acknowledgments}

We are grateful for useful discussions with A.\ V.\  Chubukov and V.\ Stanev. This work was supported by the NSF CAREER award under
Grant No. DMR-0955561, NSF Cooperative Agreement No. DMR-0654118, and the State of Florida.

\appendix

\section {Eigenvalues for Bogolyubov-deGennes Hamiltonians}
\label{AppBdGeigenvalues}

In Section \ref {SecSuperconductivity}, the spectrum of superconducting states was determined from the eigenvalues of
the Bogolyubov-deGennes Hamiltonian. Once spin-orbit coupling is included at either the $\Gamma$ point or the $\bM$ point,
such Hamiltonian matrix can be casted in the form
\be
  H_{\rm BdG} = \sum_\bk \Psi^\dagger (\bk) \Big \lbrack \tau_3 \otimes \left ( A_\bk + \vec B_\bk \cdot \vec \sigma \right ) \nonumber \\
    +\tau_1 \otimes \left ( C_\bk + \vec D_\bk \cdot \vec \sigma \right ) \Big \rbrack \Psi (\bk), \label{HBdGappendix}
\ee
where for each $\bk$, $A_\bk$ and $C_\bk$ are scalars, while $\vec B_\bk$ and $\vec D_\bk$ are vectors ``dotted'' into
the three Pauli matrices $\vec{\sigma}=(\sigma_1,\sigma_2,\sigma_3)$.
Here we demonstrate a simple method for obtaining the eigenvalues for such a 4$\times$4 matrix Hamiltonian.

Because the Eq.\ \eqref {HBdGappendix} anticommutes with $\tau_2 \otimes \bbone$, for each eigenstate at $\bk$
with an eigenvalue $E_{\bk}$, there is another eigenstate at $\bk$ with the eigenvalue $-E_{\bk}$. Therefore, solving
the secular polynomial reduces to finding zeros of a quadratic function.

Squaring the matrix in Eq.\ \eqref {HBdGappendix}, and using the well-known commutation relation $\left[\sigma_a,\sigma_b\right]=2i\epsilon_{abc}\sigma_c$, gives
\begin {align}
  \hat H^2 =& \bbone_4\left ( A_\bk^2 + \vec B_\bk^2 + C_\bk^2 + \vec D_\bk^2 \right )  \nonumber \\
  & \ + \bbone \otimes \vec \sigma \cdot \left ( 2A_\bk \vec B_\bk + 2C_\bk \vec D_\bk \right ) \cdot  \nonumber \\
  & \ - 2 \tau_2 \otimes \vec \sigma\cdot \left ( \vec B_\bk \times \vec D_\bk \right ). \label{H2appendix}
\end {align}
Choosing a basis in which $\tau_2$ is diagonal, the eigenvalues of the matrix in Eq.\ \eqref {H2appendix}, $E^2_{\bk}$,  can be readily read off.
Because $\vec B_\bk \times \vec D_\bk$ is orthogonal to both $\vec B_{\bk}$ and to $\vec D_{\bk}$, the result is
independent of whether the eigenvalue of $\tau_2$ is $+1$ or $-1$. We thus find
\begin {align}
  E_\bk^2 =& A_\bk^2 + \vec B_\bk ^2 + C_\bk^2 + \vec D_\bk^2  \nonumber \\
    & \pm 2\sqrt{ \left(A_\bk \vec B_\bk + C_\bk\vec D_\bk\right)^2+ {\vec B}^2_\bk \vec D^2_\bk -\left(\vec B_\bk\cdot \vec D_\bk\right)^2}. \label{HBdGspectrumappendix}
\end {align}
Therefore, the four eigenvalues are $\pm \sqrt{E_\bk^2}$ with uncorrelated $\pm$ signs inside, and in front of, the overall square root.

\section {Group $\bP_\bM$ and the construction of its irreducible representations}
\label{AppPM}

Since the space group of iron-pnictides lattice symmetries, $P4/nmm$, is non-symmorphic, there is a particular procedure, due to Herring,
for the construction of the space group irreducible representations at the Brillouin zone boundary \cite {Herring1942}.
We follow the procedure as illustrated in Ref.\ \onlinecite {InuiTanabeOnodera}.

The main idea is to take the subgroup of `even' translations,
\be
  {\mathcal T}_{\bM} = \left \{ \left \{ e | \bt \right \} | \exp (i \bM \cdot \bt) = 1 \right \}, \label{TM}
\ee
and decompose the space group into cosets
\be
  {\mathcal G} = R_1 {\mathcal T}_{\bM} + R_2 {\mathcal T}_{\bM} + \ldots. \label{GdecomposeM}
\ee
Since ${\mathcal T}_\bM$ is an invariant group of ${\mathcal G}$, the cosets form factor group
$\bP_\bM = {\mathcal G} / {\mathcal T}_{\bM}$.  The odd translation, $\{ e | 10 \}$, is among the
elements of $\bP_\bM$, and for any irreducible representation at $\bM$, this symmetry operation must be represented by
\be
  D (\{ e | 10 \}) = e^{i a \bM \cdot \hat {\bf x}} = - \bbone. \label{DM10}
\ee
The irreducible representations at the $\bM$-point are therefore all irreducible representations of $\bP_\bM$
for which Eq.\ \eqref {DM10} holds. All the other irreducible representations of $\bP_\bM$ are unphysical and should be disregarded.
Our task here is to identify group $\bP_\bM$ and select its physical irreducible representations.

The group $\bP_\bM$ has 32 elements; in Seitz notation these are 16 operations, $\{ g | \btau \}$, already present in $\bP_\Gamma$,
and an additional 16 operations obtained by multiplying the 16 elements of $\bP_\Gamma$ by an `odd' translation: $\{ e | 10 \} \{ g | \btau \} = \{ g | 10 + \btau \}$.
Note that the multiplications rules are not the same as in $\bP_\Gamma$, as the two symmetry operations $\{ g | \bt_1 \}$ and $\{ g | \bt_2 \}$
are equivalent in $\bP_\bM$ only if $\exp \left (i \bM \cdot (\bt_1 - \bt_2) \right ) = 1$. For example,
$\{ \sigma^x | 00 \} \{ i | \thalf \thalf \} = \{ C_2^x | \bar \thalf \thalf \}$.
We found that $\bP_\bM$ is not isomorphic to any crystallographic point group in three dimensions \cite{Tinkham}.
Therefore we were left with the task of deriving the irreducible representations for $\bP_\bM$.

In constructing irreducible representations of an unknown group, a good head start is gained by finding the largest possible invariant subgroup
whose irreducible representations we already know. In the case of $\bP_\bM$, we notice that four elements of the group, $\{ \sigma^X | \thalf \thalf \}$,
$\{ \sigma^Y | \thalf \thalf \}$, $\{ \sigma^z | \thalf \thalf \}$, and $\{ e | 10 \}$ commute with each other,
and at the same time, each of these elements combined with the identity operation forms a cyclic subgroup isomorphic
to ${\bf C}_2$. Therefore, a set of operations obtained by successive action of these four elements forms a group
which is a direct product of four cyclic groups, ${\mathcal H} \cong ({\bf C}_2)^{\otimes 4}$, and is an invariant
subgroup of ${\mathcal G}$. This subgroup has 16 elements and is Abelian. Each cyclic group ${\bf C}_2$ has two one-dimensional
irreducible representations, one even and one odd. The irreducible representations of ${\mathcal H}$ are then obtained by multiplying the
irreducible representations of each cyclic subgroup. There are 16 one-dimensional representations given by
\begin {align}
  D^{(\xi \eta \zeta \tau)} & \left ( \{ \sigma^X | \thalf \thalf \}^m \{ \sigma^Y | \thalf \thalf \}^n \{ \sigma^z | \thalf \thalf \}^p
    \{ e | 10 \}^q \right ) \nonumber\\
    &= \xi^m \eta^n \zeta^p \tau^q, \label{DH}
\end{align}
where each $\xi$, $\eta$, $\zeta$, and $\tau$ can take either $+1$ or $-1$ value corresponding to a representation that is
even/odd under $\{ \sigma^X | \thalf \thalf \}$, $\{ \sigma^Y | \thalf \thalf \}$, $\{ \sigma^z | \thalf \thalf \}$, and
$\{ e | 10 \}$, respectively. Here, $m$, $n$, $p$, and $q$ are integers, although it is sufficient to consider values
$0$ and $1$ to find all the elements of ${\mathcal H}$. For brevity, we use only the sign, $+$ or $-$, and drop $1$'s
in the designation of the irreducible representations of ${\mathcal H}$.

The next step is finding the induced representations. We write ${\mathcal G} = R_1 {\mathcal H} + R_2 {\mathcal H}$,
where $R_1 = \{ e | 00 \}$, while we choose $R_2 = \{ \sigma^x | 00 \}$ for simplicity. 
Taking an irreducible representation of ${\mathcal H}$, $D^{(\xi \eta \zeta \tau )}$, the
induced representation, $D^{(\xi \eta \zeta \tau)} \uparrow {\mathcal G}$ is defined as the $2 \times 2$ matrix
\begin {align}
  \left ( D^{(\xi \eta \zeta \tau )}  \uparrow {\mathcal G}\right)&_{ij} (P) \nonumber \\
  &= \left \lbrace \begin {array}{l c l}     D^{(\xi \eta \zeta \tau )} \left ( R_i^{-1} P R_j \right ), & \ & R_i^{-1} P R_j \in {\mathcal H} \\
    0, & \ & R_i^{-1} P R_j \notin {\mathcal H}
    \end {array} \right . , \label{Dinduced}
\end {align}
where $P \in {\mathcal G}$. The induced representations are, in general, reducible.

To determine whether given $D^{(\xi \eta \zeta \tau)} \uparrow {\mathcal G}$ is reducible or not, for each $R_i$, we
create an irreducible representation, $D^{(\xi \eta \zeta \tau), (i)}$, defined as
\be
  D^{(\xi \eta \zeta \tau), (i)} (S) = D^{(\xi \eta \zeta \tau)} \left ( R_i^{-1} S R_i \right ), \label{Di}
\ee
with $S \in {\mathcal H}$. These irreducible representations are then separated into classes of equivalent representations. However, since in our problem
there are only two irreducible representations  created this way, we can either have the two of them being equivalent and belonging to the same class,
or being inequivalent and each defining its own class. If the two irreducible representations defined in Eq.\ \eqref {Di} are inequivalent,
$D^{(\xi \eta \zeta \tau)} \uparrow {\mathcal G}$ is an irreducible 2-dimensional representation of ${\mathcal G}$ and
we name it $E^{(\xi \eta \zeta \tau)}$. Otherwise,
$D^{(\xi \eta \zeta \tau)} \uparrow {\mathcal G}$ is reducible. In such a case, the irreducible representations are constructed
by finding the unitary matrix $U$ which connects the two equivalent irreducible representations, $A_1$ and $A_2$,
\be
  D^{(\xi \eta \zeta \tau), (2)} (S) = U^{-1} D^{(\xi \eta \zeta \tau), (1)} (S) U. \label{DisUDU}
\ee
Since all $D^{(\xi \eta \zeta \tau)}$'s are one-dimensional, the unitary matrix is just a scalar of modulus unity
$U = \exp ( i \varphi )$, and  we choose it to be $1$. The two irreducible representations obtained from $D^{(\xi \eta \zeta \tau)}$ are
then given by
\be
  A_1^{(\xi \eta \zeta \tau)} (S) &=& \phantom {+} D^{(\xi \eta \zeta \tau)} (S),\nonumber\\
  A_1^{(\xi \eta \zeta \tau)} (R_2 S) &=& + D^{(\xi \eta \zeta \tau)} (S), \label{Arep} \\
  A_2^{(\xi \eta \zeta \tau)} (S) &=& \phantom {-}D^{(\xi \eta \zeta \tau)} (S),\nonumber\\
  A_2^{(\xi \eta \zeta \tau)} (R_2 S) &=& - D^{(\xi \eta \zeta \tau)} (S). \label{Brep}
\ee

For our problem we need first to find how $\{ \sigma^x | 00 \}$ conjugates the elements of ${\mathcal H}$. Using the Seitz
product rule in  $\bP_\bM$, we find
\begin {align}
  \{ \sigma^x | 00 \}^{-1} \{ \sigma^X |\thalf \thalf \} \{ \sigma^x | 00 \} =& \{ \sigma^Y | \bar \thalf \thalf \}
    = \{ e | 10 \} \{ \sigma^Y | \thalf \thalf \}, \label{sigmaXconj} \\
  \{ \sigma^x | 00 \}^{-1} \{ \sigma^Y | \thalf \thalf \} \{ \sigma^x | 00 \} =& \{ \sigma^X | \bar \thalf \thalf \}
    = \{ e | 10 \} \{ \sigma^X | \thalf \thalf \}, \label{sigmaYconj} \\
  \{ \sigma^x | 00 \}^{-1} \{ \sigma^z | \thalf \thalf \} \{ \sigma^x | 00 \} =& \{ \sigma^z | \bar \thalf \thalf \}
    = \{ e | 10 \} \{ \sigma^z | \thalf \thalf \}, \label{sigmazconj} \\
  \{ \sigma^x | 00 \}^{-1} \{ e | 10 \} \{ \sigma^x | 00 \} =& \{ e | 10 \}. \label{etconj}
\end {align}
To derive these it is useful to notice that $\{ \sigma^x | 00 \}^{-1} = \{ \sigma^x | 00 \}$.
For an arbitrary element of ${\mathcal H}$ we find
\bw
\be
  D^{(\xi \eta \zeta \tau)} \left ( \{ \sigma^x | 00 \}^{-1} \{ \sigma^X | \thalf \thalf \}^m \{ \sigma^Y | \thalf \thalf \}^n
    \{ \sigma^z | \thalf \thalf \}^p \{ e | 10 \}^q \{ \sigma^x | 00 \} \right ) &=& \nonumber \\
  D^{(\xi \eta \zeta \tau)} \left ( \{ \sigma^x | 00 \}^{-1} \{ \sigma^X | \thalf \thalf \}^m \{ \sigma^x | 00 \}
    \{ \sigma^x | 00 \}^{-1} \{ \sigma^Y | \thalf \thalf \}^n \{ \sigma^x | 00 \} \times \right . && \nonumber \\
  \left .\{ \sigma^x | 00 \}^{-1} \{ \sigma^z | \thalf \thalf \}^p \{ \sigma^x | 00 \} \{ \sigma^x | 00 \}^{-1} \{ e | 10 \}^q
    \{ \sigma^x | 00 \} \right ) &=& \nonumber \\
  D^{(\xi \eta \zeta \tau)} \left ( \left ( \{ e | 10 \} \{ \sigma^Y | \thalf \thalf \} \right )^m
    \left ( \{ e | 10 \} \{ \sigma^X | \thalf \thalf \} \right )^n \left ( \{ e | 10 \} \{ \sigma^z | \thalf \thalf \} \right )^p
    \{ e | 10 \}^q \right ) &=& \nonumber  \\
  D^{(\xi \eta \zeta \tau)} \left ( \{ \sigma^X | \thalf \thalf \}^n \{ \sigma^Y | \thalf \thalf \}^m \{ \sigma^z | \thalf \thalf \}^p
    \{ e | 10 \}^{m+n+p+q} \right ) &=& \xi^n \eta^m \zeta^p \tau^{m+n+p+q}. \label{Dconj}
\ee
\ew

This equation defines irreducible representation $D^{(\xi \eta \zeta \tau), (2)}$. This representation is equivalent to
$D^{(\xi \eta \zeta \tau), (1)} = D^{(\xi \eta \zeta \tau)}$, provided that $\xi^m \eta^n \zeta^p \tau^{q} =
\xi^n \eta^m \zeta^p \tau^{m+n+p+q}$ for any $m$, $n$, $p$, and $q$ integer.
Arranging this equation and bearing in mind
that $\xi$, $\eta$, $\zeta$, and $\tau$ only take values $\pm 1$, we find that the two irreducible representations are equivalent if
\be
 \left ( \xi \eta \tau \right )^{m+n} \tau^p = const, \label{Dequivalence}
\ee
for any integer $m$, $n$, and $p$ (also $q$, but it drops out of the condition). The left hand side of Eq.\ \eqref {Dequivalence}
is $p$-independent only if $\tau=+1$. Similarly, it is $m$ and $n$ independent provided that
$\xi \eta= +1$. There are, therefore, four irreducible representations of ${\mathcal H}$ that each induce two one-dimensional irreducible representations of ${\mathcal G}$:
\be
  \left ( D^{(++++)} \uparrow {\mathcal G} \right ) &=& A_1^{(++++)} \oplus A_2^{(++++)}, \nonumber \\
  \left ( D^{(++-+)} \uparrow {\mathcal G} \right ) &=& A_1^{(++-+)} \oplus A_2^{(++-+)}, \nonumber \\
  \left ( D^{(--++)} \uparrow {\mathcal G} \right ) &=& A_1^{(--++)} \oplus A_2^{(--++)}, \nonumber \\
  \left ( D^{(---+)} \uparrow {\mathcal G} \right ) &=& A_1^{(---+)} \oplus A_2^{(---+)}. \label{DintoAB}
\ee
These irreducible representations of $\bP_\bM$ are defined as
\begin {align}
  & A_1^{(\xi \eta \zeta \tau)} \left ( \{ \sigma^X | \thalf \thalf \}^m \{ \sigma^Y | \thalf \thalf \}^n \{ \sigma^z | \thalf \thalf \}^p \{ e | 10 \}^q \right ) \nonumber \\
  &= A_1^{(\xi \eta \zeta \tau)} \left ( \{ \sigma^x | 00 \} \{ \sigma^X | \thalf \thalf \}^m \{ \sigma^Y | \thalf \thalf \}^n \{ \sigma^z | \thalf \thalf \}^p \{ e | 10 \}^q \right ) \nonumber \\
  &= A_2^{(\xi \eta \zeta \tau)} \left ( \{ \sigma^X | \thalf \thalf \}^m \{ \sigma^Y | \thalf \thalf \}^n \{ \sigma^z | \thalf \thalf \}^p \{ e | 10 \}^q \right ) \nonumber \\
  &= - A_2^{(\xi \eta \zeta \tau)} \left ( \{ \sigma^x | 00 \} \{ \sigma^X | \thalf \thalf \}^m \{ \sigma^Y | \thalf \thalf \}^n \{ \sigma^z | \thalf \thalf \}^p \{ e | 10 \}^q \right ) \nonumber \\
  &= \xi^m \eta^n \zeta^p \tau^q, \label{A1A2definition}
\end {align}
where the only allowed combinations of $\xi$, $\eta$, $\zeta$, and $\tau$ are those  present in Eqs.\ \eqref {DintoAB}.

The remaining twelve irreducible representations of ${\mathcal H}$ induce two-dimensional irreducible representations of ${\mathcal G}$.
These come in six pairs of equivalent irreducible representations:
\be
  E^{(-++-)} \cong E^{(-+--)}, && E^{(+-+-)} \cong E^{(+---)}, \nonumber \\
  E^{(++--)} \cong E^{(--+-)}, && E^{(+++-)} \cong E^{(----)}, \nonumber \\
  E^{(+-++)} \cong E^{(-+++)}, && E^{(+--+)} \cong E^{(-+-+)}. \label{DintoE}
\ee

The matrices for the induced representations, i.e., the representations in Eq.\ \eqref {DintoE}, are
\begin {align}
  E^{(\xi \eta \zeta \tau)} & \left ( \{ \sigma^X | \thalf \thalf \}^m \{ \sigma^Y | \thalf \thalf \}^n
    \{ \sigma^z | \thalf \thalf \}^p \{ e | 10 \}^q \right )  \nonumber \\
  &= \left ( \begin {array}{c c} \xi^m \eta^n \zeta^p \tau^q & 0 \\
    0 & \xi^n \eta^m \zeta^p \tau^{m+n+p+q} \end {array} \right ), \label{Es1} \\
  E^{(\xi \eta \zeta \tau)} &\left ( \{ \sigma^x | 00 \} \{ \sigma^X | \thalf \thalf \}^m \{ \sigma^Y | \thalf \thalf \}^n
    \{ \sigma^z | \thalf \thalf \}^p \{ e | 10 \}^q \right ) \nonumber \\
  &= \left ( \begin {array}{c c} 0 & \xi^n \eta^m \zeta^p \tau^{m+n+p+q} \\
    \xi^m \eta^n \zeta^p \tau^q & 0 \end {array} \right ). \label{Es2}
\end {align}
The characters are
\begin {align}
  \chi^{(\xi \eta \zeta \tau)} & \left ( \{ \sigma^X | \thalf \thalf \}^m \{ \sigma^Y | \thalf \thalf \}^n
    \{ \sigma^z | \thalf \thalf \}^p \{ e | 10 \}^q \right ) \nonumber \\
  &= \xi^m \eta^n \zeta^p \tau^q +  \xi^n \eta^m \zeta^p \tau^{m+n+p+q} \nonumber \\
  &= \xi^m \eta^n \zeta^p \tau^q \left \lbrack 1 + \left (\xi \eta \tau \right )^{m+n} \tau^p \right \rbrack, \label{chiE1} \\
  \chi^{(\xi \eta \zeta \tau)} & \left ( \{ \sigma^x | 00 \} \{ \sigma^X | \thalf \thalf \}^m \{ \sigma^Y | \thalf \thalf \}^n
    \{ \sigma^z | \thalf \thalf \}^p \{ e | 10 \}^q \right  ) = 0. \label{chiE2}
\end {align}
Hereby we have finished the construction of the IR's for the unknown group $\bP_\bM$.

From Eqs.\ \eqref {A1A2definition}, it follows that for all one-dimensional irreducible representations,
\be
  A_1^{(\xi \eta \zeta \tau)} ( \{ e | 10 \} ) = A_2^{(\xi \eta \zeta \tau)} ( \{ e | 10 \} ) = +1.
\ee
Hence, all one-dimensional irreducible representations of $\bP_\bM$ are unphysical. Similarly,
we use Eqs.\ \eqref {Es1}-\eqref {Es2} to determine the representation for the odd translation to be
\be
  E^{(\xi \eta \zeta \tau)} (\{ e | 10 \}) = (-1)^\tau \bbone
\ee
in two-dimensional irreducible representations in Eqs.\ \eqref{DintoE}. Therefore, the irreducible representations for
which $\tau = -1$, i.e., those in the last line of Eq.\ \eqref {DintoE},
are unphysical. The remaining four irreducible representations in Eqs.\ \eqref {DintoE} are physical and we relabel them
\be
  E_{\bM 1} = E^{(-+--)}, && E_{\bM 2} = E^{(+---)}, \nonumber \\
  E_{\bM 3} = E^{(++--)}, && E_{\bM 4} = E^{(----)}. \label{EMdefinition}
\ee
Let us just note here that the first member of each $E_{\bM}$ doublet  is odd, while the second member is always even
under $\{ \sigma^z | \thalf \thalf \}$. There are four combinations for the parity under $\{ \sigma^X | \thalf \thalf \}$ and $\{ \sigma^Y | \thalf \thalf \}$
and they are all present in Eq.\ \eqref {EMdefinition}. Per construction, all the physical irreducible representations at the $\bM$-point
are odd under $\{ e | 10 \}$. Finally, the $\{ \sigma^x | 00 \}$ mirror is represented by $\sigma_1$ Pauli matrix in each
$E_{\bM}$ irreducible representations, i.e., it swaps the two components.

The irreducible representation table, Table \ref{tab:M generators irreps}, as well as the irreducible representation multiplication
tables, Tables \ref {TableExE} and \ref {TableExD4h} in the main text, follow directly from Eqs.\ \eqref {Es1} and \eqref {Es2}.

The unphysical irreducible representations we construct here are, in fact, not entirely unrelated to the space group. Noticing
that for all of them, $D (\{ e | 10\}) = + \bbone$, it follows that these irreducible representations are physical at the $\Gamma$-point.
Each irreducible representation of ${\bf D}_{4h}$ is, therefore, equivalent to one of the (unphysical at $\bM$) irreducible representations
of $\bP_\bM$ derived here:
\be
  A_{1g} = A_1^{(++++)}, && A_{1u} = A_2^{(---+)}, \nonumber \\
  A_{2g} = A_2^{(--++)}, && A_{2u} = A_1^{(++-+)}, \nonumber \\
  B_{1g} = A_1^{(--++)}, && B_{1u} = A_2^{(++-+)}, \nonumber \\
  B_{2g} = A_2^{(++++)}, && B_{2u} = A_1^{(---+)}, \nonumber \\
  E_g = E^{(+--+)}, && E_u = E^{(-+++)}. \label{D4h2IRPM}
\ee

\section {Particle-hole bilinears in the low-energy effective theory}
\label{AppGammas}

\begin {table}[ht]
\begin {tabular}{| >$c<$ | >$c<$ | >$c<$ | }
\hline
 i & \mbox {I.R.}^{\rm TRS} &  \Gamma_{i, 1}^{(m)},  \Gamma_{i, 2}^{(m)}, \ldots  \Gamma_{i, M_i}^{(m)} \\
\hline
  1 & A_{1g}^+ & \lambda_\Gamma \bbone_2, \lambda_\bM \bbone_2, \lambda_\bM \sigma_3 \\
  2 & A_{1u}^- & \lambda_2 \frac {\bbone_2 - \sigma_3}{\sqrt 2} \\
  3 & A_{2g}^- & \lambda_\Gamma \sigma_2 \\
  4 & A_{2u}^+ & \lambda_1 \frac {\bbone_2 + \sigma_3}{\sqrt 2} \\
  5 & B_{1g}^+ & \lambda_\Gamma \sigma_1 \\
  6 & B_{1u}^- & \lambda_2 \frac{\bbone_2 + \sigma_3}{\sqrt 2}  \\
  7 & B_{2g}^+ & \lambda_\Gamma \sigma_3, \lambda_3 \bbone_2, \lambda_3 \sigma_3 \\
  8 & B_{2u}^+ & \lambda_1 \frac{\bbone_2 - \sigma_3}{\sqrt 2} \\
  9 & E_g^+ & \left ( \frac {\lambda_1 \sigma_1 + \lambda_2 \sigma_2}{\sqrt 2}, \frac {- \lambda_1 \sigma_1 + \lambda_2 \sigma_2}{\sqrt 2} \right ) \\
  10 & E_u^+ & \left ( \frac {-\lambda_3 + \lambda_\bM}{\sqrt 2} \sigma_1, \frac {\lambda_3 + \lambda_\bM}{\sqrt 2} \sigma_1 \right ) \\
  11 & E_g^- & \left (  \frac {- \lambda_1 \sigma_2 + \lambda_2 \sigma_1}{\sqrt 2}, \frac {\lambda_1 \sigma_2 + \lambda_2 \sigma_1}{\sqrt 2} \right ) \\
  12 & E_u^- & \left ( \frac { - \lambda_3 + \lambda_\bM}{\sqrt 2} \sigma_2, \frac {\lambda_3 + \lambda_\bM}{\sqrt 2} \sigma_2 \right ) \\
  13 & E_{\bM 1}^+ &  \left ( \lambda_4 \frac {\bbone_2 - \sigma_3}{\sqrt 2}, \frac {\lambda_6 \sigma_1 + \lambda_7 \sigma_2}{\sqrt 2} \right ) \\
  14 & E_{\bM 1}^- & \left ( \lambda_5 \frac {\bbone_2 - \sigma_3}{\sqrt 2}, \frac {-\lambda_6 \sigma_2 + \lambda_7 \sigma_1}{\sqrt 2} \right ) \\
  15 & E_{\bM 2}^+ & \left ( \frac {\lambda_4 \sigma_1 + \lambda_5 \sigma_2}{\sqrt 2}, \lambda_6 \frac {\bbone_2 - \sigma_3}{\sqrt 2} \right ) \\
  16 & E_{\bM 2}^- & \left ( \frac {-\lambda_4 \sigma_2 + \lambda_5 \sigma_1}{\sqrt 2}, \lambda_7 \frac {\bbone_2 - \sigma_3}{\sqrt 2} \right ) \\
  17 & E_{\bM 3}^+ & \left ( \lambda_4 \frac { \bbone_2 + \sigma_3}{\sqrt 2}, \frac {\lambda_6 \sigma_1 - \lambda_7 \sigma_2}{\sqrt 2} \right ) \\
  18 & E_{\bM 3}^- & \left ( \lambda_5 \frac {\bbone_2 + \sigma_3}{\sqrt 2},  \frac {\lambda_6 \sigma_2 + \lambda_7 \sigma_1}{\sqrt 2} \right ) \\
  19 & E_{\bM 4}^+ & \left (  \frac {\lambda_4 \sigma_1 - \lambda_5 \sigma_2}{\sqrt 2},  \lambda_6 \frac { \bbone_2 + \sigma_3}{\sqrt 2} \right ) \\
  20 & E_{\bM 4}^- & \left ( \frac {\lambda_4 \sigma_2 + \lambda_5 \sigma_1}{\sqrt 2}, \lambda_7 \frac {\bbone_2 + \sigma_3}{\sqrt 2} \right ) \\
\hline
\end {tabular}
\end {table}
Here we list the matrices $\Gamma_{i, j}^{(m)}$ according to their symmetry  properties.
The eight $\lambda_j$'s are ``Gell-Mann'' matrices defined as
\be
  \lambda_1 = \begin{pmatrix} 0 & 1 & 0 \\ 1 & 0 & 0 \\ 0 & 0 & 0 \end{pmatrix}, &~&  \lambda_2 = \begin{pmatrix} 0 & -i & 0 \\ i & 0 & 0 \\ 0 & 0 & 0 \end{pmatrix}, \\
  \lambda_3 = \begin{pmatrix} 1 & 0 & 0 \\ 0 & -1 & 0 \\ 0 & 0 & 0 \end{pmatrix}, && \lambda_4 = \begin{pmatrix} 0 & 0 & 1 \\ 0 & 0 & 0 \\ 1 & 0 & 0 \end{pmatrix}, \\
  \lambda_5 = \begin{pmatrix} 0 & 0 & -i \\ 0 & 0 & 0 \\ i & 0 & 0 \end{pmatrix}, && \lambda_6 = \begin{pmatrix} 0 & 0 & 0 \\ 0 & 0 & 1 \\ 0 & 1 & 0 \end{pmatrix}, \\
  \lambda_7 = \begin{pmatrix} 0 & 0 & 0 \\ 0 & 0 & -i \\ 0 & i & 0 \end{pmatrix}, && \lambda_8 = \frac{1}{\sqrt{3}} \begin{pmatrix} 1 & 0 & 0 \\ 0 & 1 & 0 \\ 0 & 0 & -2 \end{pmatrix}.
\ee
We also used their combinations
\be
  \lambda_\bM &=& \frac 1{\sqrt {3}} \lambda_8 + \sqrt{\frac 2 3} \bbone_3 = \diag (1,1,0), \label{lambdaM} \\
    \lambda_\Gamma &=& - \sqrt {\frac 2 3} \lambda_8 + \frac 1{\sqrt 3} \bbone_3 = \diag (0,0,\sqrt 2). \label{lambdaGamma}
\ee

\section {The classification of all pairing terms in the low-energy effective theory} \label{AppSCpairs}

In Section \ref{SecSuperconductivity}, general forms for pairing bilinears  in the low-energy effective theory are
presented in Eqs.\ \eqref{DeltaSCM} and \eqref{DeltaSCMtriplet}. The focus of that section is, however, on
the $A_{1g}$ pairing terms. Here, we tabulate, according to their symmetry properties, all other spin singlet  pairing terms
that may appear in the low-energy effective theory.

A general spin singlet pairing term in our low-energy effective model has form
\be
  \sum_{\bk} \psi_\downarrow^T (-\bk) {\mathcal M} \psi_\uparrow (\bk),
\ee
The symmetry properties of the pairing term are determined by the matrix
\be
  {\mathcal M} = \left \lbrack \begin {array} {c c} {\mathcal M}_\Gamma & {\mathcal M}_{\rm PDW}  \\  {\mathcal M}_{\rm PDW}^T & {\mathcal M}_\bM \end {array} \right \rbrack. \label{mathcalM}
\ee

The $2\times 2$ matrix ${\mathcal M}_\Gamma$ corresponds to the pairs formed by the hole band  states. Its elements
are classified according to
\be
  E_g \otimes E_g &=& A_{1g} \oplus A_{2g} \oplus B_{1g} \oplus B_{2g}, \label{EgxEgAppendix}
\ee
where the orbital antisymmetric combination $A_{2g}$ should be dropped. Similarly, pairing of states in the  electron bands
is specified through the $4 \times 4$ matrix ${\mathcal M}_\bM$.  Its elements are classified according to
\be
  E_{\bM 1} \otimes E_{\bM 1} &=& A_{1g}  \oplus B_{2g} \oplus A_{2u} \oplus B_{1u}, \label{EM1xEM1Appendix}  \\
  E_{\bM 3} \otimes E_{\bM 3} &=& A_{1g}  \oplus B_{2g} \oplus A_{1u} \oplus B_{2u}, \label{EM3xEM3Appendix}  \\
  E_{\bM 1} \otimes E_{\bM 3} &=& E_g \oplus E_u, \label{EM1xEM3Appendix}
\ee
where we drop the antisymmetric orbital combinations, $A_{2u}$, and $A_{1u}$ in the first two equations respectively.

The $2\times 4$ matrix ${\mathcal M}_{\rm PDW}$ corresponds to pairing terms between a hole and an electron state. Such pairs
have momentum $\bM$, and their symmetry properties are given by one of the $E_{\bM}$ irreducible representations found in the decomposition
\be
  E_g \otimes E_{\bM 1} &=& E_{\bM 3} \otimes E_{\bM 4}, \label{EgxEM1Appendix} \\
  E_g \otimes E_{\bM 3} &=& E_{\bM 1} \otimes E_{\bM 2}. \label{EgxEM3Appendix}
\ee

The pairing terms with no overall momentum, classified according to their symmetry properties, are defined by the following ${\mathcal M}$
matrices:
\begin {align}
  A_{1g}:& \left \lbrack \begin {array}{c c} 1 & 0 \\ 0 & 1 \end {array} \right \rbrack, \quad
    \left \lbrack \begin {array}{c c c c} 1 & 0 & 0 & 0 \\ 0 & 0 & 0 & 0 \\ 0 & 0 & 1 & 0 \\ 0 & 0 & 0 & 0 \end {array} \right \rbrack, \quad
    \left \lbrack \begin {array}{c c c c}  0 & 0 & 0 & 0 \\ 0 & 1 & 0 & 0 \\ 0 & 0 & 0 & 0 \\ 0 & 0 & 0 & 1 \end {array} \right \rbrack, \label{SCsinglet_A1g} \\
  B_{1g}:& \left \lbrack \begin {array}{c c} 0 & 1 \\ 1 & 0 \end {array} \right \rbrack, \label{SCsinglet_B1g} \\
  B_{2g}:& \left \lbrack \begin {array}{c c} 1 & 0 \\ 0 & -1 \end {array} \right \rbrack, \quad
    \left \lbrack \begin {array}{c c c c} 1 & 0 & 0 & 0 \\ 0 & 0 & 0 & 0 \\ 0 & 0 & -1 & 0 \\ 0 & 0 & 0 & 0 \end {array} \right \rbrack, \quad
    \left \lbrack \begin {array}{c c c c}  0 & 0 & 0 & 0 \\ 0 & 1 & 0 & 0 \\ 0 & 0 & 0 & 0 \\ 0 & 0 & 0 & -1 \\\end {array} \right \rbrack, \label{SCsinglet_B2g} \\
  E_g:& \left ( \left \lbrack \begin {array}{c c c c}  0 & 0 & 0 & 0 \\ 0 & 0 & 1 & 0 \\ 0 & 1 & 0 & 0  \\ 0 & 0 & 0 & 0 \\ \end {array} \right \rbrack,
    \left \lbrack \begin {array}{c c c c}  0 & 0 & 0 & -1 \\ 0 & 0 & 0 & 0 \\ 0 & 0 & 0 & 0  \\ -1 & 0 & 0 & 0 \\ \end {array} \right \rbrack  \right ), \label{SCsinglet_Eg} \\
  A_{2u}:& \left \lbrack \begin {array}{c c c c}  0 & 0 & 1 & 0 \\ 0 & 0 & 0 & 0 \\ 1 & 0 & 0 & 0  \\ 0 & 0 & 0 & 0 \\ \end {array} \right \rbrack,   \label{SCsinglet_A2u} \\
  B_{2u}:& \left \lbrack \begin {array}{c c c c}  0 & 0 & 0 & 0 \\ 0 & 0 & 0 & 1 \\ 0 & 0 & 0 & 0  \\ 0 & 1 & 0 & 0 \\ \end {array} \right \rbrack,   \label{SCsinglet_B2u} \\
  E_u:& \left ( \left \lbrack \begin {array}{c c c c}  0 & 0 & 0 & 0 \\ 0 & 0 & 0 & 0 \\ 0 & 0 & 0 & 1  \\ 0 & 0 & 1 & 0 \\ \end {array} \right \rbrack,
    \left \lbrack \begin {array}{c c c c}  0 & 1 & 0 & 0 \\ 1 & 0 & 0 & 0 \\ 0 & 0 & 0 & 0  \\ 0 & 0 & 0 & 0 \\ \end {array} \right \rbrack  \right ). \label{SCsinglet_Eu}
\end {align}
Any $2\times 2$ matrix should be substituted in Eq.\ \eqref {mathcalM} in place of ${\mathcal M}_\Gamma$ with the other block matrices
being identical to zero; any $4 \times 4$ matrix in this list should be substituted in Eq.\ \eqref {mathcalM} in place of ${\mathcal M}_\bM$
with the other block matrices substituted by blocks of zeros.

In Section \ref {SecSuperconductivity}, we discussed the  $A_{1g}$ pairing terms, Eq.\ \eqref {DeltaSCM}, and
their consequences on the low-energy effective model quasi-particle spectrum. Similarly, one could construct  $B_{2g}$
pairing terms which  would be described by three parameters, each multiplying one of the pairing terms following from
Eq.\ \eqref{SCsinglet_B2g}. The $B_{2g}$ superconducting order leads to a gap formation
on both hole and electron Fermi surfaces of the low-energy effective model. The symmetry properties of $B_{2g}$
dictate that quasi-particle gap is odd under mirror reflection operations $\{ \sigma^x | 00 \}$ and $\{ \sigma^y | 00 \}$,
implying that the gap on the hole Fermi surfaces  has nodes for $k_x=0$ or $k_y=0$.

For any other irreducible representation in the list of the spin singlet pairing terms with no overall momentum, there is always just a single term
present. The superconducting order corresponding to one of these irreducible representations would therefore produce
a quasi-particle gap on either the hole or the electron bands, but it would never affect the spectrum of both hole
and electron  excitations. This suggests that pairing into any of these channels is likely irrelevant for the iron based
superconductors.

Here we list the complete set of the symmetry adapted matrices ${\mathcal M}_{\rm PDW}$:
\begin {align}
    E_{\bM 1}:& \left ( \left \lbrack \begin {array}{c c c c} 0 & 0 & 0 & 1 \\ 0 & 0 & 0 & 0 \end {array} \right \rbrack,
      \left \lbrack \begin {array}{c c c c} 0 & 0 & 0 & 0 \\ 0 & -1 & 0 & 0 \end {array} \right \rbrack \right ), \label{SCsinglet_FFLO_M1} \\
    E_{\bM 2}:& \left ( \left \lbrack \begin {array}{c c c c} 0 & 0 & 0 & 0 \\ 0 & 0 & 0 & 1 \end {array} \right \rbrack,
      \left \lbrack \begin {array}{c c c c} 0 & -1 & 0 & 0 \\ 0 & 0 & 0 & 0 \end {array} \right \rbrack \right ), \label{SCsinglet_FFLO_M2} \\
    E_{\bM 3}:& \left ( \left \lbrack \begin {array}{c c c c} 0 & 0 & 0 & 0 \\ 0 & 0 & 1 & 0 \end {array} \right \rbrack,
      \left \lbrack \begin {array}{c c c c} -1 & 0 & 0 & 0 \\ 0 & 0 & 0 & 0 \end {array} \right \rbrack \right ), \label{SCsinglet_FFLO_M3} \\
    E_{\bM 4}:& \left ( \left \lbrack \begin {array}{c c c c} 0 & 0 & 1 & 0 \\ 0 & 0 & 0 & 0 \end {array} \right \rbrack,
      \left \lbrack \begin {array}{c c c c} 0 & 0 & 0 & 0 \\ -1 & 0 & 0 & 0 \end {array} \right \rbrack \right ). \label{SCsinglet_FFLO_M4}
\end {align}
There is precisely one doublet of pair-density wave bilinears for each irreducible representation at the $\bM$-point. Therefore,
any spin singlet pair-density wave superconductivity in iron-based superconductors built upon the low-energy effective theory can be described by
a single two-component $\bk$-independent order parameter. An on-site pair-density wave located on iron sites corresponds to one
of the bilinears in Eq.\ \eqref {SCsinglet_FFLO_M4}; similarly, an on-site pair-density wave on pnictide sites is represented
by bilinears in Eq.\ \eqref {SCsinglet_FFLO_M2}.


\begin{thebibliography}{99}
\bibitem{note:blgHeff} We used a similar procedure previously in bilayer graphene, for details see Refs.\ \onlinecite {bilayer}.
\end{thebibliography}
\end{document}